\pdfoutput=1
\documentclass[conference, twocolumn]{IEEEtran}

\usepackage{listings}
\usepackage{graphicx}
\usepackage{subcaption}
\usepackage[table,xcdraw]{xcolor}
\usepackage{amsmath}
\usepackage{amssymb}
\usepackage{pifont}
\definecolor{mydarkblue}{rgb}{0,0.08,0.65}
\usepackage[colorlinks=true,
            linkcolor=mydarkblue,
            citecolor=mydarkblue,
            filecolor=mydarkblue,
            urlcolor=mydarkblue]{hyperref}
\usepackage{cleveref}
\usepackage{soul}
\usepackage[shortlabels]{enumitem}
\usepackage{url}
\usepackage{color}
\usepackage{tikz}
\usepackage{balance}
\usepackage{multirow}
\usepackage{comment}
\usepackage{wrapfig,lipsum,booktabs}
\usepackage[frozencache,cachedir=.]{minted}
\usepackage[symbol]{footmisc}
\usepackage{tablefootnote}
\usepackage{tabularx}
\usepackage{placeins}
\usepackage{algorithm}
\usepackage{algpseudocode}
\usepackage{cuted}
\usepackage{float}
\usepackage{caption}
\usepackage{multicol}
\usepackage{dblfloatfix}
\usepackage{natbib}

\usepackage{microtype}
\usepackage{xurl}
\usetikzlibrary{arrows.meta,calc,positioning}

\newcommand{\corr}{\textsuperscript{*}}

\definecolor{codegreen}{rgb}{0,0.6,0}
\definecolor{codegray}{rgb}{0.5,0.5,0.5}
\definecolor{codepurple}{rgb}{0.58,0,0.82}
\definecolor{backcolour}{rgb}{0.95,0.95,0.92}
\lstdefinestyle{mystyle}{
    backgroundcolor=\color{backcolour},
    commentstyle=\color{codegreen},
    keywordstyle=\color{magenta},
    numberstyle=\tiny\color{codegray},
    stringstyle=\color{codepurple},
    basicstyle=\ttfamily\footnotesize,
    breakatwhitespace=false,
    breaklines=true,
    captionpos=b,
    keepspaces=true,
    numbers=left,
    numbersep=5pt,
    showspaces=false,
    showstringspaces=false,
    showtabs=false,
    tabsize=2,
}
\makeatletter
\def\blfootnote{\xdef\@thefnmark{}\@footnotetext}
\makeatother
\AtBeginDocument{%
  \providecommand\BibTeX{{%
    \normalfont B\kern-0.5em{\scshape i\kern-0.25em b}\kern-0.8em\TeX}}}

\IEEEoverridecommandlockouts

\graphicspath{{gen/figures/}{figures/}}
\input{gen/paper/numbers.tex}
\input{figures/fig_numbers.tex} 
\expandafter\def\csname zn@lyr/alone/deepseek/bf16/max\endcsname{1.01}
\expandafter\def\csname zn@lyr/alone/deepseek/bf16/min\endcsname{0.99}
\expandafter\def\csname zn@lyr/alone/deepseek/fp8/max\endcsname{1.77}
\expandafter\def\csname zn@lyr/alone/deepseek/fp8/min\endcsname{1.74}
\expandafter\def\csname zn@lyr/alone/deepseek/paperlayer\endcsname{1.762}
\expandafter\def\csname zn@lyr/alone/mistral/bf16/max\endcsname{1.03}
\expandafter\def\csname zn@lyr/alone/mistral/bf16/min\endcsname{1.02}
\expandafter\def\csname zn@lyr/alone/mistral/fp8/max\endcsname{1.20}
\expandafter\def\csname zn@lyr/alone/mistral/fp8/min\endcsname{1.16}
\expandafter\def\csname zn@lyr/alone/mistral/paperlayer\endcsname{1.160}
\expandafter\def\csname zn@lyr/alone/qwen3/bf16/max\endcsname{1.02}
\expandafter\def\csname zn@lyr/alone/qwen3/bf16/min\endcsname{0.98}
\expandafter\def\csname zn@lyr/alone/qwen3/fp8/max\endcsname{1.17}
\expandafter\def\csname zn@lyr/alone/qwen3/fp8/min\endcsname{1.16}
\expandafter\def\csname zn@lyr/alone/qwen3/paperlayer\endcsname{1.171}
\expandafter\def\csname zn@lyr/deepseek/bf16/16k/max\endcsname{1.01}
\expandafter\def\csname zn@lyr/deepseek/bf16/16k/med\endcsname{1.00}
\expandafter\def\csname zn@lyr/deepseek/bf16/16k/min\endcsname{0.99}
\expandafter\def\csname zn@lyr/deepseek/bf16/32k/max\endcsname{1.01}
\expandafter\def\csname zn@lyr/deepseek/bf16/32k/med\endcsname{1.00}
\expandafter\def\csname zn@lyr/deepseek/bf16/32k/min\endcsname{0.99}
\expandafter\def\csname zn@lyr/deepseek/bf16/8k/max\endcsname{1.01}
\expandafter\def\csname zn@lyr/deepseek/bf16/8k/med\endcsname{1.00}
\expandafter\def\csname zn@lyr/deepseek/bf16/8k/min\endcsname{0.99}
\expandafter\def\csname zn@lyr/deepseek/fp8/16k/max\endcsname{1.76}
\expandafter\def\csname zn@lyr/deepseek/fp8/16k/med\endcsname{1.75}
\expandafter\def\csname zn@lyr/deepseek/fp8/16k/min\endcsname{1.72}
\expandafter\def\csname zn@lyr/deepseek/fp8/32k/max\endcsname{1.81}
\expandafter\def\csname zn@lyr/deepseek/fp8/32k/med\endcsname{1.77}
\expandafter\def\csname zn@lyr/deepseek/fp8/32k/min\endcsname{1.75}
\expandafter\def\csname zn@lyr/deepseek/fp8/8k/max\endcsname{1.91}
\expandafter\def\csname zn@lyr/deepseek/fp8/8k/med\endcsname{1.88}
\expandafter\def\csname zn@lyr/deepseek/fp8/8k/min\endcsname{1.83}
\expandafter\def\csname zn@lyr/deepseek/rescale/l0\endcsname{21}
\expandafter\def\csname zn@lyr/docrho/deepseek\endcsname{0.61}
\expandafter\def\csname zn@lyr/docrho/mistral\endcsname{0.80}
\expandafter\def\csname zn@lyr/docrho/qwen3\endcsname{0.93}
\expandafter\def\csname zn@lyr/inside/deepseek/aiter/max\endcsname{7}
\expandafter\def\csname zn@lyr/inside/deepseek/aiter/min\endcsname{6}
\expandafter\def\csname zn@lyr/inside/deepseek/zyfa/max\endcsname{8}
\expandafter\def\csname zn@lyr/inside/deepseek/zyfa/min\endcsname{6}
\expandafter\def\csname zn@lyr/inside/mistral/aiter/max\endcsname{12}
\expandafter\def\csname zn@lyr/inside/mistral/aiter/min\endcsname{11}
\expandafter\def\csname zn@lyr/inside/mistral/zyfa/max\endcsname{4}
\expandafter\def\csname zn@lyr/inside/mistral/zyfa/min\endcsname{2}
\expandafter\def\csname zn@lyr/inside/qwen3/aiter/max\endcsname{11}
\expandafter\def\csname zn@lyr/inside/qwen3/aiter/min\endcsname{8}
\expandafter\def\csname zn@lyr/inside/qwen3/zyfa/max\endcsname{4}
\expandafter\def\csname zn@lyr/inside/qwen3/zyfa/min\endcsname{0}
\expandafter\def\csname zn@lyr/mistral/bf16/16k/max\endcsname{1.03}
\expandafter\def\csname zn@lyr/mistral/bf16/16k/med\endcsname{1.03}
\expandafter\def\csname zn@lyr/mistral/bf16/16k/min\endcsname{1.02}
\expandafter\def\csname zn@lyr/mistral/bf16/32k/max\endcsname{1.03}
\expandafter\def\csname zn@lyr/mistral/bf16/32k/med\endcsname{1.03}
\expandafter\def\csname zn@lyr/mistral/bf16/32k/min\endcsname{1.02}
\expandafter\def\csname zn@lyr/mistral/bf16/8k/max\endcsname{1.04}
\expandafter\def\csname zn@lyr/mistral/bf16/8k/med\endcsname{1.03}
\expandafter\def\csname zn@lyr/mistral/bf16/8k/min\endcsname{1.01}
\expandafter\def\csname zn@lyr/mistral/fp8/16k/max\endcsname{1.29}
\expandafter\def\csname zn@lyr/mistral/fp8/16k/med\endcsname{1.26}
\expandafter\def\csname zn@lyr/mistral/fp8/16k/min\endcsname{1.25}
\expandafter\def\csname zn@lyr/mistral/fp8/32k/max\endcsname{1.27}
\expandafter\def\csname zn@lyr/mistral/fp8/32k/med\endcsname{1.24}
\expandafter\def\csname zn@lyr/mistral/fp8/32k/min\endcsname{1.23}
\expandafter\def\csname zn@lyr/mistral/fp8/8k/max\endcsname{1.35}
\expandafter\def\csname zn@lyr/mistral/fp8/8k/med\endcsname{1.27}
\expandafter\def\csname zn@lyr/mistral/fp8/8k/min\endcsname{1.22}
\expandafter\def\csname zn@lyr/mistral20/clock\endcsname{1.30}
\expandafter\def\csname zn@lyr/mistral20/cycles\endcsname{1.51}
\expandafter\def\csname zn@lyr/mistral20/sclk/aiter\endcsname{2108}
\expandafter\def\csname zn@lyr/mistral20/sclk/zyfa\endcsname{1624}
\expandafter\def\csname zn@lyr/mistral20/time\endcsname{1.16}
\expandafter\def\csname zn@lyr/paper/deepseek\endcsname{1.762}
\expandafter\def\csname zn@lyr/paper/mistral\endcsname{1.158}
\expandafter\def\csname zn@lyr/paper/qwen3\endcsname{1.171}
\expandafter\def\csname zn@lyr/qwen3/bf16/16k/max\endcsname{1.03}
\expandafter\def\csname zn@lyr/qwen3/bf16/16k/med\endcsname{1.02}
\expandafter\def\csname zn@lyr/qwen3/bf16/16k/min\endcsname{0.98}
\expandafter\def\csname zn@lyr/qwen3/bf16/32k/max\endcsname{1.03}
\expandafter\def\csname zn@lyr/qwen3/bf16/32k/med\endcsname{1.02}
\expandafter\def\csname zn@lyr/qwen3/bf16/32k/min\endcsname{0.98}
\expandafter\def\csname zn@lyr/qwen3/bf16/8k/max\endcsname{1.03}
\expandafter\def\csname zn@lyr/qwen3/bf16/8k/med\endcsname{1.01}
\expandafter\def\csname zn@lyr/qwen3/bf16/8k/min\endcsname{0.96}
\expandafter\def\csname zn@lyr/qwen3/bf16/first3/aiter\endcsname{0}
\expandafter\def\csname zn@lyr/qwen3/bf16/first3/rmax\endcsname{0.98}
\expandafter\def\csname zn@lyr/qwen3/bf16/first3/rmin\endcsname{0.98}
\expandafter\def\csname zn@lyr/qwen3/bf16/first3/zyfa\endcsname{4}
\expandafter\def\csname zn@lyr/qwen3/fp8/16k/max\endcsname{1.27}
\expandafter\def\csname zn@lyr/qwen3/fp8/16k/med\endcsname{1.25}
\expandafter\def\csname zn@lyr/qwen3/fp8/16k/min\endcsname{1.21}
\expandafter\def\csname zn@lyr/qwen3/fp8/32k/max\endcsname{1.23}
\expandafter\def\csname zn@lyr/qwen3/fp8/32k/med\endcsname{1.22}
\expandafter\def\csname zn@lyr/qwen3/fp8/32k/min\endcsname{1.19}
\expandafter\def\csname zn@lyr/qwen3/fp8/8k/max\endcsname{1.32}
\expandafter\def\csname zn@lyr/qwen3/fp8/8k/med\endcsname{1.26}
\expandafter\def\csname zn@lyr/qwen3/fp8/8k/min\endcsname{1.22}
\expandafter\def\csname zn@lyr/qwen3/fp8/first3/aiter\endcsname{2}
\expandafter\def\csname zn@lyr/qwen3/fp8/first3/rmax\endcsname{1.22}
\expandafter\def\csname zn@lyr/qwen3/fp8/first3/rmin\endcsname{1.21}
\expandafter\def\csname zn@lyr/qwen3/fp8/first3/zyfa\endcsname{5}
\expandafter\def\csname zn@lyr/qwen3/fp8/rescalecorr\endcsname{-0.71}
\expandafter\def\csname zn@lyr/qwen3/rescale/first3/max\endcsname{53}
\expandafter\def\csname zn@lyr/qwen3/rescale/first3/min\endcsname{38}
\expandafter\def\csname zn@lyr/qwen3/rescale/rest/med\endcsname{4}
\newif\ifzdraft
\zdrafttrue        

\newcommand{\sys}{ZYFA}
\newcommand{\dqk}{d_{\mathrm{qk}}}
\newcommand{\dv}{d_{\mathrm{v}}}
\newcommand{\x}{$\times$}
\definecolor{znpend}{rgb}{0.85,0.40,0.00}
\definecolor{znnote}{rgb}{0.10,0.35,0.75}

\makeatletter
\renewcommand{\zn}[1]{\ifcsname zn@#1\endcsname\csname zn@#1\endcsname\else
  \typeout{ZN-MISSING: #1}\textbf{\textcolor{red}{??#1}}\fi}
\newcommand{\zne}[1]{\ifcsname zn@#1\endcsname
  \expandafter\expandafter\expandafter\zn@sci\csname zn@#1\endcsname ee\relax
  \else\typeout{ZN-MISSING: #1}\textbf{\textcolor{red}{??#1}}\fi}
\def\zn@sci#1e#2e#3\relax{\ifx\relax#2\relax #1\else
  \ensuremath{#1{\times}10^{\number#2}}\fi}
\newcommand{\znp}[2]{\ifcsname zn@#1\endcsname\csname zn@#1\endcsname\else
  \typeout{ZN-PENDING: #1 = #2}\textcolor{znpend}{#2}\fi}
\newcommand{\znpe}[2]{\ifcsname zn@#1\endcsname
  \expandafter\expandafter\expandafter\zn@sci\csname zn@#1\endcsname ee\relax
  \else\typeout{ZN-PENDING: #1 = #2}\textcolor{znpend}{\zn@sci#2ee\relax}\fi}
\makeatother

\newcommand{\znpctless}[1]{\ifcsname zn@#1\endcsname\fpeval{round(100*(1-\csname zn@#1\endcsname),0)}\else\zn{#1}\fi}

\zdraftfalse 

\begin{document}

\title{ZYFA: Speed and Accuracy of FP8 FlashAttention on AMD CDNA4}

\author{
Quentin Anthony\corr $\quad$
Ganesh Bikshandi $\quad$
Praneeth Medepalli $\quad$
Stephen Ebert \\[0.6em]
{\textbf{Zyphra}} \\[0.2em]
{San Francisco, CA}
}

\maketitle

\blfootnote{\textsuperscript{*}Corresponding author: \texttt{quentin@zyphra.com}}

\renewcommand{\thefootnote}{\arabic{footnote}}

\begin{abstract}
Attention is a core building block of frontier models and often dominates compute-bound prefill at long context. FP8 matrix instructions offer twice the peak throughput of BF16, but they accelerate only matrix multiplication, so a GPU kernel's total speedup from FP8 shrinks as other operations take a larger share of the overall kernel time. For example in attention, the softmax between the two matrix multiplications is on vector units which FP8 does not accelerate. On the AMD Instinct MI355X, the vendor's AITER kernels (AMD's library of inference kernels) reach \zn{peakpct/mha/fp8-d128/32k/aiter}\% of peak FP8 matrix throughput at $32$K tokens, compared with \zn{peakpct/mha/bf16-d128/32k/aiter}\% in BF16. Unlike NVIDIA's Hopper and Blackwell GPUs, the MI355X also does not track the completion of matrix instructions in hardware, so fine-grained instruction order must be carried out via more manual methods like assembly instruction ordering, which is vital for attention kernels such as overlapping the softmax with the matrix multiplications. Finally, attention is more numerically sensitive to quantization than feed-forward layers, so considerable numerical analysis must be done to ensure fidelity. To overcome these challenges, we present \sys{}, prefill kernels for the MI355X built on three techniques: (1)~computing each FP8 probability directly from its score with a multiply-add and an integer conversion instead of an exponential, since an FP8 bit pattern approximates a scaled base-2 logarithm of its raw value; (2)~a ping-pong schedule in which one group of threads computes its softmax while another executes its matrix multiplications; and (3)~scaling each block of probabilities up by a power of two before rounding, so that far fewer small probabilities round to zero, and normalizing by the sum of the rounded probabilities, so that the attention weights still sum to one. On attention inputs captured from three representative models and timed at sustained clock speeds (the lower, steady clock that the GPU settles at after a few seconds of back-to-back launches), \sys{} achieves \zn{hl/fp8/real/ratio/min}--\zn{hl/fp8/real/ratio/max}\x{} the throughput of AITER. On the random Gaussian inputs of kernel benchmarks at $32$K tokens, \sys{} achieves \zn{t7/gqa/fp8-d128/32k}--\zn{t7/mha/fp8-d128/32k}\x{} the throughput of AITER at head dimension $128$ and \zn{t7/mha/fp8-d192/32k}\x{} for DeepSeek-style multi-head latent attention. The sustained clock of the MI355X depends on operand values, so throughput on random inputs does not reliably predict throughput on real model activations, where \sys{} takes fewer clock cycles than AITER's kernel but runs at a lower per-cycle clock speed. At head dimension $128$, \sys{}'s numerical error is within a few percent of that of AMD's Composable Kernel and Triton kernels and \znpctless{num/qwen3/fp8/zyfa/vsaiter}--\znpctless{num/mistral/fp8/zyfa/vsaiter}\% lower than that of AITER's.
The source code is available at \url{https://github.com/Zyphra/zyfa}.
\end{abstract}

\section{Introduction}
\label{sec:intro}

The attention operation remains both widespread and computationally expensive. The longer the input context to a model, the larger the share of prefill time that attention takes, and long contexts are becoming increasingly common: coding agents work across large repositories~\citep{jimenez2024swebench}, and genomic models read long DNA sequences~\citep{nguyen2024evo}. At these lengths attention often dominates the prefill phase of autoregressive inference. IO-aware algorithms like FlashAttention perform the same arithmetic as standard attention, but stream blocks of keys and values through on-chip memory and keep a running softmax for each query, so that the score matrix never reaches the comparatively slow high-bandwidth memory (HBM) of the GPU~\citep{dao2022flash,dao2023flash2}. The speed of attention then depends on how well its GPU kernel can keep the arithmetic units of the GPU busy, and because the throughput of matrix or tensor cores has increased significantly faster in subsequent GPU generations than all other hardware elements, the kernel engineer's primary focus has shifted to keeping them fed. FlashAttention-3 hides the softmax behind asynchronous matrix instructions on Hopper GPUs~\citep{shah2024flash3}, and FlashAttention-4 approximates portions of the exponentials with polynomials on Blackwell GPUs, whose matrix hardware units grew faster than their exponential units~\citep{shah2026flash4}. Low precision worsens this hardware gap further. Recent GPUs execute 8-bit floating-point (FP8) matrix instructions at twice the theoretical BF16 rate~\citep{micikevicius2022fp8,cdna4-wp}, so FP8 attention could, in principle, take half the time of BF16 attention if it were purely bound by its matrix multiplications.

However, on the AMD Instinct MI355X, FP8 attention gains much less than this in practice. At $32$K tokens on random Gaussian inputs, the FP8 attention kernel of AITER, AMD's library of inference kernels~\citep{aiter}, reaches \zn{peakpct/mha/fp8-d128/32k/aiter}\% of the FP8 datasheet peak, while its BF16 kernel reaches \zn{peakpct/mha/bf16-d128/32k/aiter}\% of the BF16 peak. Accuracy is a second concern. DeepSeek-V3, for example, trains its linear layers, including the experts, in FP8 but keeps attention in higher precision because attention is more numerically sensitive~\citep{deepseekv3}. This paper asks what limits the FP8 attention operation on the MI355X GPU, how to work around these limits, and how the error of FP8 attention can be sourced and mitigated.

The first limit is the softmax. FP8 doubles the speed of the matrix products but not of the softmax operation. Each score still needs an exponential, and an FP8 kernel must also convert each probability to 8 bits before the second matrix product, and both of these operations are performed on vector units whose speed is the same in both FP8 and BF16. For $32$ queries against a block of $128$ keys at head dimension $128$, the two matrix products take \zn{budget/cyc/bf16mm} matrix-unit cycles in BF16 and \zn{budget/cyc/fp8mm} in FP8, while the softmax takes \zn{budget/cyc/exactp} vector-unit cycles in FP8, even more than the \zn{budget/cyc/exactp-bf16} it takes in BF16, because converting a probability to FP8 costs more than converting it to BF16 (see Section~\ref{sec:limits-sol}). Having nearly the same clock cycles spent in vector units and matrix units is troubling for the programmer, since one must overlap the softmax with matrix products in order to keep the matrix cores busy. FP8 therefore requires almost perfect overlap (\zn{budget/cyc/fp8mm} matrix cycles against \zn{budget/cyc/exactp} vector cycles), but BF16 has considerable slack (\zn{budget/cyc/bf16mm} against \zn{budget/cyc/exactp-bf16}) that makes implementation easier. In our thread traces of AITER's FP8 kernel (cycle-by-cycle records of every instruction that a wave issues), we see the matrix unit idle for \zn{att/aiter-fp8-d128/idlebar}\% of cycles while the softmax runs. NVIDIA's Hopper and Blackwell GPUs provide asynchronous matrix instructions whose completion events are tracked by hardware~\citep{shah2024flash3,shah2026flash4}, so the hardware tells a kernel when a matrix product is ready and overlap is much easier to organize by the kernel programmer. On the MI355X, however, matrix instructions keep running while the wave that issued them (a group of $64$ threads that issue instructions together) issues other instructions, and the hardware does not track their completion, so the overlap depends entirely on the order in which the kernel issues its instructions, whose order must also ensure that each matrix instruction has finished before its result is read (Section~\ref{sec:bg-gpu}). We find this often necessitates raw GPU assembly kernels, whose development can be eased by generator functions written in higher-level languages like Python.

The clock speed is the second limit one encounters when implementing attention. Under sustained load, the MI355X lowers its clock, and since each bit flip draws current, the clock rate falls as more operand bits change from one instruction to the next. This clock behavior is an important factor for attention, since as we remove softmax as a bottleneck, the matrix cores can do more arithmetic per cycle, which is more bit flips and therefore a lower clock. This effect is further complicated by the fact that real-world inputs have very different distributions from classical synthetic kernel inputs, such as random Gaussian values, which one uses for development or kernel comparison. Speed and numerics are therefore closely intertwined. For example, we found that AITER lets more probabilities underflow to zero on some real model inputs, and zeros switch fewer bits in subsequent arithmetic, so AITER will have a higher clock rate but worse numerics. Achieving both high performance and low numerical error is therefore the primary goal of our work.

We measure the low-level instruction costs and the clock behavior of the MI355X and use them to design \sys{}, causal forward kernels in BF16 and FP8. They support head dimension $128$ and the head shape of multi-head latent attention (MLA) in DeepSeek-V2~\citep{deepseekv2}, with query-key dimension $192$ and value dimension $128$, the per-head form of MLA that prefill uses.\footnote{MLA caches one compressed latent vector per token instead of per-head keys and values. Prefill expands the latent vectors into per-head keys and values and runs standard attention with these dimensions. Decoding instead folds the expansion of the keys into the query and attends over the latent vectors directly, with a query-key dimension of $576$ and a value dimension of $512$ that all heads share.} Our FP8 kernels attempt to shrink the number of cycles spent in softmax by exploiting the FP8 format to perform a similar numerical operation to the exponential, specifically by using the fact that the exponent bits of an FP8 number already encode an exponential. Read as an integer, the 8-bit bit pattern (or \textit{code}) of a positive e4m3 number, the FP8 format with $4$ exponent and $3$ mantissa bits, grows by $8$ each time the value doubles, so it is close in effect to a scaled and shifted base-2 logarithm of the value being represented. The code of a probability $p=2^x$, where $x$ is the score minus the row maximum in base-2 units, is therefore close to $8x$ plus a constant, the same relation that Schraudolph's fast exponential uses for IEEE floating-point numbers~\citep{schraudolph1999exp}. At head dimension $128$ the FP8 kernel computes this code with one multiply-add and one integer conversion, which we call \emph{lean P}. Lean P replaces both the exponential and the floating-point conversion, and lowers the vector cycles per block from \zn{budget/cyc/exactp} to \zn{budget/cyc/leanp}. It is an approximation that raises the output error of one attention call to \zn{lean/err/vsexact}\x{} that of exact P, which evaluates the exponential, but the kernel's error stays below that of AITER's kernel, at \zn{lean/err/vsaitergm}\x{} in geometric mean (Section~\ref{sec:leancost}). Essentially, lean P trades a small amount of numerical accuracy for speed. Lean P is optional, and \sys{} uses it by default at head dimension $128$. The kernel at the MLA shape keeps exact P only because there lean P falls just below the signal-to-noise threshold of our correctness test in \zn{d192lean/snrfails} of \zn{d192lean/snrcells} configurations, although it matches exact P end to end (Section~\ref{sec:leancost}).

Two further changes keep the FP8 probabilities as accurate as possible. In a long row of the attention matrix most probabilities lie far below the largest one, often below $2^{-6}$, where e4m3 values lose precision, or below $2^{-10}$, where they underflow to zero and we lose information. The kernels therefore multiply each block of $64$ probabilities by a power of two so that its largest value is close to $1$, and the scaled FP8 matrix instruction of CDNA4 undoes the factor at no cost. The kernels also keep each output as a weighted average whose weights sum to one. The output divides $\sum_j p_j v_j$ by $\sum_j p_j$, and the numerator uses the probabilities after rounding to FP8. If the denominator sums the unrounded probabilities, as in AITER's kernel, probabilities that underflow to zero still count in the denominator, so every output row shrinks slightly, which accounts for much of the larger error of AITER's FP8 kernel (Section~\ref{sec:res-kernel}). \sys{} divides by the sum of the rounded probabilities instead, and it computes this sum with one extra matrix instruction that multiplies the FP8 probabilities by a vector of ones. The extra instruction adds \zn{lean/rowsum/matrix}\% to the matrix work and removes vector additions from the vector unit, which is the principal bottleneck in FP8.

Finally, our assembly schedule pairs two waves, AMD's counterpart of warps, on each SIMD unit and runs the softmax of one wave during the matrix instructions of the other. With lean P the wave that computes the softmax finishes its work before the wave that issues the matrix instructions does, so the matrix instructions determine the speed of the loop over blocks of keys.

The contributions of this work are as follows.
\begin{enumerate}[nosep,leftmargin=*]
\item A characterization of the MI355X hardware in the context of attention (Section~\ref{sec:limits}), covering the vector cost of each class of instruction, the overlap capability of vector and matrix work on a single SIMD, and clock dependencies on operand values, expanding on the detail provided by the official instruction set architecture (ISA) documentation.
\item \sys{}, whose FP8 kernels achieve \zn{hl/fp8/real/ratio/min}--\zn{hl/fp8/real/ratio/max}\x{} the throughput of AITER's default kernels on activations captured from real models at sustained clocks and \zn{t7/gqa/fp8-d128/32k}--\zn{t7/mha/fp8-d192/32k}\x{} on Gaussian inputs at $32$K tokens (Sections~\ref{sec:design} and~\ref{sec:res-speed}).
\item A measurement of the error of single attention calls (Section~\ref{sec:res-kernel}), which shows that at head dimension $128$ \sys{}'s FP8 error is within a few percent of that of Composable Kernel and Triton and \znpctless{num/qwen3/fp8/zyfa/vsaiter}--\znpctless{num/mistral/fp8/zyfa/vsaiter}\% lower than that of AITER's assembly kernel.
\item An account of FP8 attention error in three full models (Section~\ref{sec:res-e2e}). Most of it comes from quantizing the inputs rather than the probabilities and is spread over most heads, so keeping the most sensitive heads in BF16 is not a useful middle ground. On Mistral-7B and Qwen3-8B, even $90\%$ of heads in BF16 does not sufficiently match BF16 accuracy, and by then attention takes as long as all-BF16. A Hadamard rotation of $Q$ and $K$ instead roughly halves the error from quantizing the inputs at almost no cost.
\end{enumerate}
Table~\ref{tab:terms} defines the terms that the rest of the paper uses. The source code is available at \url{https://github.com/Zyphra/zyfa}.

\begin{figure}[t]
\centering
\includegraphics[width=\columnwidth]{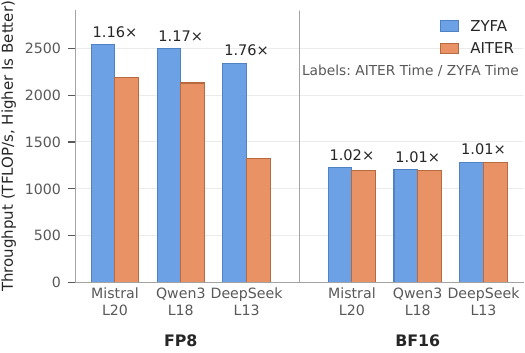}
\caption{Throughput of \sys{} and AITER's default kernels on the queries, keys, and values captured from one layer per model at sequence length $16$K, timed at sustained clocks (the steady clock after warmup launches). L20 is layer~20. The label above each pair is AITER's time divided by \sys{}'s time. Mistral-7B and Qwen3-8B have head dimension $128$, where the FP8 kernel uses lean P, which writes each FP8 probability with one multiply-add instead of an exponential. DeepSeek-V2-Lite has the MLA shape, where the FP8 kernel uses exact P, which evaluates the exponential (Section~\ref{sec:fp8p}), and AITER uses Composable Kernel (CK) in FP8 and its OPUS kernel, written with AITER's C++ template library, in BF16. Table~\ref{tab:real} gives the clock of these runs.}
\label{fig:headline}
\end{figure}

\section{Background}
\label{sec:background}

\subsection{The MI355X GPU}
\label{sec:bg-gpu}

The AMD Instinct MI355X (CDNA4 architecture) has $256$ compute units (CUs), analogous to the streaming multiprocessors of NVIDIA GPUs. The CUs are distributed across eight chiplets (XCDs), each with its own $4$\,MB L2 cache. At the maximum shader clock (SCLK), $2.4$\,GHz, the datasheet peak dense matrix throughput is $2.5$\,PFLOP/s in BF16 and $5.0$\,PFLOP/s in FP8~\citep{cdna4-wp} (Table~\ref{tab:gpu}). Each CU has four SIMD units, which issue the instructions of the threads that run on them, and $160$\,KB of shared memory called the local data share (LDS). Threads run in groups of $64$ called waves, analogous to the $32$-thread warps of NVIDIA GPUs, and we call the position of a thread within its wave its lane. Waves in the same workgroup, a group of waves launched together like a thread block on NVIDIA GPUs, run on one CU and share LDS and workgroup barriers. A wave stays on one SIMD and issues its instructions in program order. Each SIMD has a register file and two execution units, which we call pipes, a vector pipe for elementwise arithmetic (VALU) and a matrix pipe for matrix fused multiply-add (MFMA) instructions~\citep{cdna4-isa}.\footnote{An MFMA instruction multiplies two small matrix tiles held in the registers of a wave and adds the product to an accumulator tile. For example, \texttt{v\_mfma\_f32\_32x32x64\_f8f6f4} multiplies a $32{\times}64$ FP8 tile by a $64{\times}32$ FP8 tile and accumulates the $32{\times}32$ result in FP32.} When two waves share a SIMD, the vector instructions of one wave can execute while a matrix instruction of the other occupies the matrix pipe (Section~\ref{sec:limits-valu}). Unlike NVIDIA Hopper GPUs, CDNA4 has no tensor memory accelerator (a unit that copies tiles between global and shared memory), no hardware barrier between producer and consumer waves, and no hardware tracking of matrix instructions in flight, so kernels that use these features to overlap work, such as FlashAttention-3~\citep{shah2024flash3}, do not port directly. On CDNA4 the order of instructions in the kernel therefore decides how memory, vector, and matrix work overlap.

\begin{table}[t]
\centering
{\footnotesize
\begin{tabular}{ll}
\toprule
Property & Value per GPU \\
\midrule
Compute units & $256$ in $8$ XCDs, $4$ SIMDs each \\
Wave, VGPRs & $64$ lanes, $512$ VGPRs per lane per SIMD \\
LDS & $160$\,KB per CU \\
L2 cache & $4$\,MB per XCD \\
Infinity Cache & $256$\,MB, shared by the XCDs \\
HBM3E & $288$\,GB, $8$\,TB/s \\
Maximum clock & $2.4$\,GHz \\
Dense matrix peak & BF16 $2.5$, FP8 $5.0$, FP4 $10$\,PFLOP/s \\
Total board power (TBP) & $1400$\,W \\
\bottomrule
\end{tabular}}
\caption{Properties of one MI355X from~\citep{cdna4-wp,cdna4-isa}. The peaks assume the maximum clock. VGPRs are the vector registers of a wave, XCDs are the eight chiplets, and LDS is the shared memory of a compute unit (CU). Our system caps the board power at the TBP.}
\label{tab:gpu}
\end{table}

\begin{table}[t]
\centering
{\footnotesize
\setlength{\tabcolsep}{4pt}
\begin{tabular}{@{}l@{\hspace{6pt}}p{0.70\columnwidth}@{}}
\toprule
Term & Meaning \\
\midrule
CU, SIMD & A compute unit (CU) has four SIMD units, each of which issues the instructions of its waves \\
wave & $64$ threads that issue instructions together, AMD's counterpart of a $32$-thread NVIDIA warp \\
MFMA & Matrix fused multiply-add, the matrix instruction that runs on the matrix pipe, the SIMD's matrix unit \\
VALU & The vector pipe, the SIMD's vector unit, which runs elementwise arithmetic such as the softmax \\
LDS & Local data share, the $160$\,KB shared memory of a CU \\
XCD & One of the eight chiplets of the GPU, each with its own L2 cache \\
SCLK & Shader clock, the clock of the CUs, at most $2.4$\,GHz \\
e4m3 & The FP8 format with $4$ exponent and $3$ mantissa bits, whose bit pattern, read as an integer, we call its code \\
binade & A range $[2^k, 2^{k+1})$, so a value $n$ binades below another is about $2^{-n}$ times it \\
AITER & AMD's library of inference kernels and the baseline of this paper \\
CK & Composable Kernel, AMD's kernel library, which AITER uses for FP8 attention at $\dqk{=}192$ \\
exact P & Computing each FP8 probability with an exponential and a conversion to FP8 \\
lean P & Writing the code of each FP8 probability with one multiply-add and an integer conversion, without an exponential \\
kernel-only timing & Device time of the kernel alone, excluding host launch time \\
sustained clocks & The lower, steady clock that the GPU settles at after seconds of back-to-back launches \\
captured activations & The $Q$, $K$, and $V$ that a real model passes to attention, recorded during a forward pass \\
Gaussian inputs & Random $Q$, $K$, and $V$ with standard deviation $0.1$, as in kernel benchmarks \\
BF16 noise band & KL divergences from the BF16 model no larger than that between the models with two correct BF16 kernels, \sys{}'s and AITER's \\
ge2 & Gradient-weighted error of a head, which ranks heads by their effect on the model output \\
\bottomrule
\end{tabular}}
\caption{Terms used throughout the paper.}
\label{tab:terms}
\end{table}

\subsection{Tiled Attention}
\label{sec:bg-attn}

For one head, causal attention computes $O=\mathrm{softmax}(c\,QK^{\top} + M)V$ where $Q,K\in\mathbb{R}^{S\times\dqk}$, $V\in\mathbb{R}^{S\times\dv}$, $S$ is the sequence length, $c$  is the softmax scale, and $M$ is the causal mask. FlashAttention-2~\citep{dao2023flash2} assigns a tile of query rows to a group of threads, which loops over blocks of keys. For each block, the kernel computes scores in base-2 units, $t=c\,q k^{\top}\log_2 e$,  updates the running maximum $m$ of each query row, and forms unnormalized softmax weights $p=2^{t-m}$.\footnote{Subtracting the maximum keeps every $p\le1$, so the exponential cannot overflow, and it does not change the softmax, because the same factor $2^{-m}$ multiplies the numerator and the denominator.} When $m$ grows, the kernel rescales the output accumulator and the row sum $\ell$ by $2^{m_{\mathrm{old}}-m}$, and it then accumulates $pV$ and $\sum p$.\footnote{The rescale keeps every term of the two sums relative to the same maximum. Each $p$ accumulated before the maximum grew was computed as $2^{t-m_{\mathrm{old}}}$, relative to the old maximum, and multiplying that $p$ by $2^{m_{\mathrm{old}}-m}$ turns it into $2^{t-m}$, its value relative to the new maximum $m$.} The kernel returns the output $O=\mathrm{acc}/\ell$ and the log-sum-exp (LSE), which lets results computed over different key ranges be merged. With batch size $B$ and $H$ query heads, a causal forward pass has
\begin{equation}
F = B\,H\,S^2(\dqk+\dv)
\label{eq:flops}
\end{equation}
floating-point operations, half the count without the causal mask. We report throughput as $F$ divided by kernel time.

\subsection{FP8 Attention}
\label{sec:bg-fp8}

The FP8 kernels studied here store $Q$, $K$, and $V$ in the e4m3 format, which has a 4-bit exponent with bias 7 and a 3-bit mantissa. We call the bit pattern of an e4m3 value, read as an 8-bit unsigned integer, its code. Its largest finite value is $448$. A binade is the range $[2^k,2^{k+1})$ of values that share one exponent, so a probability $n$ binades below the row maximum is roughly $2^{-n}$ times the row maximum. Values of at least $2^{-6}$ are normal values, which keep all $3$ mantissa bits, and smaller values are subnormal values, which keep fewer.\footnote{As in other floating-point formats, a normal e4m3 value is $1.f\times2^{e}$, with an implicit leading $1$ and a $3$-bit fraction $f$, so each binade holds $8$ values spaced $1/8$ to $1/16$ of their size apart. Below $2^{-6}$, the smallest normal value, the format drops the implicit $1$ and stores subnormal values as $0.f\times2^{-6}$, spaced $2^{-9}$ apart. A subnormal value therefore keeps fewer significant bits the smaller it is, and values below $2^{-10}$, half of that spacing, round to zero.} Each of $Q$, $K$, and $V$ has one scale factor, which the kernel folds into $c$ and into the output, and $p$ is converted to e4m3 before the $PV$ product~\citep{micikevicius2022fp8,shah2024flash3}. Because $p\le1$, a probability converted to e4m3 without scaling lies in the lower part of the format's range, where values below $2^{-6}$ are subnormal and values below $2^{-10}$ underflow, that is, round to zero.

\section{Performance Effects of the MI355X}
\label{sec:limits}

\subsection{Cost of Vector Instructions}
\label{sec:limits-valu}
\label{app:micro}

We time each individual instruction with microbenchmarks comprised of inline-assembly loops over a fixed block of instructions. The compiler emits each block as written, without inserting or reordering instructions. The block names its registers explicitly and declares them clobbered (overwritten by the block), so the compiler keeps no other values in them. One workgroup with $96$\,KB of LDS runs on each of the $256$ CUs, and every wave reads \texttt{s\_memtime}, which counts cycles of the shader clock (SCLK), and \texttt{s\_memrealtime}, which counts time at a constant rate. The operands change on every instruction, because the clock depends on how many bits switch (Section~\ref{sec:limits-clock}). To measure the cost of one instruction, two waves on one SIMD issue independent copies of it.

We give the cost of an instruction as the number of cycles for which it occupies the vector pipe, during which the pipe accepts no other vector instruction, and this number therefore sets the throughput of the instruction. An FP32 add, multiply, or fused multiply-add (FMA) occupies the pipe for $2$ cycles, since the pipe processes the $64$ lanes of a wave at $32$ lanes per cycle. Maximum, integer conversion, compare, and permute instructions occupy it for $4$ cycles, or $16$ lanes per cycle. Packed instructions, which apply one operation to two values in each lane, also occupy it for $4$ cycles, so a packed FP32 instruction does the work of two FP32 instructions in the same pipe time, while the wave issues one instruction instead of two. A wave issues at most one instruction every $4$ cycles. The exponential \texttt{v\_exp\_f32} and the packed FP8 conversion \texttt{v\_cvt\_pk\_fp8\_f32} occupy it for $8$ cycles. The exponential therefore runs at a quarter of the FMA rate, as in earlier CDNA GPUs. Lower precision does not make the exponential cheaper. The FP16 exponential also occupies the vector pipe for $8$ cycles, and the ISA has no packed exponential that would evaluate two values per lane.
Two waves share each SIMD, and a vector instruction from one wave can execute at the same time as a matrix instruction from the other wave scheduled to the SIMD. Next to 32$\times$32 MFMAs a wave of $8$-cycle vector instructions takes \zn{valu/coissue/slow}\x{} as long as in isolation, and the matrix wave takes at most \zn{valu/coissue/mfmaslow}\x{} as long. Next to BF16 16$\times$16 MFMAs the vector wave reaches \zn{valu/coissue/keep16}\% of its throughput in isolation. Packed FP32 instructions are the exception.\footnote{Because it halves the number of instructions that the softmax issues, packed FP32 would be the natural choice for a softmax if it could run next to MFMAs.} They must wait while an MFMA executes on the same SIMD, so a wave of packed FP32 instructions next to a matrix-bound wave advances only when the matrix pipe is idle. In our test such a wave took \zn{valu/pk/starve}\x{} as long as alone, which is the largest slowdown that the test can resolve. The softmax in \sys{}'s kernels therefore uses non-packed FP32 instructions. Packed FP16 instructions, by contrast, run next to MFMAs at their standalone rate. The ISA documentation~\citep{cdna4-isa} describes neither this behavior of packed FP32 instructions nor the $8$-cycle cost of the packed FP8 conversion.

Two further rules of the hardware affect hand scheduling. The matrix pipe serves the older of two waves first, so a younger matrix-bound wave waits until the older wave stops issuing MFMAs. An FP8 MFMA of $16$ passes, where a pass is $4$ cycles of the matrix pipe, writes its $16$ destination registers one after another without an interlock, so the last registers are valid only $20$ wait states after issue.\footnote{Without an interlock the hardware does not hold back an instruction that reads a destination register before the MFMA has written it. The kernel must place enough independent instructions, or explicit wait states, between the MFMA and the first read. A wait state is the ISA's unit of delay between dependent instructions, which independent instructions or \texttt{s\_nop} instructions provide.}

\subsection{Pipe Cycles per Block}
\label{sec:limits-sol}
\label{app:limitsmore}
\label{app:att}

Table~\ref{tab:budget} counts the pipe cycles for one wave that computes $32$ queries against one $128$-key block. In FP8 the softmax needs about as many vector cycles as the two matrix products need matrix cycles, \zn{budget/cyc/exactp} against \zn{budget/cyc/fp8mm}, so the matrix pipe stays busy only if the vector work of one wave runs entirely during the MFMAs of the other wave on the same SIMD. In BF16 the matrix products take \zn{budget/cyc/bf16mm} cycles and the softmax \zn{budget/cyc/exactp-bf16}, so the softmax fits within the time of the matrix work with room to spare, and a schedule that overlaps the two imperfectly still keeps the matrix pipe busy.

\begin{table}[t]
\centering
{\footnotesize
\begin{tabular}{lr}
\toprule
work for one wave and one block & cycles \\
\midrule
\multicolumn{2}{l}{\emph{matrix pipe}} \\
$QK^{\top}$ and $PV$ in FP8 & \zn{budget/cyc/fp8mm} \\
$QK^{\top}$ and $PV$ in BF16 & \zn{budget/cyc/bf16mm} \\
row sum of the FP8 probabilities & \zn{budget/cyc/rowsum} \\
\midrule
\multicolumn{2}{l}{\emph{vector pipe}} \\
row maximum ($32$ three-input maxima) & \zn{budget/cyc/max} \\
exact P in FP8 ($64$ FMA, $64$ exp, $32$ packed cvt) & \zn{budget/cyc/exactp} \\
exact P in BF16 ($64$ FMA, $64$ exp, $32$ packed cvt) & \zn{budget/cyc/exactp-bf16} \\
lean P ($64$ FMA, $64$ integer cvt) & \zn{budget/cyc/leanp} \\
\bottomrule
\end{tabular}}
\caption{In FP8 the softmax needs about as many vector-pipe cycles as the matrix products need matrix-pipe cycles, so the two must overlap almost perfectly to keep the matrix pipe busy, while in BF16 the matrix products take more than twice as long as the softmax. The table gives pipe cycles for one wave (a group of $64$ threads) that computes $32$ queries against one $128$-key block at $\dqk{=}\dv{=}128$, and each of its $64$ lanes holds $64$ of the $4096$ scores. Matrix cycles, on the pipe that runs matrix (MFMA) instructions, follow from the matrix rate per cycle, and vector cycles, on the pipe that runs elementwise instructions, follow from the measured costs in Section~\ref{sec:limits-valu}. Exact P, the standard method, evaluates the exponential and converts the result to e4m3 or BF16, and the packed BF16 conversion occupies the vector pipe for $4$ cycles instead of $8$. Lean P, \sys{}'s method, writes the e4m3 bit pattern of each probability with one multiply-add and one integer conversion (Section~\ref{sec:lean}). Rescaling the output adds vector work when the row maximum grows.}
\label{tab:budget}
\end{table}

\begin{figure*}[tp]
\centering
\begin{subfigure}[t]{0.8\textwidth}
\centering
\includegraphics[width=\linewidth]{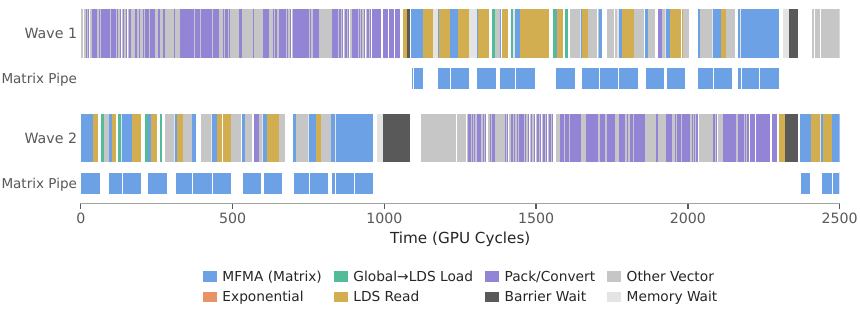}
\caption{\sys{}}
\end{subfigure}\\[4pt]
\begin{subfigure}[t]{0.8\textwidth}
\centering
\includegraphics[width=\linewidth]{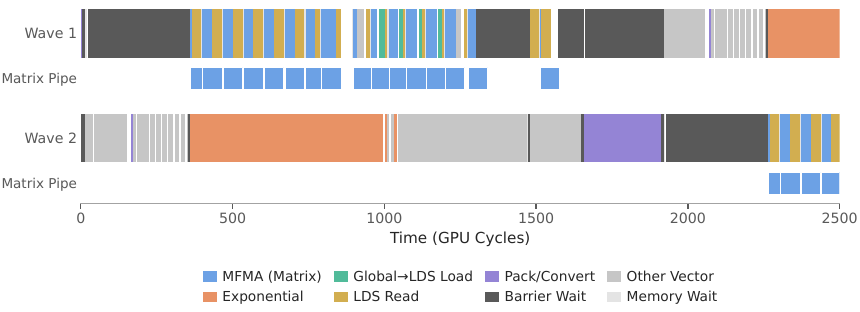}
\caption{AITER}
\end{subfigure}
\caption{Thread traces showing that AITER's FP8 kernel leaves the matrix pipe idle while its waves compute exponentials and wait at barriers, and that \sys{}'s FP8 kernel keeps the matrix pipe busy most of the time. A thread trace records every instruction that a wave issues, cycle by cycle, here with the thread-trace mode of AMD's \texttt{rocprofv3} profiler. Each panel shows the two waves that share one SIMD over $2500$ cycles in the middle of the loop over key blocks, where every block repeats the same work, in the FP8 kernels at $\dqk{=}128$ ($S{=}8$K tokens, $H{=}32$ heads, multi-head attention). Each wave row shows the class of each instruction that the wave issues, and the matrix-pipe row below it shows when that wave's MFMAs run. A vector instruction runs on the vector pipe during its bar in the wave row, so the wave rows already show when the vector pipe works. An MFMA keeps running on the matrix pipe for $32$ to $64$ cycles after its short bar in the wave row, while the wave issues other instructions, so the matrix-pipe rows show that work separately.}
\label{fig:att-fp8}
\end{figure*}

Thread traces, cycle-by-cycle records of every instruction that a wave issues, show what happens when the overlap is imperfect. We record them with \texttt{rocprofv3} advanced thread traces for AITER's FP8 assembly kernel and for the base kernel, an earlier version of \sys{}'s FP8 kernel written in HIP~\citep{hip}, AMD's C++ kernel language and the counterpart of CUDA. Table~\ref{tab:sol} counts their pipe cycles at $S{=}32$K from their dynamic instruction mixes (the instructions that the waves execute, as opposed to those in the kernel binary) and the instruction costs of Section~\ref{sec:limits-valu}. In both kernels the vector pipe is busy for about as many cycles as the matrix pipe. A kernel that ran its matrix work and its vector work one after the other would take the sum of the two, the serialized row of the table. AITER takes fewer cycles than this sum because the two waves on each of its SIMDs overlap part of their matrix and vector work. AITER alternates the two waves of a SIMD between a matrix phase and a softmax phase, and each phase ends at a workgroup barrier, where every wave waits until all waves of the workgroup arrive. AITER's softmax phase takes longer than its matrix phase, so for \zn{att/aiter-fp8-d128/idlebar}\% of SIMD cycles the matrix pipe idles while one wave computes probabilities and the other waits at the barrier (Figure~\ref{fig:att-fp8}). The base kernel takes about the serialized sum. All of its pipes are idle in \zn{att/base-fp8-d128/allidle}\% of cycles, and in \zn{att/base-fp8-d128/idleboth}\% of cycles both waves of a SIMD execute vector instructions at the same time, so neither issues MFMAs.

\begin{table}[t]
\centering
{\footnotesize
\begin{tabular}{lrr}
\toprule
cycles per SIMD & base & AITER \\
\midrule
matrix pipe (packed FP32) & \zn{sol/base/matrix} (\zn{sol/base/packed}) & \zn{sol/aiter/matrix} (\zn{sol/aiter/packed}) \\
vector pipe & \zn{sol/base/valu} & \zn{sol/aiter/valu} \\
instruction issue & \zn{sol/base/issue} & \zn{sol/aiter/issue} \\
LDS & \zn{sol/base/lds} & \zn{sol/aiter/lds} \\
serialized (matrix + vector) & \zn{sol/base/serial} & \zn{sol/aiter/serial} \\
\textbf{achieved} & \zn{sol/base/achieved} & \zn{sol/aiter/achieved} \\
SCLK (GHz) & \zn{sol/base/sclk} & \zn{sol/aiter/sclk} \\
\bottomrule
\end{tabular}}
\caption{Both kernels keep the vector pipe about as busy as the matrix pipe, and neither overlaps the two fully. The table gives pipe cycles per SIMD at $S{=}32$K for FP8 multi-head attention at $\dqk{=}128$, for two units of work of one wave on $32$ queries and $64$ keys each, which together equal the work of one wave on one $128$-key block in Table~\ref{tab:budget}. AITER is AITER's FP8 assembly kernel. Base is an earlier HIP version of \sys{}'s FP8 kernel with a Schraudolph exponential (an exponential that writes the bits of an FP32 number directly from the score) and an FP32 row sum, and its instruction mix comes from $S{=}8$K traces. The matrix row includes the cycles in which packed FP32 instructions block the matrix pipe, given in parentheses. The serialized row is the sum of matrix and vector cycles, which the work would take if the two never overlapped, and the achieved row gives the cycles that each kernel actually takes. SCLK is the shader clock.}
\label{tab:sol}
\end{table}

How far can the vector work shrink? Each score needs at least one instruction that subtracts the row maximum and applies the scale, and one that converts the result to $8$ bits. A kernel that takes the maximum of three values per instruction and computes the row sum on the matrix unit therefore needs at least \zn{sol/floor/valu/listed} vector cycles per block, against \zn{sol/aiter/valu} in AITER's kernel. Section~\ref{sec:fp8p} describes how \sys{} reaches this minimum.

\subsection{Clock Under Sustained Load}
\label{sec:limits-clock}

The matrix pipe completes the datasheet number of operations per cycle in every run, so throughput differs between runs only through the shader clock (SCLK). We measure SCLK while every SIMD runs the same FP8 MFMA loop and change only the values in the operand registers. The clock is \zn{ceil/fp8-32x32-const/sclk}\,GHz with constant operands, \zn{ceil/fp8-32x32/sclk}\,GHz with operands that change every instruction, as $K$ and $V$ do in attention, and \zn{ceil/fp8-32x32-randbits/sclk}\,GHz with random bit patterns. Adding \texttt{v\_fma\_f32} instructions between the MFMAs lowers it further (Table~\ref{tab:ceil}). The clock thus falls as the values in the arithmetic units change more from one instruction to the next.\footnote{Each bit that switches between $0$ and $1$ charges or discharges a small capacitance in the circuit, so the current that the chip draws grows with the number of bits that change per cycle. Constant operands switch few bits in the multipliers, and random bit patterns switch about half of them on every instruction.} On NVIDIA GPUs, inputs that switch more bits also draw more power~\citep{gregersen2024input} and run at a lower clock once the board reaches its power limit~\citep{he2024predictable}. On the MI355X, board power stays below the $1400$\,W cap in all of these runs, so the board power limit does not set the clock. Instead, the throttle counters of amd-smi, AMD's GPU monitoring tool, point to a fast-acting limit on current that the telemetry does not identify (Section~\ref{app:sclk}).

Because the clock depends on the operands, we measure empirical peaks under sustained load with operands that change every instruction, over about $2.5$\,s of back-to-back launches on all $256$ CUs (Table~\ref{tab:ceil}). A loop of 16$\times$16 MFMAs, the shape with the higher throughput, sustains \zn{ceil/bf16/pf}\,PFLOP/s in BF16 and \zn{ceil/fp8/pf}\,PFLOP/s in FP8. A synthetic attention loop with the per-tile counts of MFMAs, LDS reads, loads into LDS, and barriers of \sys{}'s FP8 kernel at $\dqk{=}128$ and the vector instructions of lean P sustains \zn{ceil/fp8-d128/mixpf}\,PFLOP/s. The kernel itself issues two more row-sum MFMAs per tile and more vector-pipe cycles than this loop. A synthetic loop with the instructions of the BF16 kernels runs slower than the kernels themselves, so the table leaves it out.

\begin{table}[t]
\centering
{\footnotesize
\setlength{\tabcolsep}{4pt}
\begin{tabular}{lrrr}
\toprule
workload & GHz & kW & PF/s \\
\midrule
BF16 MFMA 32$\times$32 & \zn{ceil/bf16-32x32/sclk} & \zn{ceil/bf16-32x32/kw} & \zn{ceil/bf16-32x32/pf} \\
BF16 MFMA 16$\times$16 & \zn{ceil/bf16-16x16/sclk} & \zn{ceil/bf16-16x16/kw} & \zn{ceil/bf16/pf} \\
FP8 MFMA 32$\times$32 & \zn{ceil/fp8-32x32/sclk} & \zn{ceil/fp8-32x32/kw} & \zn{ceil/fp8-32x32/pf} \\
FP8 MFMA 16$\times$16 & \zn{ceil/fp8-16x16/sclk} & \zn{ceil/fp8-16x16/kw} & \zn{ceil/fp8/pf} \\
FP4 MFMA 16$\times$16 & \zn{ceil/fp4-16x16/sclk} & \zn{ceil/fp4-16x16/kw} & \zn{ceil/fp4-16x16/pf} \\
VALU only & \zn{ceil/valu/sclk} & \zn{ceil/valu/kw} & -- \\
LDS reads only & \zn{ceil/lds/sclk} & \zn{ceil/lds/kw} & -- \\
FP8 attention loop (synthetic) & -- & -- & \zn{ceil/fp8-d128/mixpf} \\
\midrule
datasheet BF16/FP8/FP4 & 2.40 & -- & 2.5/5.0/10 \\
\bottomrule
\end{tabular}}
\caption{Under sustained load the MFMA loops run below the $2.4$\,GHz clock that the datasheet peaks assume, so each reaches less than its datasheet throughput. The table gives empirical peaks, the sustained throughput of microbenchmark loops on all $256$ compute units (CUs) with two waves per SIMD and operands that change every instruction, over about $2.5$\,s of back-to-back launches. The columns give the shader clock (SCLK) in GHz, the board power in kW, and the throughput in PFLOP/s. The synthetic attention loop has the per-tile counts of MFMAs, reads from the on-chip shared memory (LDS), loads into it (DMA), and barriers of \sys{}'s FP8 kernel at $\dqk{=}128$. The BF16 rows use \texttt{v\_mfma\_f32\_32x32x16\_bf16} and \texttt{v\_mfma\_f32\_16x16x32\_bf16}, the FP8 rows \texttt{v\_mfma\_f32\_32x32x64\_f8f6f4} and \texttt{v\_mfma\_f32\_16x16x128\_f8f6f4} with e4m3 operands, and the FP4 row \texttt{v\_mfma\_f32\_16x16x128\_f8f6f4} with FP4 operands, all in their unscaled forms. The datasheet peaks assume $2.4$\,GHz.}
\label{tab:ceil}
\end{table}

\begin{table}[t]
\centering
{\footnotesize
\begin{tabular}{@{}lp{0.74\columnwidth}@{}}
\toprule
Attention & Causal forward pass, without sliding-window, document, or block masks \\
Heads & MHA and GQA ($H_q$ a multiple of $H_{kv}$) \\
Layouts & BHSD and BSHD (strided) \\
Batch & Any $B$ with equal sequence lengths, without packed variable-length batches \\
Lengths & $M{=}N$, a multiple of $256$ (no KV-cache prefix or decoding) \\
Inputs & BF16, or e4m3 with one FP32 scale per tensor \\
Outputs & BF16 $O$ and FP32 LSE \\
\bottomrule
\end{tabular}}
\caption{Configurations that the \sys{} kernels support. MHA is multi-head attention and GQA grouped-query attention, in which $H_q/H_{kv}$ query heads share each key-value head. BHSD and BSHD give the order of the batch, head, sequence, and head-dimension axes in memory, $M$ and $N$ are the numbers of queries and keys, and LSE is the log-sum-exp of each row of scores.}
\label{tab:support}
\end{table}

\section{Kernel Design}
\label{sec:design}
\label{app:support}

\sys{} has four causal forward kernels. They cover BF16 and FP8, each at $\dqk{=}\dv{=}128$ and at the MLA shape with $\dqk{=}192$ and $\dv{=}128$, which they compute without padding the query-key dimension to $256$. All four accept multi-head and grouped-query attention and return the LSE (Table~\ref{tab:support}). The BF16 kernels are written in HIP~\citep{hip}, and the FP8 kernels are CDNA4 assembly emitted by a generator (Section~\ref{sec:asm}). The BF16 kernels also run the BF16 heads of the mixed-precision policies in Section~\ref{sec:res-e2e}, which run some heads or layers of a model in BF16 and the rest in FP8.

\subsection{Structure Shared by All Kernels}
\label{sec:common}

A workgroup has eight waves, two per SIMD, and each wave owns $32$ query rows. The kernels compute $KQ^{\top}$ instead of $QK^{\top}$, so that in the output layout of the matrix instruction each lane holds the scores of a single query. The maximum and the sum of a row then need only one exchange between the two lanes that share its query, and rescaling the output takes one scalar per lane. Blocks of $K$ and $V$ pass through LDS, and the kernel loads the next blocks while it processes the current blocks.

\paragraph{Causal pairing}
In causal attention, where each query attends only to the keys up to its own position, the queries of tile $j$ (the $j$-th group of consecutive query rows, counting from $0$) attend to $j{+}1$ blocks of keys, so tiles near the end of the sequence have the most work. A workgroup runs tile $j$ and tile $\mathit{last}{-}j$ back to back, so every workgroup does the same amount of work (tile $j$ has $j{+}1$ blocks and tile $\mathit{last}{-}j$ has $\mathit{last}{-}j{+}1$, so every pair has $\mathit{last}{+}2$). A tile has \zn{kernel/qtile/rows} query rows, which gives \zn{kernel/pairs/16k} pairs per query head at $16$K. With grouped-query attention and an even group size, each pair covers two query heads with tiles of half the size, and the number of pairs per query head stays the same. A persistent grid of up to $256$ workgroups, one per CU, works through the pairs (each workgroup loops over many pairs instead of processing one and exiting).

\begin{figure}[t]
\centering
\includegraphics[width=\columnwidth]{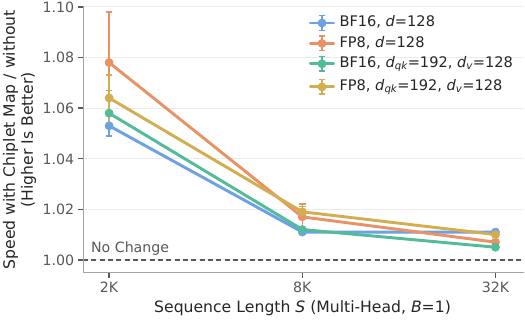}
\caption{The chiplet map helps most at short sequences. The figure gives the throughput of \sys{} with the chiplet map divided by that of the same kernel without it. The map sends consecutive tile pairs of one head to the same chiplet (XCD), so that they share its L2 cache. Multi-head attention, batch size $B{=}1$, kernel-only timing on Gaussian inputs as in Table~\ref{tab:t7}.}
\label{fig:xcd}
\end{figure}

\paragraph{Chiplet placement}
\label{app:xcd}
The hardware places workgroup $i$ on chiplet $i \bmod 8$, and each chiplet has its own L2 cache. In the default order of the grid the index of the tile pair varies fastest. Consecutive pairs of one head then run on different chiplets, and each chiplet loads the $K$ and $V$ of that head into its own L2 cache. Similar to the chiplet-aware grid order of HipKittens~\citep{hipkittens}, we remap the grid so that each chiplet receives a contiguous run of tile pairs from the same head, which then share $K$ and $V$ in one L2 cache. Figure~\ref{fig:xcd} gives the speed with the map relative to the speed without it. The gain is largest at $2$K and falls with the sequence length. Each element of $K$ and $V$ is used by a number of queries that grows with $S$, so the extra loads take a smaller share of the time at longer sequences.

\paragraph{Lazy rescaling}
When the running maximum of every row in a wave grows by less than a threshold $\tau$, the kernel keeps the old maximum and skips the rescale of the output accumulator~\citep{shah2026flash4,hipkittens}. The output stays the same because the accumulator and $\ell$ use the same maximum.\footnote{The stale maximum is at most $\tau$ below the true one, so the probabilities exceed $1$ by at most a factor of $2^{\tau}$. The same factor enters the accumulator and $\ell$, and the division $\mathrm{acc}/\ell$ cancels it.} \sys{} uses $\tau{=}0.125$ in FP8, $0.5$ in BF16 at $\dqk{=}128$, and $8$ in BF16 at $\dqk{=}192$, in base-2 units of the score.

\subsection{FP8 Probabilities}
\label{sec:fp8p}

\subsubsection{FP8 Probabilities from One Multiply-Add}
\label{sec:lean}

Write $x=t-m\le0$ for the base-2 score of a key relative to the row maximum, so that $p=2^{x}$. A standard FP8 kernel computes $x$ with one FMA, evaluates $2^x$ with \texttt{v\_exp\_f32}, and rounds the result to e4m3 with \texttt{v\_cvt\_pk\_fp8\_f32}. We call this exact P. It occupies the vector pipe for $14$ cycles per score, $2$ for the FMA, $8$ for the exponential, and $4$ for its half of a packed conversion (Section~\ref{sec:limits-valu}).

The code of an e4m3 value is $k=8E+F$, with exponent field $E$ and mantissa field $F$. For $E\ge1$ the value is $2^{E-7}(1+F/8)$. The value doubles every $8$ steps of $k$ and grows linearly between powers of two.\footnote{For example, the codes of $1$, $2$, and $4$ are $56$, $64$, and $72$. The code $60$, halfway between $56$ and $64$, stands for $1.5$, close to $2^{0.5}\approx1.41$. Adding $8$ to a code doubles its value, so the code behaves like $8\log_2$ of the value plus a constant.} The code of $2^{x}$ is therefore close to $8x$ plus a constant, which is the observation behind Schraudolph's fast exponential~\citep{schraudolph1999exp}. Lean P computes
\begin{equation}
y \;=\; 8\,(x+\sigma+7) - \beta
\label{eq:lean}
\end{equation}
and rounds $y$ to the nearest integer. The result is the e4m3 code of an approximation of $2^{x+\sigma}$. Because $x$ is a linear function of the raw score, Equation~\ref{eq:lean} is one FMA per score, whose coefficients fold in the softmax scale, the input scales, and the row maximum. The shift $\sigma{=}6$ multiplies every probability by $2^6$, which moves small probabilities $6$ binades further from the bottom of the e4m3 range. The factor $2^6$ cancels in the output, because the row sum is computed from the same scaled probabilities (Section~\ref{sec:rowsum}). The bias $\beta{=}0.35$ centers the interpolation error around zero (a straight line between two powers of two lies above the curve $2^x$, so without the bias every probability would be slightly too large). The conversion \texttt{v\_cvt\_pk\_u8\_f32} rounds $y$ to the nearest integer, clamps it to $[0,255]$, and writes the byte into a packed register, so a negative $y$ becomes code $0$. Lean P occupies the vector pipe for $6$ cycles per score, $2$ for the FMA and $4$ for the conversion, and it lowers the vector cycles per block in Table~\ref{tab:budget} from \zn{budget/cyc/exactp} to \zn{budget/cyc/leanp}. Compared with an earlier HIP kernel that used a Schraudolph exponential in FP32 and \texttt{v\_cvt\_pk\_fp8\_f32}, lean P reduces the number of $8$-cycle vector instructions in the FP8 $\dqk{=}128$ kernel from \zn{lean/c8/before} to \zn{lean/c8/after} and the number of instructions in its loops by \zn{lean/issue/drop}\% (counted in the kernel binary rather than as executed).

\begin{figure}[t]
\centering
\includegraphics[width=\columnwidth]{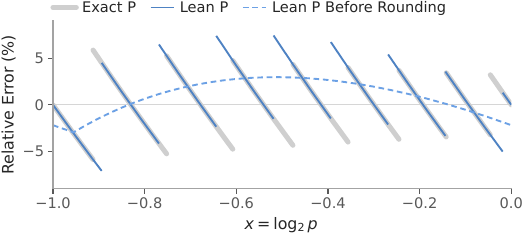}
\caption{Lean P, which computes each FP8 probability with one multiply-add instead of an exponential, stays within \zn{lean/pwl/relerr}\% of the true probability, against \zn{leanmath/exact/max}\% for rounding the exact value. The figure gives the relative error of the e4m3 probability with exact P and lean P over one binade (one factor of two) of $p$, plotted against $x=\log_2 p$. Exact P rounds $2^x$ to the nearest e4m3 value. Lean P rounds $y$ from Equation~\ref{eq:lean} (with shift $\sigma{=}6$ and bias $\beta{=}0.35$) to the nearest integer and uses it as the e4m3 bit pattern, and the dashed line is the error of $y$ before rounding. Over the normal e4m3 range the error depends on the fractional part of $x$ alone, so one binade shows all of it.}
\label{fig:leanp}
\end{figure}

Lean P is less accurate than exact P. The linear interpolation between powers of two differs from $2^x$ by up to \zn{lean/pwl/relerr/pwl}\% with the bias and by up to \zn{lean/pwl/relerr/unbiased}\% without it. After rounding to a code, lean P is within \zn{lean/pwl/relerr}\% of $p$, and rounding the exact $2^x$ to the nearest code is within \zn{leanmath/exact/max}\% (Figure~\ref{fig:leanp}). Probabilities more than $12$ binades (factors of two) below the row maximum fall into the subnormal range of e4m3, where the value grows linearly with the code instead of doubling every $8$ codes, so lean P underestimates them by up to $77\%$. Section~\ref{sec:res-kernel} measures the effect on the output.

\subsubsection{Per-Block Scaling}
\label{sec:pblock}

\begin{figure}[t]
\centering
\includegraphics[width=\columnwidth]{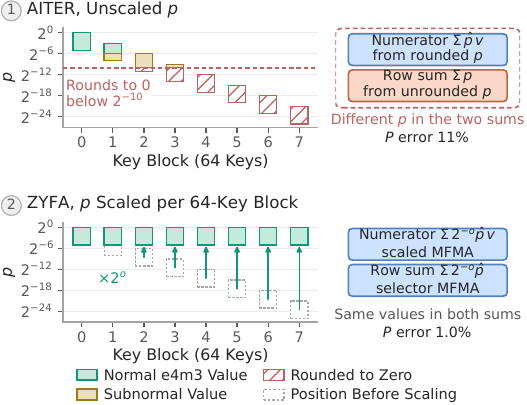}
\caption{Scaling each block of probabilities before rounding keeps them in the range where FP8 is accurate, and dividing by the sum of the same rounded values keeps the attention weights summing to one. The schematic shows the probabilities $p$ of one query row by $64$-key block. (1) AITER's FP8 kernel rounds $p$ to e4m3 relative to the row maximum. Blocks more than $6$ binades (factors of two) below the maximum become subnormal, with fewer significant bits, and values more than $10$ binades below underflow to zero in the numerator. The row sum adds the unrounded $p$ in FP32. (2) \sys{} multiplies each block by $2^{o}$ so that its largest probability gets the same e4m3 value as $p{=}1$, and the scaled matrix instruction (MFMA) that computes $PV$ divides the factor out at no cost. The numerator and the row sum use the same rounded values $\hat p$. The $P$ error is the relative error that rounding $p$ adds to the attention output on Mistral-7B (Figure~\ref{fig:errflow}).}
\label{fig:pscale}
\end{figure}

An unscaled e4m3 probability keeps its 3 mantissa bits down to $6$ binades below the row maximum and underflows to zero more than $10$ binades below it. In Mistral-7B layer~20 at $16$K, the block maximum is more than $8$ binades below the row maximum for \zn{pcut/mistral-l20/t8}\% of the (row, block) pairs that the causal mask does not hide. The block maximum is more than $16$ binades below the row maximum for \zn{pcut/mistral-l20/t16}\%. For each query and each $64$-key block, both FP8 kernels therefore multiply the probabilities of the block by $2^{o}$, where $o$ is the number of whole binades between the block maximum and the row maximum. In lean P this adds $8o$ to $y$. The block maximum then gets the code that $p{=}1$ would get (Figure~\ref{fig:pscale}). The scaled FP8 matrix instruction of CDNA4, which multiplies its operands by power-of-two block scales as in the microscaling formats~\citep{rouhani2023mx}, removes the factor with a block scale of $2^{-o}$ in the $PV$ product.\footnote{In the microscaling formats, each block of $32$ values shares one 8-bit power-of-two exponent. The scaled MFMA applies such exponents to its operands as part of the product, so removing the factor $2^{o}$ costs no extra instruction.} Probabilities more than $13$ binades (lean P) or $16$ binades (exact P) below the maximum of their own block still underflow. In the Mistral-7B layer they carry at most \zn{lean/blocktail/mass}\% of the mass of a row. In HIP versions of the kernels, per-block scaling cost \zn{mxq/cost/d128}\% of the throughput at $\dqk{=}128$ and \zn{mxq/cost/d192}\% at $\dqk{=}192$.

\subsubsection{Row Sum on the Matrix Unit}
\label{sec:rowsum}

The output divides $\sum_j p_j v_j$ by $\ell=\sum_j p_j$. Let $\hat p$ denote the e4m3-rounded probability. If the numerator uses $\hat p$ and $\ell$ uses the unrounded $p$, the two sums differ by the rounding error of every term. Underflow always removes mass, so the difference accumulates over a long row.\footnote{Dividing by the unrounded sum treats every probability that underflowed to zero as a key whose value vector is zero, so each output shrinks toward zero by the weight of the dropped keys. Dividing by the sum of the rounded probabilities instead gives a proper weighted average, with weights that sum to one, of the values whose probabilities survive rounding.} \sys{} computes $\ell$ from the same rounded probabilities $\hat p$ that multiply $V$. The row sum is the product of these FP8 values with a vector of ones. For each $64$-key block one extra 16$\times$16$\times$128 FP8 MFMA computes it, with a $0/1$ matrix as second operand that gives each lane the sum for its own query. The block sums are multiplied by $2^{-o}$ and accumulated in FP32. The extra MFMA adds \zn{lean/rowsum/matrix}\% to the matrix cycles per block and removes the vector additions. The kernels return the LSE of the rounded probabilities. A split-key merge (as in split-KV decoding or context-parallel attention, where different workgroups or GPUs process different ranges of keys) multiplies each partial output by the exponential of its LSE to recover the unnormalized sum $\sum_j p_j v_j$, so the LSE has to come from the same rounded probabilities as the output. With lean P this LSE differs from the fp64 LSE by at most \zn{lean/lse/err} nats in the worst row of the synthetic inputs.

\paragraph{Non-finite inputs}
\label{app:nan}
The row sum on the matrix unit multiplies the FP8 probabilities of the other queries by $0$, and $0\times\mathrm{NaN}$ is NaN. In the FP8 $\dqk{=}192$ kernel a NaN in query row $r$ therefore also makes row $r{+}16$ NaN, and a NaN key reaches rows that cannot see it. In every FP8 kernel a NaN in $V$ reaches rows of the same $256$-row tile that cannot see it. With lean P a NaN key converts to code $0$, the FP8 value zero, and is dropped. With finite inputs the extra products are zero and change no output row.

\subsection{Instruction Schedule of the FP8 Kernels}
\label{sec:asm}

\begin{figure}[t]
\centering
\includegraphics[width=\columnwidth]{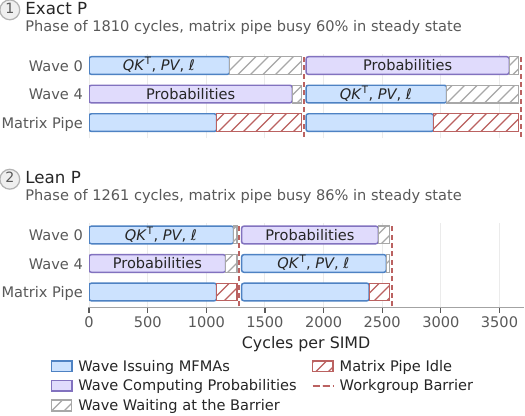}
\caption{With lean P the matrix pipe idles for a much smaller part of each phase, because the probabilities are ready before the matrix work ends. The figure shows one SIMD of \sys{}'s FP8 kernel at $\dqk{=}128$ over two phases, in which its two waves take turns, one issuing the matrix instructions (MFMAs) while the other computes the FP8 probabilities, and each processes one $128$-key block. Waves 0 and 4 share SIMD~0. Bar widths are the mean cycles from the start of a phase until each wave reaches the workgroup barrier that ends it, taken from thread traces. The busy cycles of the matrix pipe are drawn together at the start of each phase. (1) With exact P, in an earlier build of the same kernel, the wave that computes probabilities is the critical wave, the wave that reaches the barrier last and so sets the length of the phase, and the matrix pipe idles until the barrier. (2) With lean P the wave that issues the MFMAs is the critical wave. $\ell$ denotes the row-sum MFMAs.}
\label{fig:schedule}
\end{figure}

A Python generator emits the FP8 kernels as CDNA4 assembly, one kernel per head dimension. The generator gives us the fine-grained control of assembly together with the ease of development of Python. The two waves of each SIMD take turns in two phases per $128$-key block, with a workgroup barrier after each phase (Figure~\ref{fig:schedule}). FlashAttention-3 calls this arrangement ping-pong scheduling~\citep{shah2024flash3}, HipKittens uses it on CDNA GPUs~\citep{hipkittens}, and AITER uses it as well (Section~\ref{sec:limits-sol}). In each phase one wave, the MFMA wave, issues the matrix instructions of its block. These instructions are two row-sum MFMAs, eight $QK^{\top}$ MFMAs for the next block, and eight $PV$ MFMAs for the current block. Between these MFMAs the MFMA wave computes the row maximum and the rescale of the next block. The other wave, the softmax wave, computes the FP8 probabilities. In the next phase the two waves swap roles, so the matrix pipe always has one wave that feeds it. The wave that reaches the barrier last sets the length of a phase, and we call it the critical wave.

Both waves issue from the same SIMD, so the vector instructions of the softmax wave and the instructions of the MFMA wave compete for issue slots. In thread traces each vector instruction delays the critical wave by about \zn{asm/issue/valu} cycles, and exponentials and FP8 conversions delay it longer. With exact P the softmax wave is the critical wave, and the matrix pipe idles at the end of every phase. In steady state, which we take as $40$ consecutive phases in the middle of the key loop, the matrix pipe is busy for \zn{asm/exact/mfmabusy}\% of the cycles. With lean P the softmax wave issues \zn{asm/lean/smvalu} vector instructions per block and reaches the barrier before the MFMA wave. The MFMA wave becomes the critical wave, and the matrix pipe is busy for \zn{asm/lean/mfmabusy}\% of the cycles in steady state. Over the whole key loop, which includes the transitions between query tiles, the matrix pipe is busy for \zn{asm/exact/mfmabusy/loop}\% of the cycles with exact P and \zn{asm/lean/mfmabusy/loop}\% with lean P. All pipes of the assembly kernel are idle in \zn{att/asm-fp8-d128/allidle}\% of cycles, against \zn{att/base-fp8-d128/allidle}\% in the base kernel of Section~\ref{sec:limits-sol} (an earlier HIP version of \sys{}'s FP8 kernel), and the vector pipe of the assembly kernel is busy for \zn{sol/zyfa-asm/valu} cycles in the units of Table~\ref{tab:sol}. The HIP kernel with lean P remains the fallback for tensors larger than $32$-bit offsets allow, and on captured activations at sustained clocks it needs \zn{asm/hip/morecycles}\% more cycles than the assembly kernel.

\subsection{BF16 Kernels}
\label{sec:bf16}

\begin{figure*}[tp]
\centering
\begin{subfigure}[t]{0.49\textwidth}
\centering
\includegraphics[width=\linewidth]{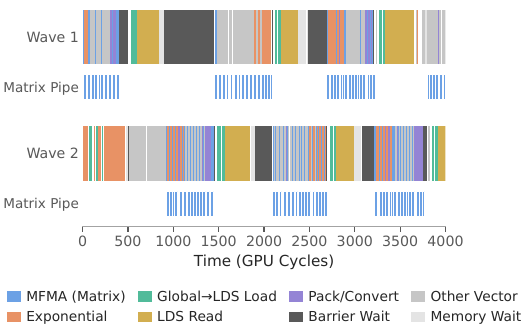}
\caption{\sys{}, $\dqk{=}128$}
\end{subfigure}\hfill
\begin{subfigure}[t]{0.49\textwidth}
\centering
\includegraphics[width=\linewidth]{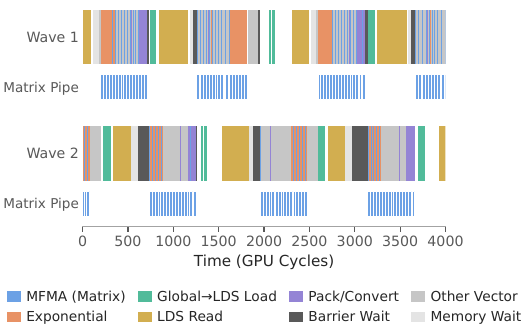}
\caption{AITER, $\dqk{=}128$}
\end{subfigure}\\[4pt]
\begin{subfigure}[t]{0.49\textwidth}
\centering
\includegraphics[width=\linewidth]{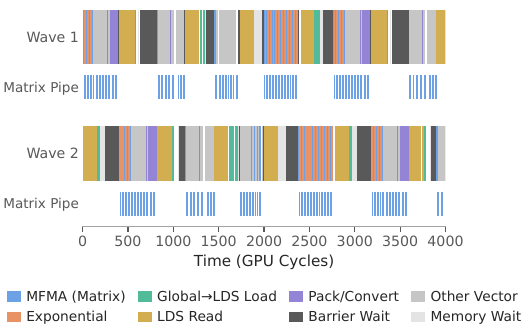}
\caption{\sys{}, $\dqk{=}192$}
\end{subfigure}\hfill
\begin{subfigure}[t]{0.49\textwidth}
\centering
\includegraphics[width=\linewidth]{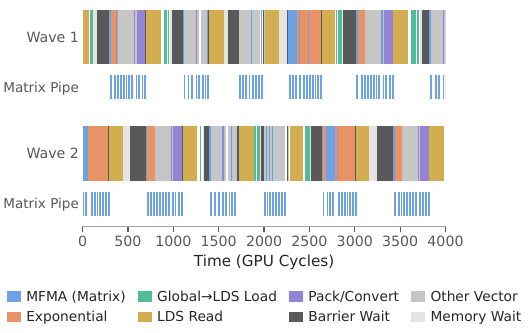}
\caption{AITER, $\dqk{=}192$}
\end{subfigure}
\caption{Thread traces showing that in BF16 the matrix products set the time of all four kernels of \sys{} and AITER, because they take more than twice as many cycles as the softmax (Table~\ref{tab:budget}). A thread trace records every instruction that a wave issues, cycle by cycle. Each panel shows the two waves that share one SIMD over $4000$ cycles in the middle of the loop over key blocks, where every block repeats the same work ($S{=}8$K tokens, $H{=}32$ heads, multi-head attention). Each wave row shows the class of each instruction that the wave issues, and the matrix-pipe row below it shows when that wave's MFMAs run. A vector instruction runs on the vector pipe during its bar in the wave row, whereas an MFMA keeps running on the matrix pipe after its short bar, while the wave issues other instructions.}
\label{fig:att-bf16}
\end{figure*}

The BF16 kernel at $\dqk{=}128$ keeps the 8-wave ping-pong schedule of HipKittens~\citep{hipkittens} and adds the structure of Section~\ref{sec:common}. The BF16 kernel at $\dqk{=}192$ follows the eight-stage pipeline of AITER's kernel for this shape, with a workgroup barrier after each stage. Memory stages read half of an operand into registers, and compute stages issue the matching MFMAs next to the softmax of the previous block. Waves 4--7 run one stage behind waves 0--3, so on each SIMD the MFMAs of one wave overlap the LDS reads of the other. $K$ is stored in LDS in groups of $8$ keys followed by $16$ bytes of padding, which makes every read of an MFMA operand free of bank conflicts.\footnote{LDS is divided into banks, and lanes that read different addresses in the same bank in one instruction take turns. The padding shifts consecutive groups of keys to different banks.} Figure~\ref{fig:att-bf16} shows thread traces of both BF16 kernels next to those of AITER's kernels.

\section{Experimental Setup}
\label{sec:experiments}
\label{app:methods}

\subsection{System and Baselines}
\label{app:system}

All experiments run on one node with eight AMD Instinct MI355X GPUs (gfx950), with the default automatic clock policy without pinned clocks, a $1400$\,W power cap, and a maximum shader clock (SCLK) of $2.4$\,GHz. Each job runs on one GPU, selected by its PCI bus address, and a job does not start on a GPU that runs another process. The software is ROCm~7.2.3 with HIP~7.2.53211 and AMD Clang~22.0.0, PyTorch~2.12, AITER~0.1.19, Triton~3.7.1, and transformers~5.15.1 in the vLLM ROCm container v0.28.0 with the amdgpu driver~6.14.14. All kernels are built from one source commit.

Throughput follows the clock of the slowest chiplet, which differs between GPUs.\footnote{The eight chiplets can run at different clocks, and a kernel ends only when the slowest chiplet finishes its share of the work.} From $4$K tokens up, the throughput of one configuration differs between GPUs by a median of \zn{gpu/spread/tfs/median}\% and at most \zn{gpu/spread/tfs}\%, and the ratio of \sys{}'s to AITER's throughput by a median of \zn{gpu/spread/ratio/median} and at most \zn{gpu/spread/ratio}, the most for FP8 at $\dqk{=}192$ and $32$K. These differences come from the configurations that were measured on more than one GPU. At $2$K tokens and below, a few configurations differ as much between sessions on one GPU as between GPUs, by up to \zn{gpu/session/short/tfs}\% in throughput and \zn{gpu/session/short/ratio} in the ratio to AITER. The largest difference is in AITER's FP8 kernel at $\dqk{=}192$ and $1$K, which runs at one of two speeds in each session. Each precision and attention type therefore keeps all its kernel-only jobs on one GPU, so every comparison within a table or figure uses runs on one GPU.

The baseline is the kernel that AITER~0.1.19~\citep{aiter}, AMD's library of inference kernels, selects by default: an assembly kernel for BF16 and FP8 at $\dqk{=}128$, a HIP kernel written with OPUS, AITER's C++ template library, for BF16 at $\dqk{=}192$, and Composable Kernel (CK)~\citep{ck} for FP8 at $\dqk{=}192$, where AITER has no assembly kernel. CK issues the CDNA3 matrix (MFMA) instruction \texttt{v\_mfma\_f32\_32x32x16\_fp8\_fp8} for both matrix products. On the MI355X it completes \zn{isa/mfma/fp8legacy/flopclk} FLOP per clock per SIMD, the same as BF16 and \zn{isa/mfma/fp8legacy/rate}\x{} the rate of \texttt{v\_mfma\_f32\_32x32x64\_f8f6f4}, the instruction that AITER's assembly kernel and \sys{} use.
Table~\ref{tab:fp8mfma} lists the FP8 matrix instructions that each FP8 kernel issues, read from the disassembled kernel binaries that the timing runs launch, with their rates measured on all $256$ CUs.

\begin{table*}[tp]
\centering
{\footnotesize\setlength{\tabcolsep}{3pt}
\begin{tabular}{lllrr}
\toprule
kernel & product & instruction & per kernel & FLOP/clock/SIMD \\
\midrule
\sys{}, FP8 $d{=}128$ & row sum of $P$ & \texttt{v\_mfma\_f32\_16x16x128\_f8f6f4} & 20 & 2041 \\
 & $QK^{\top}$ & \texttt{v\_mfma\_f32\_32x32x64\_f8f6f4} & 64 & 2048 \\
 & $PV$ & \texttt{v\_mfma\_scale\_f32\_32x32x64\_f8f6f4} & 80 & 2048 \\
\midrule
\sys{}, FP8 $d_{qk}{=}192$ & row sum of $P$ & \texttt{v\_mfma\_f32\_16x16x128\_f8f6f4} & 10 & 2041 \\
 & $QK^{\top}$ & \texttt{v\_mfma\_f32\_32x32x64\_f8f6f4} & 48 & 2048 \\
 & $PV$ & \texttt{v\_mfma\_scale\_f32\_32x32x64\_f8f6f4} & 40 & 2048 \\
\midrule
AITER assembly, FP8 $d{=}128$ & $QK^{\top}$ and $PV$ & \texttt{v\_mfma\_f32\_32x32x64\_f8f6f4} & 128 & 2048 \\
\midrule
Composable Kernel, FP8 $d_{qk}{=}192$ & $QK^{\top}$ and $PV$ & \texttt{v\_mfma\_f32\_32x32x16\_fp8\_fp8} & 80 & 1020 \\
\bottomrule
\end{tabular}
}
\caption{FP8 matrix (MFMA) instructions of each FP8 kernel, read from the disassembled kernel binaries. The count is the number of times each instruction appears in the kernel binary, and the rate is in FLOP per clock per SIMD unit, measured with two waves (groups of $64$ threads) per SIMD on all $256$ compute units (CUs). \sys{}'s $PV$ instruction applies the per-block scale of Section~\ref{sec:pblock}, a power of two for each block of $64$ probabilities. Composable Kernel (CK) is AITER's FP8 kernel at $\dqk{=}192$.}
\label{tab:fp8mfma}
\end{table*}

Before every timed run a correctness test compares each kernel with AITER on every reported configuration. In BF16 it requires a maximum absolute difference below $2{\times}10^{-3}$, and in FP8 a maximum absolute difference below $0.03$ and a signal-to-noise ratio (SNR) of at least $30$\,dB, which means that the RMS of the difference is at most about $3\%$ of the RMS of the output. The reference is itself an FP8 kernel, so this test catches defects rather than measuring accuracy, which Section~\ref{sec:res-kernel} measures against fp64.

\subsection{Kernel-Only Timing}
\label{app:timing}

Kernel-only timing counts only the time that the GPU spends executing the attention kernel. As in HipKittens~\citep{hipkittens}, a library of attention and matrix kernels for AMD GPUs, we report the mean device time of $100$ launches after $500$ warm-up launches. Device times come from \texttt{rocprofv3} kernel traces and exclude the time that the host takes to launch a kernel. The time of one launch is the total device time of every kernel that the operation dispatches (launches on the GPU) in the timed range, divided by the number of launches. We use $B{=}1$, $H{=}\max(32, 262144/S)$ query heads so that every call has at least $2^{18}$ query rows, and $H/4$ key-value heads for grouped-query attention (GQA). The inputs, which we call the Gaussian inputs, are $q$, $k$, and $v$ drawn from a normal distribution with standard deviation $0.1$, which makes the attention probabilities of each query row nearly uniform (Section~\ref{sec:res-real}). The GPU idles for $5$\,s before each configuration with $S\ge16$K. Each configuration runs at least three times, the order of kernels rotates between repetitions, and the reported spread is the range over repetitions. Throughput is $F$ from Equation~\ref{eq:flops}, which counts the FLOPs of $QK^{\top}$ and $PV$ in the causal half of the score matrix, divided by the time.

\subsection{Sustained Timing, Clock, and Power}
\label{app:sustained}

Kernel-only timing on the Gaussian inputs of kernel benchmarks measures a kernel at the clock that a short burst of launches gets. Because the clock depends on the data and settles under sustained load (Section~\ref{sec:limits-clock}), we also time kernels on model activations in long runs. We capture the post-RoPE $q$, $k$, and $v$ (after the rotary position embedding, as the attention kernel receives them) of one layer each of Mistral-7B-v0.3~\citep{jiang2023mistral}, Qwen3-8B~\citep{qwen3}, and DeepSeek-V2-Lite~\citep{deepseekv2} on WikiText-2~\citep{merity2017wikitext} at $S{=}16$K. We call these captured activations, and Gaussian inputs of the same shape serve as a control. Under sustained load the GPU lowers its shader clock (SCLK) and then holds it at a lower, steady value, which we call the sustained clock. Timing at sustained clocks means timing only after this clock has settled. In a sustained run one kernel launches back to back on a GPU that runs nothing else. After $2$\,s of idle time each kernel runs $1$\,s of warm-in launches and then a timed window of at least $3$\,s, so that the clock settles as it would in a long prefill. The comparisons of exact and lean P on the Mistral-7B layer (Section~\ref{sec:res-real}) and of the HIP and assembly kernels (Section~\ref{sec:asm}) instead run warm-in launches until two consecutive $0.5$\,s chunks agree in time within $0.3\%$, and then time a window of $4$\,s. Each sustained run is repeated three times with alternating kernel order, and we report the median and the range. SCLK is the median, over the samples of a run, of the clock of the slowest chiplet. Board power and the limiter status come from \texttt{amd-smi}. Energy efficiency in TFLOP/J is the throughput in TFLOP/s divided by the board power in watts.

\subsection{Error of One Attention Call}
\label{app:errmethod}

To measure the error that a kernel itself adds, we compare its output with attention computed in fp64 on the same inputs the kernel receives: $Q$, $K$, and $V$ rounded to BF16, or to e4m3 with one power-of-two scale per tensor.\footnote{The error of quantizing $Q$, $K$, and $V$ is the same for every FP8 kernel that uses the same scales. Comparing against the quantized inputs removes it, so the comparison isolates the arithmetic of the kernel. Section~\ref{sec:res-e2e} measures the input error separately.} The error of a kernel is the relative RMS error and the maximum absolute error of its output against this reference. We compute it on every layer of the three models, from the post-RoPE $q$, $k$, and $v$ of the first $4095$ tokens of the WikiText-2 test set after the BOS token, and report means over layers, or over random seeds for synthetic inputs.

\subsection{End-to-End Accuracy}
\label{app:e2emethods}

For end-to-end accuracy, we run each model in BF16 and replace its attention, per layer and per head, by the kernel under test. The study uses Mistral-7B-v0.3 ($\dqk{=}128$, $32$ query and $8$ KV heads), Qwen3-8B ($\dqk{=}128$, $32$ query and $8$ KV heads, with QK normalization), and DeepSeek-V2-Lite (MLA with $\dqk{=}192$, $\dv{=}128$, and $16$ heads), all with BF16 weights. The documents come from the PG-19 test books~\citep{rae2020pg19} in a seeded shuffle. A document is the text of one book after its first $2000$ characters, cut to $8$K, $16$K, or $32$K tokens. Each model is evaluated with teacher forcing (the model reads the true text, and we score its prediction of each next token) on \zn{pareto/docs} documents from \zn{pareto/books} books, one third at each length. A separate calibration set of \zn{calib/docs} documents from \zn{calib/books} other books, one third at each length, is used to select policies and to set static FP8 scales.

\paragraph{Metrics and the BF16 noise band}
We report the per-token KL divergence of the next-token distribution from that of the model with \sys{}'s BF16 attention, which measures how far the predictions move, and the change in negative log-likelihood of the true next token (dNLL), both pooled per token over documents. Their $95\%$ confidence intervals are percentile bootstraps over books with $1000$ resamples.\footnote{A bootstrap interval resamples the books with replacement and recomputes the statistic for each resample. Resampling whole books rather than tokens accounts for the correlation between the tokens of one book.} A paired difference between two variants resamples the per-document sums of both variants together, with $2000$ resamples. Two correct BF16 kernels give slightly different model outputs because they round in different orders. The \emph{BF16 noise band} runs from zero to the upper end of the $95\%$ bootstrap interval of the KL divergence between the models with \sys{}'s and AITER's BF16 kernels. A policy inside the band is therefore as close to the BF16 model as a second, equally correct BF16 kernel would be. Attention in fp64 lies just inside the BF16 noise band on Mistral-7B, so the band reflects rounding order alone and is a strict criterion.

\paragraph{FP8 inputs and policies}
FP8 inputs use per-tensor power-of-two descales, computed after RoPE and after any rotation or K smoothing.\footnote{A descale is the factor that converts a scaled FP8 value back to its original range. A power of two changes only the exponent, so applying it adds no rounding error.} Per-call scales, which we also call dynamic scales, use the absolute maximum of each tensor in each call. Static scales are set per layer, transform, and tensor from the largest absolute maximum over the calibration documents, taken with BF16 attention in every layer. A policy assigns each head or layer to BF16 or FP8 and may rotate the FP8 inputs. Policies and static scales are selected on the calibration documents.

\paragraph{Error of each stage of one call}
To attribute the FP8 error of one call to its stages, we divide that error into three terms. The reference for each call is fp64 attention on the BF16 $q$, $k$, and $v$ of the call. The input term is the error of fp64 attention on the e4m3-quantized inputs. The $P$ term is the error that quantizing $p$ as the kernel does adds to the input term, emulated with the exact row maximum and FP32 accumulation. The last term is the error that the kernel itself adds, which comes from the accumulation order, the running maximum, and the rounding of the output. Each term is a relative RMS error pooled over all heads of the first \zn{budget/docs} evaluation documents, one third at each length.

\paragraph{Rotation and K smoothing}
After RoPE, $Q$ and $K$ are multiplied in FP32 by the same orthogonal matrix $R=\mathrm{diag}(s)\,(C_m\otimes H_{D/m})$. Here $s$ is a vector of random signs with seed $0$, $H$ is the normalized Sylvester--Hadamard matrix, and $C_m$ is the orthonormal DCT-II of size $m$. We use $m{=}1$ for $D{=}128$ and $m{=}3$ for $D{=}192$, because Sylvester's construction gives only sizes that are powers of two. Since $R$ is orthogonal, $(QR)(KR)^{\top}=QK^{\top}$, and the rotation changes only the rounding error. One matrix serves every layer and head. FP8 scales are computed after the rotation, and $V$ and $O$ are not rotated. K smoothing~\citep{zhang2024sage} subtracts the mean key of each head over the sequence before the rotation. The subtraction lowers every score of a query row by the same value, which the softmax cancels.\footnote{Subtracting a constant from every score of a row multiplies every $p$ of the row by the same factor, which the division by the row sum removes.}

\paragraph{Timing inside the model}
Attention time inside the model comes from kernel-only windows, intervals between two GPU timing events that contain only the attention kernels of one call. A short GPU busy-wait before each start event keeps the host launch time out of the window.\footnote{Without the busy-wait, the GPU can record the start event and then wait for the host to launch the kernel, so the window would include the launch overhead.} Adding the busy-wait changed the time of policies with mixed layers by up to about $2\%$, so we treat smaller differences as unresolved. Quantization, the output descale, and the copies that gather the heads of each precision in a mixed layer and scatter their outputs back into $O$ lie outside the window. Times with a fused quantization pass add the time of an existing one-pass kernel that reads $Q$ in BF16 and writes e4m3, plus one pass over $K$ when $K$ is rotated or smoothed. $K$ and $V$ are quantized when they are written to the cache, so no separate pass is added for them. Table~\ref{tab:mixcopies} gives the time of the gather and scatter copies, which every mixed-precision time in Section~\ref{sec:res-pareto} excludes.

\section{Results}
\label{sec:results}
\label{app:results}

\subsection{Speed}
\label{sec:res-speed}
\label{sec:res-recipe}
\label{app:cells}

\begin{table}[t]
\centering
{\footnotesize
\setlength{\tabcolsep}{4.2pt}
\begin{tabular}{lrrrrrrr}
\toprule
& \multicolumn{3}{c}{multi-head} & \multicolumn{3}{c}{grouped-query} & \\
\cmidrule(lr){2-4}\cmidrule(lr){5-7}
kernel & 2K & 16K & 32K & 2K & 16K & 32K & spread \\
\midrule
BF16 $128$ & \zn{t7/mha/bf16-d128/2k} & \zn{t7/mha/bf16-d128/16k} & \zn{t7/mha/bf16-d128/32k} & \zn{t7/gqa/bf16-d128/2k} & \zn{t7/gqa/bf16-d128/16k} & \zn{t7/gqa/bf16-d128/32k} & \zn{t7/mha/bf16-d128/rng32k} \\
BF16 $192$ & \zn{t7/mha/bf16-d192/2k} & \zn{t7/mha/bf16-d192/16k} & \zn{t7/mha/bf16-d192/32k} & \zn{t7/gqa/bf16-d192/2k} & \zn{t7/gqa/bf16-d192/16k} & \zn{t7/gqa/bf16-d192/32k} & \zn{t7/mha/bf16-d192/rng32k} \\
FP8 $128$ & \zn{t7/mha/fp8-d128/2k} & \zn{t7/mha/fp8-d128/16k} & \zn{t7/mha/fp8-d128/32k} & \zn{t7/gqa/fp8-d128/2k} & \zn{t7/gqa/fp8-d128/16k} & \zn{t7/gqa/fp8-d128/32k} & \zn{t7/mha/fp8-d128/rng32k} \\
FP8 $192$ & \zn{t7/mha/fp8-d192/2k} & \zn{t7/mha/fp8-d192/16k} & \zn{t7/mha/fp8-d192/32k} & \zn{t7/gqa/fp8-d192/2k} & \zn{t7/gqa/fp8-d192/16k} & \zn{t7/gqa/fp8-d192/32k} & \zn{t7/mha/fp8-d192/rng32k} \\
\bottomrule
\end{tabular}}
\caption{Throughput of \sys{} divided by that of AITER's default kernel at batch size $B{=}1$, so values above $1$ mean that \sys{} takes less time. Inputs are Gaussian (random $q$, $k$, and $v$ with standard deviation $0.1$), and times are kernel-only (the device time of the kernel alone), averaged over repetitions. Rows give the precision and the query-key head dimension $\dqk$. At $\dqk{=}192$ the FP8 baseline is Composable Kernel (CK). The last column is the spread of \sys{}'s multi-head throughput over repetitions at $32$K. Figure~\ref{fig:throughput} gives all sequence lengths, and Table~\ref{tab:b4} gives $B{=}4$.}
\label{tab:t7}
\end{table}

\begin{figure*}[tp]
\centering
\begin{subfigure}[t]{0.49\textwidth}
\centering
\includegraphics[width=\linewidth]{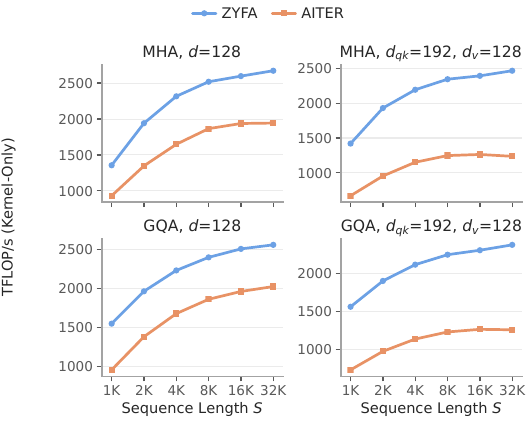}
\caption{FP8, $B{=}1$}
\label{fig:fp8b1}
\end{subfigure}\hfill
\begin{subfigure}[t]{0.49\textwidth}
\centering
\includegraphics[width=\linewidth]{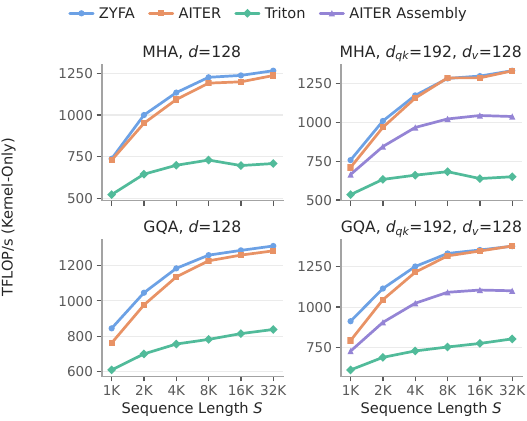}
\caption{BF16, $B{=}1$}
\label{fig:bf16b1}
\end{subfigure}\\[4pt]
\begin{subfigure}[t]{0.49\textwidth}
\centering
\includegraphics[width=\linewidth]{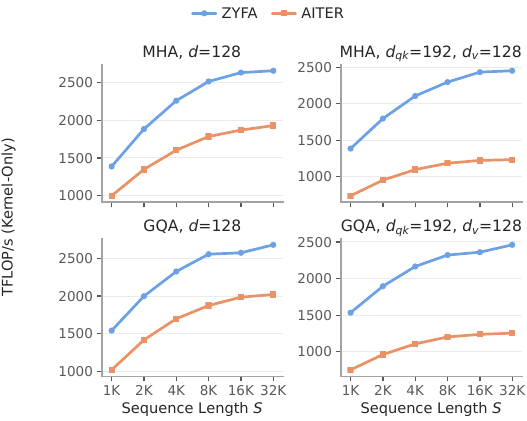}
\caption{FP8, $B{=}4$}
\label{fig:fp8b4}
\end{subfigure}\hfill
\begin{subfigure}[t]{0.49\textwidth}
\centering
\includegraphics[width=\linewidth]{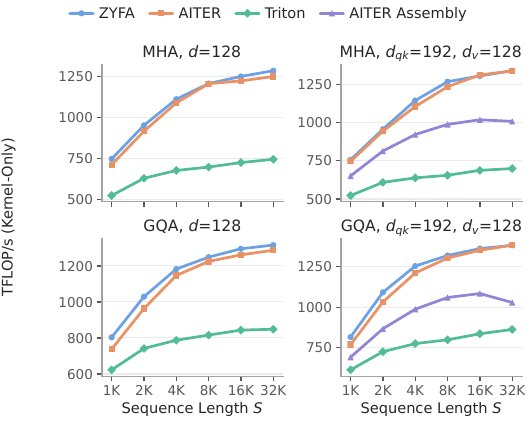}
\caption{BF16, $B{=}4$}
\label{fig:bf16b4}
\end{subfigure}
\caption{Throughput of \sys{} and AITER against sequence length on Gaussian inputs (random $q$, $k$, and $v$ with standard deviation $0.1$). Times are kernel-only, the device time of the kernel alone without the time that the host takes to launch it. AITER is AITER's default kernel for each configuration, which at $\dqk{=}192$ is its OPUS kernel in BF16 and Composable Kernel in FP8. The BF16 panels also show AITER's Triton kernel and, at $\dqk{=}192$, AITER's assembly kernel.}
\label{fig:throughput}
\end{figure*}

\begin{figure}[t]
\centering
\includegraphics[width=\columnwidth]{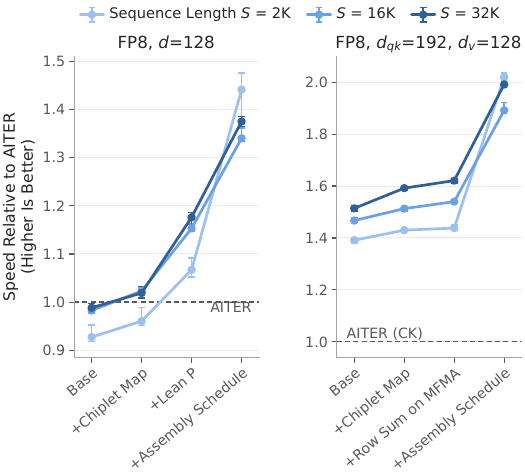}
\caption{Speed of \sys{}'s FP8 kernels relative to AITER's default kernel as design decisions are added from left to right, kernel-only on Gaussian inputs, multi-head at $B{=}1$. Error bars span the repetitions. At $\dqk{=}192$ the baseline is Composable Kernel (CK), AITER's default FP8 kernel for that shape. Step ``+Lean P'' computes each FP8 probability with a multiply-add instead of an exponential, and step ``+Row Sum on MFMA'' sums each row of probabilities on the matrix unit. Both steps also turn on per-block scaling, and step ``+Assembly Schedule'' replaces the HIP kernel with the hand-scheduled assembly kernel. The rightmost point of each line is the final kernel, measured in the same jobs as Table~\ref{tab:t7}. Section~\ref{sec:ladder} defines every step, and Figure~\ref{fig:ladder-bf16} shows the BF16 kernels.}
\label{fig:ladder}
\end{figure}

\paragraph{Gaussian inputs}
\sys{}'s FP8 kernels achieve a higher throughput than AITER's default kernels at every sequence length from $1$K to $32$K, and \sys{}'s BF16 kernels are within $3\%$ of AITER's throughput at $16$K and $32$K and achieve a higher throughput at $2$K (Table~\ref{tab:t7} and Figure~\ref{fig:throughput}). The pipe cycles of Table~\ref{tab:budget} explain the difference between the two precisions. In BF16 the matrix products are the critical path of both libraries, the chain of work that sets the kernel time, and the times of the two libraries differ by a few percent. In FP8 the softmax is the critical path of a kernel with an exact softmax, and lean P, which replaces the exponential with one multiply-add, removes more than half of its vector work. The throughput ratio of \sys{} to AITER is largest at FP8 $\dqk{=}192$, where the assembly schedule gives the largest step (Figure~\ref{fig:ladder}) and the baseline, Composable Kernel (CK), issues FP8 matrix instructions that run at the BF16 rate (Section~\ref{sec:experiments}). At $32$K the FP8 kernel at $\dqk{=}128$ reaches \zn{peakpct/mha/fp8-d128/32k/zyfa}\% of the FP8 datasheet peak and \zn{mixpct/mha/fp8-d128/32k}\% of the throughput of the synthetic attention loop of Section~\ref{sec:limits-clock}. The rest of the distance to the datasheet peak comes from the lower clock under sustained load and from vector work that does not overlap the MFMAs.

\begin{figure}[t]
\centering
\includegraphics[width=\columnwidth]{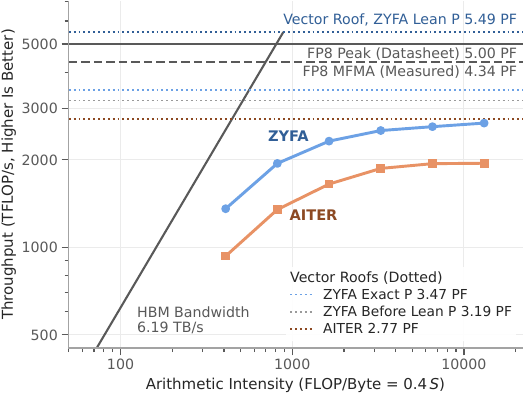}
\caption{Roofline for FP8 multi-head attention at $\dqk{=}\dv{=}128$, kernel-only on Gaussian inputs. At a given arithmetic intensity, the FLOPs per byte moved from memory, a kernel cannot exceed the sloped memory roof or the flat arithmetic roofs. The roofs are the empirical HBM bandwidth, the datasheet FP8 peak, and the empirical peak of the 32$\times$32$\times$64 FP8 matrix instruction (MFMA) that the kernels use. The 16$\times$16 shape, which has the higher throughput, reaches \zn{ceil/fp8/pf}\,PFLOP/s (Table~\ref{tab:ceil}). Each vector roof is the throughput at which all vector instructions of a kernel, packed FP32 included, fill the vector pipe, the unit that runs elementwise arithmetic such as the softmax. The roof of exact P, which evaluates the exponential, comes from a thread trace of an earlier build of the assembly kernel. With FLOPs $=BHS^2(\dqk+\dv)$ the arithmetic intensity is $0.4\,S$ FLOP/byte, so every length from $2$K up is compute-bound.}
\label{fig:roofline}
\end{figure}

Figure~\ref{fig:throughput} gives the kernel-only throughput of every configuration against sequence length at $B{=}1$ and $B{=}4$, and Table~\ref{tab:b4} compares the two batch sizes at $8$K and $32$K. Figure~\ref{fig:roofline} places the FP8 kernel at $\dqk{=}128$ on a roofline with empirical roofs.\footnote{A roofline plots throughput against arithmetic intensity, the FLOPs per byte moved from memory. Below the sloped roof, memory bandwidth limits a kernel, and below the flat roofs, arithmetic does.} The FLOPs of attention grow with $S^2$ and the bytes it moves grow with $S$, so its arithmetic intensity grows linearly with the sequence length, and every length from $2$K up is compute-bound.

\begin{table}[t]
\centering
{\footnotesize
\begin{tabular}{@{}lrrrr@{}}
\toprule
 & \multicolumn{2}{c}{$S{=}8$K} & \multicolumn{2}{c}{$S{=}32$K} \\
\cmidrule(lr){2-3}\cmidrule(lr){4-5}
kernel & $B{=}1$ & $B{=}4$ & $B{=}1$ & $B{=}4$ \\
\midrule
BF16 $\dqk{=}128$ & 1.03 & 1.00 & 1.02 & 1.03 \\
FP8 $\dqk{=}128$ & 1.35 & 1.41 & 1.37 & 1.38 \\
BF16 $\dqk{=}192$ & 1.00 & 1.03 & 1.00 & 1.00 \\
FP8 $\dqk{=}192$ & 1.88 & 1.94 & 1.99 & 1.99 \\
\bottomrule
\end{tabular}
}
\caption{Throughput of \sys{} divided by that of AITER's default kernel at batch sizes $B{=}1$ and $B{=}4$, for multi-head attention on Gaussian inputs, kernel-only. Values above $1$ mean that \sys{} takes less time.}
\label{tab:b4}
\end{table}

\begin{figure}[t]
\centering
\includegraphics[width=\columnwidth]{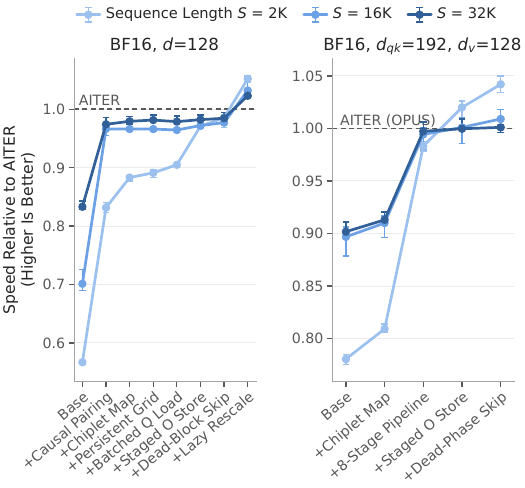}
\caption{Speed of \sys{}'s BF16 kernels relative to AITER's default kernel as design decisions are added from left to right, kernel-only on Gaussian inputs, multi-head at $B{=}1$. At $\dqk{=}192$ the baseline is AITER's OPUS kernel, a HIP kernel built with AITER's C++ template library. Section~\ref{sec:ladder} defines every step.}
\label{fig:ladder-bf16}
\end{figure}

\paragraph{Design steps}
\label{sec:ladder}
\label{app:ladder}
Figure~\ref{fig:ladder} adds the design decisions one at a time. In both FP8 kernels the assembly schedule gives the largest step, and at $\dqk{=}128$ lean P gives the second largest. At $\dqk{=}192$ the base HIP kernel, which issues full-rate FP8 matrix instructions, already reaches \zn{ladder/fp8-d192/0/32k}\x{} the throughput of CK at $32$K. The HIP and assembly kernels with lean P compute the same output, so the last step at $\dqk{=}128$ comes from the instruction schedule alone. In the BF16 kernels (Figure~\ref{fig:ladder-bf16}), causal pairing gives the largest step at $\dqk{=}128$ at every length because it balances the work across workgroups, and the eight-stage pipeline gives the largest step at $\dqk{=}192$.

Each step builds on the previous step in the same source tree, and the last step is the final kernel. The base FP8 kernels are HIP kernels without the chiplet map. They compute each probability in FP32, convert it to e4m3, and sum the rows on the vector unit. The exponential is a Schraudolph exponential at $\dqk{=}128$ and \texttt{v\_exp\_f32} at $\dqk{=}192$. At $\dqk{=}128$, step ``+Lean P'' switches to lean P. At $\dqk{=}192$, step ``+Row Sum on MFMA'' sums the rows on the matrix unit. Both steps also turn on per-block scaling, lower the lazy-rescaling threshold from $8$ to $0.125$, and multiply $p$ and $\ell$ by $2^{6}$. Step ``+Assembly Schedule'' replaces the HIP kernel with the assembly kernel of Section~\ref{sec:asm} and its two-wave ping-pong schedule. At $\dqk{=}192$ it also replaces a grid of $512$, $1024$, and $2048$ workgroups at $2$K, $16$K, and $32$K with a persistent grid of $256$.

The base BF16 kernel at $\dqk{=}128$ has the HipKittens schedule with two correctness fixes. It applies the softmax scale to the FP32 scores and rescales the output whenever a row maximum grows, while HipKittens skips increases of up to $8$. Step ``+Causal Pairing'' runs the causal pairs of Section~\ref{sec:common}, step ``+Chiplet Map'' adds the chiplet map, and step ``+Persistent Grid'' hands out the pairs from a persistent grid. In the base kernel the address descriptor of $Q$ (the base address and size of the tensor, which buffer loads read) lies in vector registers, so each of the eight loads of a $Q$ tile runs in a loop over the lanes and waits for the previous load.\footnote{Buffer loads take their address descriptor from scalar registers, which all lanes of a wave share. A descriptor in vector registers could differ between lanes, so the compiler wraps each load in a loop that serves the lanes with one descriptor value at a time.} Step ``+Batched Q Load'' keeps the descriptor in scalar registers, and the eight loads run concurrently. Step ``+Staged O Store'' writes the output tile through LDS, so each lane stores $16$ contiguous bytes of a row instead of $8$. The same step also defers these stores to the next tile and, as for $Q$, keeps their address descriptor in scalar registers. In the last three $64$-key blocks of a tile the causal mask hides every key from some of the waves. Step ``+Dead-Block Skip'' skips the MFMAs and the softmax of those waves in those blocks, while the loads into LDS and the barriers still run. Step ``+Lazy Rescale'' enables lazy rescaling with $\tau{=}0.5$. Because the base kernel rescales on every increase of the maximum, this step recovers part of a saving that HipKittens already has. The base BF16 kernel at $\dqk{=}192$ runs causal pairs without the chiplet map, and step ``+8-Stage Pipeline'' adds the eight-stage pipeline of Section~\ref{sec:bf16} and applies the softmax scale to the FP32 scores.

\begin{table}[t]
\centering
{\footnotesize
\setlength{\tabcolsep}{2.4pt}
\begin{tabular}{llrrcc}
\toprule
& & \multicolumn{2}{c}{time ratio} & \multicolumn{2}{c}{SCLK (MHz), capt./Gauss.} \\
\cmidrule(lr){3-4}\cmidrule(lr){5-6}
model & & capt. & Gauss. & \sys{} & AITER \\
\midrule
Mistral & FP8 & \zn{hl/mistral/fp8/real/ratio} & \zn{hl/mistral/fp8/recipe/ratio} & \zn{hl/mistral/fp8/real/zyfa/sclk}/\zn{hl/mistral/fp8/recipe/zyfa/sclk} & \zn{hl/mistral/fp8/real/aiter/sclk}/\zn{hl/mistral/fp8/recipe/aiter/sclk} \\
Qwen3 & FP8 & \zn{hl/qwen3/fp8/real/ratio} & \zn{hl/qwen3/fp8/recipe/ratio} & \zn{hl/qwen3/fp8/real/zyfa/sclk}/\zn{hl/qwen3/fp8/recipe/zyfa/sclk} & \zn{hl/qwen3/fp8/real/aiter/sclk}/\zn{hl/qwen3/fp8/recipe/aiter/sclk} \\
DeepSeek & FP8 & \zn{hl/deepseek/fp8/real/ratio} & \zn{hl/deepseek/fp8/recipe/ratio} & \zn{hl/deepseek/fp8/real/zyfa/sclk}/\zn{hl/deepseek/fp8/recipe/zyfa/sclk} & \zn{hl/deepseek/fp8/real/aiter/sclk}/\zn{hl/deepseek/fp8/recipe/aiter/sclk} \\
Mistral & BF16 & \zn{hl/mistral/bf16/real/ratio} & \zn{hl/mistral/bf16/recipe/ratio} & \zn{hl/mistral/bf16/real/zyfa/sclk}/\zn{hl/mistral/bf16/recipe/zyfa/sclk} & \zn{hl/mistral/bf16/real/aiter/sclk}/\zn{hl/mistral/bf16/recipe/aiter/sclk} \\
Qwen3 & BF16 & \zn{hl/qwen3/bf16/real/ratio} & \zn{hl/qwen3/bf16/recipe/ratio} & \zn{hl/qwen3/bf16/real/zyfa/sclk}/\zn{hl/qwen3/bf16/recipe/zyfa/sclk} & \zn{hl/qwen3/bf16/real/aiter/sclk}/\zn{hl/qwen3/bf16/recipe/aiter/sclk} \\
DeepSeek & BF16 & \zn{hl/deepseek/bf16/real/ratio} & \zn{hl/deepseek/bf16/recipe/ratio} & \zn{hl/deepseek/bf16/real/zyfa/sclk}/\zn{hl/deepseek/bf16/recipe/zyfa/sclk} & \zn{hl/deepseek/bf16/real/aiter/sclk}/\zn{hl/deepseek/bf16/recipe/aiter/sclk} \\
\bottomrule
\end{tabular}}
\caption{\sys{} and AITER on one layer per model at $S{=}16$K (Mistral-7B layer~20, Qwen3-8B layer~18, DeepSeek-V2-Lite layer~13). The captured columns use the $q$, $k$, and $v$ that the model produces on WikiText-2 text, and the Gaussian columns use random inputs of the same shape. Each run launches one kernel back to back for seconds and is timed only after the GPU clock has settled (sustained clocks). The time ratio is AITER's time divided by \sys{}'s time, so values above $1$ mean that \sys{} takes less time. SCLK is the shader clock of the slowest of the eight chiplets, which sets the kernel time.}
\label{tab:real}
\end{table}

\begin{figure}[t]
\centering
\includegraphics[width=\columnwidth]{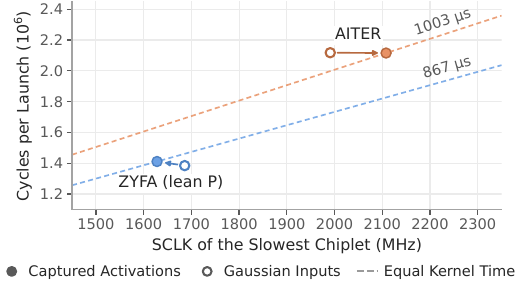}
\caption{Cycles per launch against the shader clock (SCLK) for \sys{}'s and AITER's FP8 kernels on Mistral-7B layer~20 at $S{=}16$K, at sustained clocks. Kernel time is cycles divided by clock, so a kernel can take less time by needing fewer cycles or by running at a higher clock. Cycles are kernel time multiplied by the SCLK of the slowest chiplet. Filled markers are captured activations, the $q$, $k$, and $v$ of the real model, and open markers are Gaussian inputs of the same shape. Points on a dashed line take the same time as the captured-activation point of that kernel.}
\label{fig:cyclesclock}
\end{figure}

\paragraph{Captured activations at sustained clocks}
\label{sec:res-real}
On captured activations \sys{}'s FP8 kernels lead AITER's by less than on Gaussian inputs of the same shape (Table~\ref{tab:real}). The Gaussian timing inputs have a standard deviation of $0.1$, which gives logits with a standard deviation of about $0.01$, so attention is nearly uniform, every probability lies close to the row maximum, and both kernels write normal e4m3 values. On the captured layers \sys{}'s clock falls on all three models, while the clock of AITER's assembly kernel rises on Mistral-7B and Qwen3-8B and the clock of CK on DeepSeek-V2-Lite falls by less than \sys{}'s. \sys{} still needs fewer cycles than AITER on both inputs (Figure~\ref{fig:cyclesclock}). In an exact-P version of the FP8 kernel a hardware counter of busy cycles changed by at most \zn{power/cycles/realvstwin}\% between the two inputs, so the data distribution changes the total attention time mainly by affecting the clock speed. One cause is the FP8 values of the probabilities that enter the $PV$ product. In the Mistral-7B layer, \zn{pcut/mistral-l20/t8}\% of the pairs of a row and a $64$-key block lie more than $8$ binades below the row maximum (Section~\ref{sec:pblock}), so AITER, which does not scale these blocks up, writes subnormal values or zeros there, while \sys{}'s per-block scaling keeps them normal. Nonzero values switch more bits in the $PV$ MFMAs and therefore lower the clock (Section~\ref{sec:limits-clock}). The same effect takes back part of the gain of lean P. On the Mistral-7B layer lean P needs \zn{lean/sustained/cycles}\% fewer cycles than exact P but runs at a \zn{lean/sustained/sclk}\% lower clock, so exact P takes only \zn{change/mistral16k/exactoverlean}\x{} the time of lean P. Appendix~\ref{app:layers} times every layer of each model and shows that the speedup is nearly the same in every layer, so the layers in Table~\ref{tab:real} are representative.

\begin{figure*}[tp]
\centering
\includegraphics[width=\textwidth]{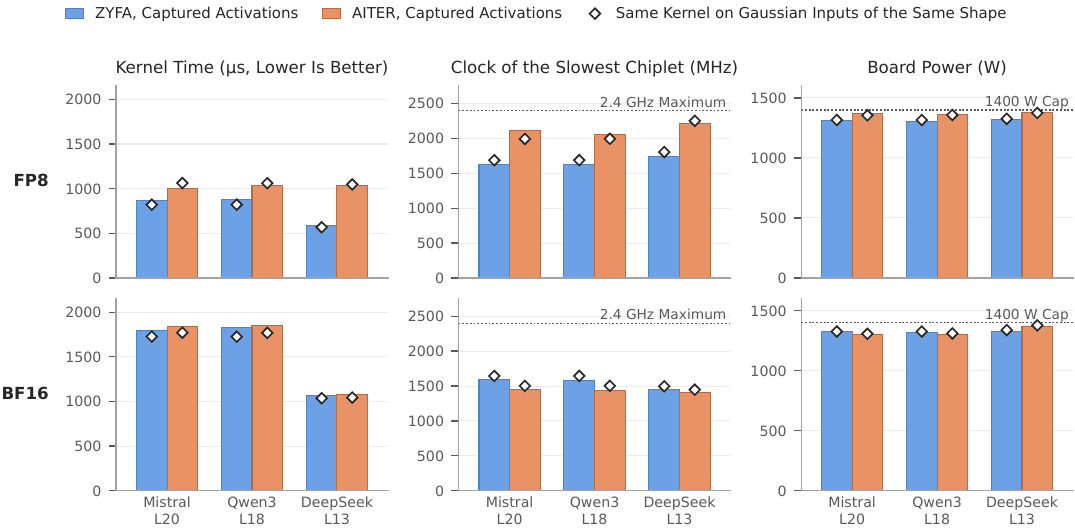}
\caption{Kernel time, shader clock (SCLK) of the slowest chiplet, and board power of \sys{}'s and AITER's kernels at $S{=}16$K. Bars are captured activations, the $q$, $k$, and $v$ of one layer per model, and diamonds are Gaussian inputs of the same shape. Every run launches one kernel back to back and is measured after the clock has settled (sustained clocks). The runs are those of Figure~\ref{fig:headline}.}
\label{fig:sclk}
\end{figure*}

\paragraph{Clock and power}
\label{sec:res-clock}
\label{app:sclk}
Figure~\ref{fig:sclk} gives the kernel time, the SCLK of the slowest chiplet, and the board power of each kernel on captured activations and on Gaussian inputs of the same shape, and Figure~\ref{fig:cyclesclock} separates the kernel time on the Mistral-7B layer into cycles and clock. To isolate the effect of the FP8 probabilities on the clock, an MFMA loop multiplies the FP8 probabilities captured from the Mistral-7B layer at $16$K, and we compare its clock with that of the same loop on the FP8 probabilities of Gaussian inputs. With \sys{}'s per-block scaled values the clock changes by \zn{power/mfma/scaled}\%, and with AITER's unscaled values by \zn{power/mfma/unscaled}\%.

The telemetry of \texttt{amd-smi} does not name the limit that sets the clock, but it rules out the two limits that it reports. Board power stays below the $1400$\,W cap for every kernel, and the thermal throttle counters do not change. The power throttle counter is active least often for \sys{}'s FP8 kernel, which runs at the lowest clock, whereas a limit on average power that set the clock would be active most often for the kernel with the lowest clock. The package-power status flag reads active for every kernel, so it does not distinguish the kernels, and \texttt{amd-smi} does not report the voltage. The clock also falls within a single launch. Together these observations point to a fast limit on current rather than on average power.

\begin{figure*}[tp]
\centering
\includegraphics[width=\textwidth]{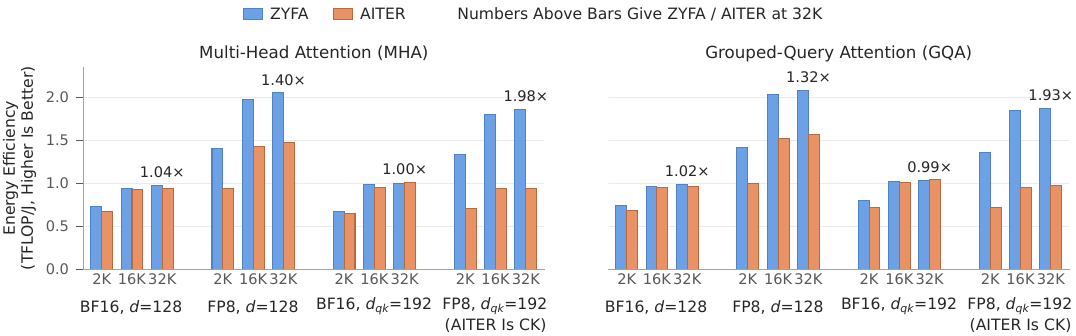}
\caption{\sys{}'s FP8 kernels compute more per joule than AITER's, and its BF16 kernels match AITER's. Each pair of bars gives the TFLOP of attention computed per joule of board energy for one configuration and sequence length, under sustained load on Gaussian inputs. For FP8 at $\dqk{=}128$ and $32$K, \sys{} reaches \zn{en/mha/fp8-d128/32k/zyfa/tfpj} and AITER \zn{en/mha/fp8-d128/32k/aiter/tfpj}\,TFLOP/J. At $\dqk{=}192$ AITER's FP8 kernel is Composable Kernel (CK).}
\label{fig:energy}
\end{figure*}

\paragraph{Energy}
\label{app:energy}
On the captured Mistral-7B and Qwen3-8B layers \sys{}'s FP8 kernel delivers \zn{hl/mistral/fp8/real/zyfa/tfpj} and \zn{hl/qwen3/fp8/real/zyfa/tfpj}\,TFLOP/J, against \zn{hl/mistral/fp8/real/aiter/tfpj} and \zn{hl/qwen3/fp8/real/aiter/tfpj}\,TFLOP/J for AITER. Figure~\ref{fig:energy} gives the energy efficiency of each configuration under sustained load on Gaussian inputs. The ratio of the energy efficiencies of two kernels is their throughput ratio divided by the ratio of their board powers. For FP8 at $\dqk{=}128$ and $32$K, \sys{} draws \zn{en/mha/fp8-d128/32k/zyfa/w}\,W and AITER \zn{en/mha/fp8-d128/32k/aiter/w}\,W, and \sys{} reaches \zn{en/mha/fp8-d128/32k/zyfa/tfpj}\,TFLOP/J and AITER \zn{en/mha/fp8-d128/32k/aiter/tfpj}\,TFLOP/J. The kernels were tuned for time only, and we have not determined the cause of the difference in power.

\subsection{Error of One Attention Call}
\label{sec:res-acc}
\label{sec:res-kernel}
\label{app:numerics}

\begin{table}[t]
\centering
{\footnotesize
\setlength{\tabcolsep}{3.0pt}
\begin{tabular}{lrrrrr}
\toprule
& \multicolumn{4}{c}{FP8} & BF16 \\
\cmidrule(lr){2-5}\cmidrule(lr){6-6}
model & AITER & \sys{} & CK & Triton & \sys{} \\
\midrule
Mistral-7B & \zne{num/mistral/fp8/aiter/relrms} & \zn{num/mistral/fp8/zyfa/vsaiter} & \zn{num/mistral/fp8/ck/vsaiter} & \zn{num/mistral/fp8/triton/vsaiter} & \zn{num/mistral/bf16/zyfa/vsaiter} \\
Qwen3-8B & \zne{num/qwen3/fp8/aiter/relrms} & \zn{num/qwen3/fp8/zyfa/vsaiter} & \zn{num/qwen3/fp8/ck/vsaiter} & \zn{num/qwen3/fp8/triton/vsaiter} & \zn{num/qwen3/bf16/zyfa/vsaiter} \\
DeepSeek-V2-Lite & \zne{num/deepseek/fp8/aiter/relrms} & \zn{num/deepseek/fp8/zyfa/vsaiter} & \zn{num/deepseek/fp8/ck/vsaiter} & -- & \zn{num/deepseek/bf16/zyfa/vsaiter} \\
\bottomrule
\end{tabular}}
\caption{Output error of one attention call, measured against attention computed in fp64 on the same quantized inputs, so it counts only the error that each kernel adds. Values are means over all layers of each model on WikiText-2 at $S{=}4$K. The AITER column gives the relative RMS error of AITER's default FP8 kernel, which is Composable Kernel (CK) at $\dqk{=}192$. The other FP8 columns give each kernel's error divided by AITER's, so values below $1$ mean a smaller error. AITER's Triton kernel does not support FP8 when $\dqk{\neq}\dv$. The BF16 column gives \sys{}'s BF16 error divided by the error of AITER's BF16 kernel.}
\label{tab:num}
\end{table}

Table~\ref{tab:num} compares each kernel with fp64 attention on the same quantized inputs, so it measures the error that the kernel itself adds. In FP8 at head dimension $128$ the errors of \sys{}, CK, and Triton~\citep{tillet2019triton} differ by a few percent. On Mistral-7B and Qwen3-8B, \sys{}'s error is \zn{num/mistral/fp8/zyfa/vsaiter} and \zn{num/qwen3/fp8/zyfa/vsaiter} times the error of AITER's assembly kernel. That kernel divides by the sum of the unrounded probabilities, so probabilities that underflow to zero in the numerator still count in the denominator, and every output row shrinks slightly (Figure~\ref{fig:pscale} and Section~\ref{sec:rowsum}). In BF16, \sys{}'s error is \zn{num/qwen3/bf16/zyfa/vsaiter}--\zn{num/mistral/bf16/zyfa/vsaiter}\x{} that of AITER on Mistral-7B and Qwen3-8B. On DeepSeek-V2-Lite \sys{}'s and AITER's BF16 kernels use the same arithmetic and have the same error.

\paragraph{BF16 kernels}
On Mistral-7B the error of \sys{}'s BF16 kernel at $\dqk{=}128$ exceeds AITER's by up to \zn{num/mistral/bf16/zyfa/worst}\% per layer. The difference comes from the order in which the kernels visit the keys. For the long-row half of each causal pair, AITER visits the key blocks from the diagonal back to key~$0$, so the attention-sink token at key~$0$ (a token that receives a large share of the attention of most queries) comes last. Each key that sets the running maximum before the sink token gets $p{=}1$, which BF16 represents exactly. \sys{} visits key~$0$ first, so the sink token sets the maximum and the probabilities of the other keys are rounded. The BF16 kernels also multiply the FP32 scores by the softmax scale $c$. Rounding $c\,q\log_2 e$ to BF16 before the MFMA instead, as the HipKittens attention kernel does, adds an error to each score that grows with the magnitude of the score, and the softmax turns it into a relative error of $p$ (Section~\ref{sec:res-e2e}). At logit standard deviations of about $4$ to $16$ this rounding raises the output error to \zn{bf16/qprescale/errvsaiter}\x{} AITER's.

\paragraph{Lean P}
\label{sec:leancost}
On \zn{lean/err/layers} real layers at $\dqk{=}128$, the output error of lean P, which computes each FP8 probability with one multiply-add (Section~\ref{sec:lean}), is \zn{lean/err/vsexact}\x{} that of exact P, which evaluates the exponential, in the same kernel, and the geometric mean of its error is \zn{lean/err/vsaitergm}\x{} the error of AITER's kernel. Across input distributions the comparison with AITER varies. On Gaussian inputs with $q$ and $k$ of standard deviation $0.5$ to $3$, the error of lean P is \zn{lean/err/gaussvsaiter}\% higher than AITER's, on ramp and spike inputs it is up to \zn{lean/err/rampvsaiter}\% higher, and on the captured layers at $\dqk{=}128$ its RMS error exceeds AITER's on \zn{lean/err/layersabove} of \zn{lean/err/layers} layers. Against a model with fp64 attention, lean P raises the KL divergence of the output by \zn{lean/e2e/mistralkl}\% relative to exact P on Mistral-7B, and on Qwen3-8B the change rounds to \zn{lean/e2e/qwen3kl}\%. At $\dqk{=}192$ lean P would take \zn{d192lean/mha/gainmin}--\zn{d192lean/mha/gainmax}\% less time, but in \zn{d192lean/snrfails} of \zn{d192lean/snrcells} configurations its signal-to-noise ratio falls \zn{d192lean/snrmiss}\,dB below the $30$\,dB threshold of the correctness test, so the kernel at $\dqk{=}192$ uses exact P, although lean P matches exact P end to end on DeepSeek-V2-Lite.

\begin{figure*}[tp]
\centering
\includegraphics[width=\textwidth]{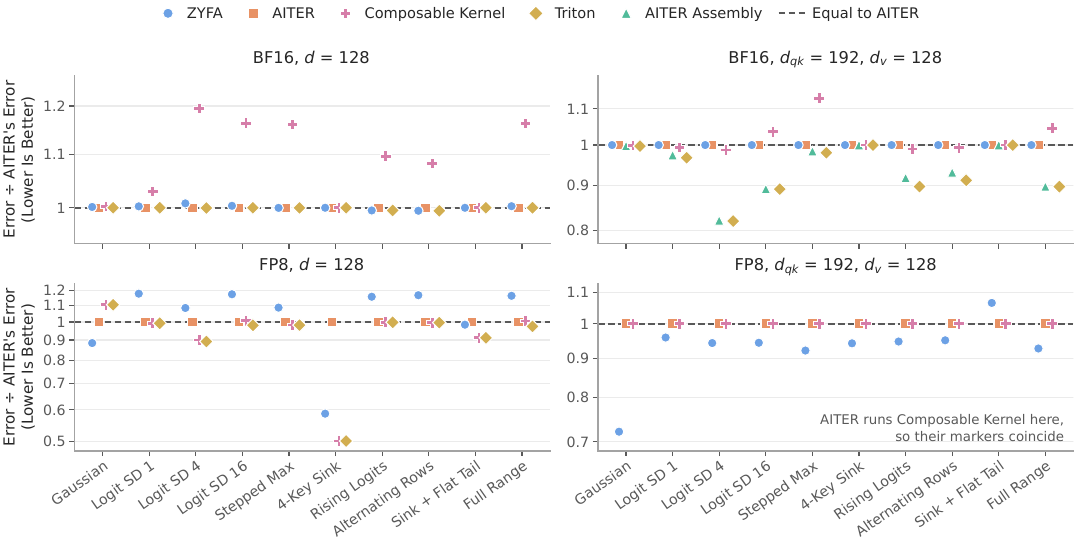}
\caption{Output error of one attention call on synthetic inputs at $S{=}8$K tokens, as each kernel's relative RMS error against attention computed in fp64 on the same inputs, divided by AITER's error (below $1$ is more accurate than AITER). Gaussian draws $q$, $k$, and $v$ from $\mathcal{N}(0, 0.1^2)$, as in the timing runs. Every other input starts from standard normal $q$, $k$, and $v$. Logit SD~1, 4, and 16 scale $q$ and $k$ by $1$, $2$, and $4$, which gives logits with these standard deviations in nats. Stepped Max raises the logits of each successive block of $64$ keys by a fixed step, so the row maximum keeps growing. 4-Key Sink scales the first four keys by $8$. Rising Logits adds a linear ramp along the keys to every row, and Alternating Rows adds it to every other row. Sink + Flat Tail puts the first key $12$ nats above a nearly uniform rest. Full Range draws magnitudes log-uniformly between $2^{-6}$ and $2^{8}$ with random signs, which spans the e4m3 range after scaling.}
\label{fig:numsynth}
\end{figure*}

\begin{figure*}[tp]
\centering
\includegraphics[width=\textwidth]{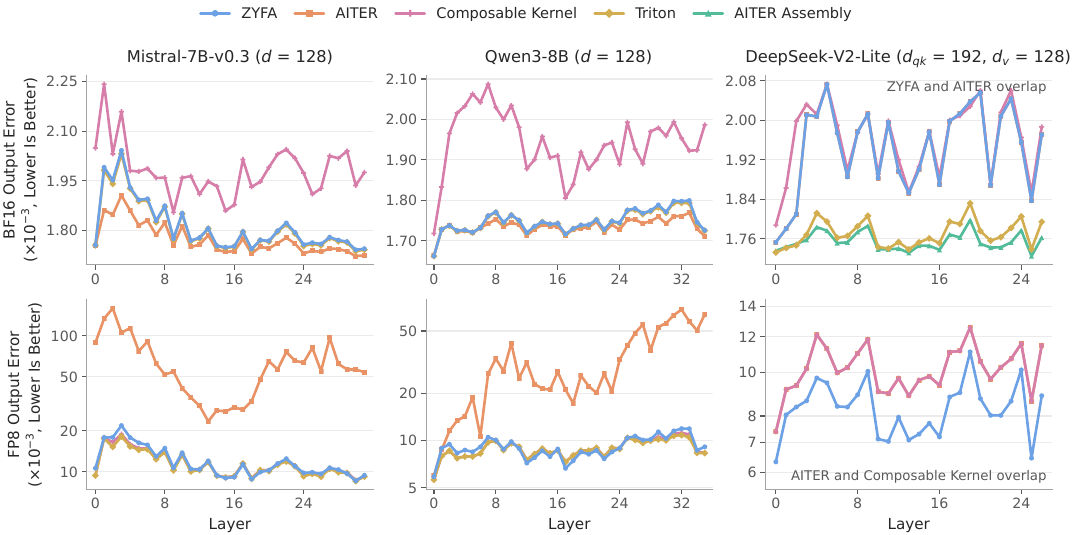}
\caption{Output error of one attention call for every layer of each model, as relative RMS error against attention computed in fp64 on the same inputs (lower is more accurate). The inputs are the $q$, $k$, and $v$ that each model produces on the first $4095$ tokens of WikiText-2 ($S{=}4$K). The FP8 row uses a logarithmic axis.}
\label{fig:numreal}
\end{figure*}

\begin{figure*}[tp]
\centering
\includegraphics[width=\textwidth]{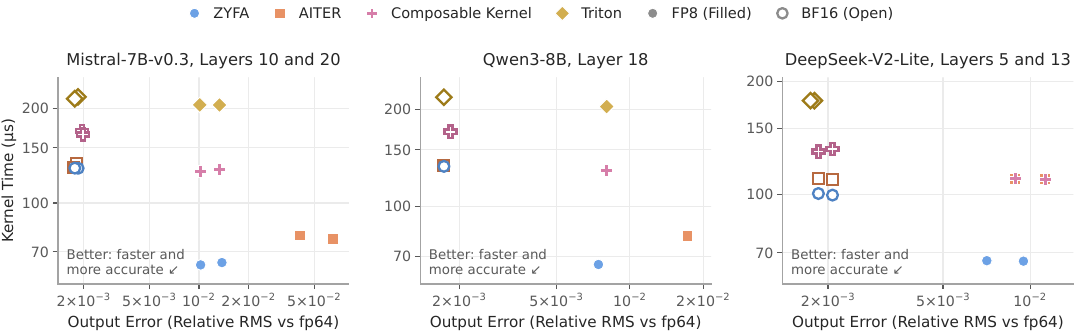}
\caption{Kernel time against output error for each kernel on captured layers at $S{=}4$K, the $q$, $k$, and $v$ that each model produces on WikiText-2. Output error is the relative RMS error against attention computed in fp64 on the same inputs. Filled markers are FP8 kernels, and open markers are BF16 kernels. Each marker is one kernel on one layer, and markers toward the lower left are faster and more accurate. At the head dimensions of DeepSeek-V2-Lite, AITER's FP8 kernel is Composable Kernel, so their FP8 markers coincide.}
\label{fig:speederr}
\end{figure*}

\paragraph{Synthetic and captured inputs}
Figure~\ref{fig:numsynth} gives the output error of each kernel on synthetic input distributions, and Figure~\ref{fig:numreal} gives it for every captured layer. Figure~\ref{fig:speederr} places each kernel by its time and its error on the captured layers.

\begin{figure*}[tp]
\centering
\includegraphics[width=\textwidth]{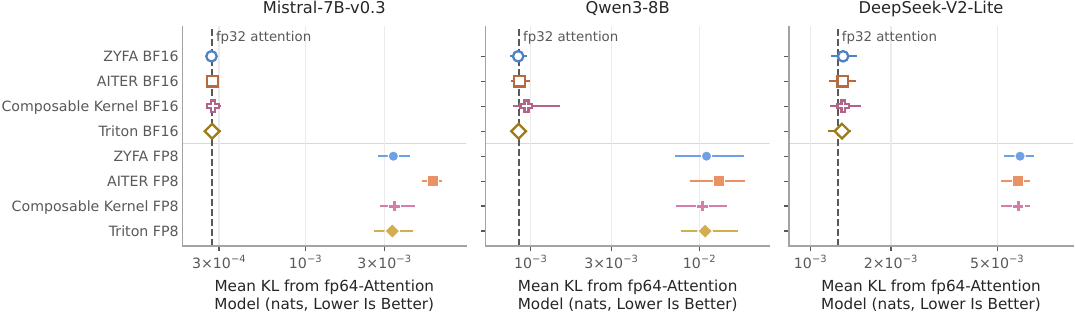}
\caption{KL divergence of each next-token distribution from that of the same model with attention computed in fp64, when every attention call uses the kernel under test, averaged over the tokens of eight WikiText-2 sequences of $4$K tokens (lower is closer to fp64). Each marker is the mean over the sequences, and its bar spans the lowest to the highest sequence. The dashed line is attention computed in fp32. Filled markers are FP8 kernels, and open markers are BF16 kernels. Triton has no FP8 kernel at the head dimensions of DeepSeek-V2-Lite.}
\label{fig:drift}
\end{figure*}

\paragraph{Error in a full model}
A single call does not show how errors accumulate over the layers of a model. Figure~\ref{fig:drift} therefore replaces every attention call of each model with the kernel under test and compares the next-token distributions with those of a model with fp64 attention on eight sequences of $4$K tokens. The FP8 kernels of \sys{}, Composable Kernel, and Triton give KL divergences on Mistral-7B of \zne{drift/mistral/zyfa_fp8/kl}, \zne{drift/mistral/ck_fp8/kl}, and \zne{drift/mistral/triton_fp8/kl}. AITER's FP8 kernel gives \zne{drift/mistral/aiter_fp8/kl}. Every BF16 kernel gives the KL divergence of FP32 attention, \zne{drift/mistral/fp32/kl}.

\paragraph{Run-to-run reproducibility}
\label{sec:res-nondet}
\label{app:nondet}
\sys{}'s kernels return identical outputs in every launch, and AITER's FP8 assembly kernel changes its output in a few launches when a GEMM (a general matrix multiplication) runs between them. Table~\ref{tab:nondet} repeats a launch of each FP8 kernel on the same captured inputs and counts the launches whose output differs bitwise from the first launch. Between launches the GPU runs nothing, a BF16 GEMM, or a BF16 attention kernel. Each kernel runs \zn{nondet/rerun/launches} launches on three captured layers. \sys{}'s output is identical in every launch. Of AITER's launches, \zn{nondet/rerun/faulty} differ, all with a GEMM between launches. The differences lie in output columns \zn{nondet/cols} of \zn{nondet/heads} heads per launch, and in \zn{nondet/cols/last32} of the \zn{nondet/rerun/faulty} launches they lie only in the last $32$ columns, a pattern that suggests a race on the last $V$ or output block that depends on timing. The differences range from one unit in the last place to wrong values. On the affected rows the relative RMS error reaches \zn{nondet/worst/relrms}, against \zn{nondet/typical/relrms} for a launch with the correct output. GEMMs run between the attention calls of a full model, so AITER's end-to-end results in Section~\ref{sec:res-e2e} may include such launches, and in our end-to-end runs two passes of AITER's FP8 kernel over the same sequence sometimes gave different logits.

\begin{table}[t]
\centering
{\footnotesize
\begin{tabular}{llrr}
\toprule
layer & between launches & AITER & \sys{} \\
\midrule
Mistral-7B L20, 4K & nothing & \zn{nondet/rerun/mistral/aiter/none}/\zn{nondet/rerun/mistral/aiter/none/n} & \zn{nondet/rerun/mistral/zyfa/none}/\zn{nondet/rerun/mistral/zyfa/none/n} \\
Mistral-7B L20, 4K & BF16 GEMM & \zn{nondet/rerun/mistral/aiter/gemm}/\zn{nondet/rerun/mistral/aiter/gemm/n} & \zn{nondet/rerun/mistral/zyfa/gemm}/\zn{nondet/rerun/mistral/zyfa/gemm/n} \\
Mistral-7B L20, 4K & BF16 attention & \zn{nondet/rerun/mistral/aiter/attn}/\zn{nondet/rerun/mistral/aiter/attn/n} & \zn{nondet/rerun/mistral/zyfa/attn}/\zn{nondet/rerun/mistral/zyfa/attn/n} \\
\midrule
Qwen3-8B L6, 4K & nothing & \zn{nondet/rerun/qwen3/aiter/none}/\zn{nondet/rerun/qwen3/aiter/none/n} & \zn{nondet/rerun/qwen3/zyfa/none}/\zn{nondet/rerun/qwen3/zyfa/none/n} \\
Qwen3-8B L6, 4K & BF16 GEMM & \zn{nondet/rerun/qwen3/aiter/gemm}/\zn{nondet/rerun/qwen3/aiter/gemm/n} & \zn{nondet/rerun/qwen3/zyfa/gemm}/\zn{nondet/rerun/qwen3/zyfa/gemm/n} \\
Qwen3-8B L6, 4K & BF16 attention & \zn{nondet/rerun/qwen3/aiter/attn}/\zn{nondet/rerun/qwen3/aiter/attn/n} & \zn{nondet/rerun/qwen3/zyfa/attn}/\zn{nondet/rerun/qwen3/zyfa/attn/n} \\
\midrule
Qwen3-8B L18, 16K & nothing & \zn{nondet/rerun/qwen316k/aiter/none}/\zn{nondet/rerun/qwen316k/aiter/none/n} & \zn{nondet/rerun/qwen316k/zyfa/none}/\zn{nondet/rerun/qwen316k/zyfa/none/n} \\
Qwen3-8B L18, 16K & BF16 GEMM & \zn{nondet/rerun/qwen316k/aiter/gemm}/\zn{nondet/rerun/qwen316k/aiter/gemm/n} & \zn{nondet/rerun/qwen316k/zyfa/gemm}/\zn{nondet/rerun/qwen316k/zyfa/gemm/n} \\
Qwen3-8B L18, 16K & BF16 attention & \zn{nondet/rerun/qwen316k/aiter/attn}/\zn{nondet/rerun/qwen316k/aiter/attn/n} & \zn{nondet/rerun/qwen316k/zyfa/attn}/\zn{nondet/rerun/qwen316k/zyfa/attn/n} \\
\bottomrule
\end{tabular}}
\caption{Number of launches, out of the total after the slash, whose output differs bitwise from the first launch on the same captured FP8 inputs. Between launches the GPU runs nothing, a BF16 GEMM (matrix multiplication), or a BF16 attention kernel. AITER is AITER's FP8 assembly kernel.}
\label{tab:nondet}
\end{table}

\subsection{FP8 Error in Full Models}
\label{sec:res-e2e}
\label{app:e2e}

Every FP8 variant we test moves the output of the three models outside the BF16 noise band, the range of KL divergence in which a model is as close to the BF16 model as it would be with a second, equally correct BF16 attention kernel (Section~\ref{app:e2emethods}). The KL divergence of all-FP8 \sys{} (FP8 attention in every head) is \zn{pareto/mistral/zyfa-fp8/klcap}, \zn{pareto/qwen3/zyfa-fp8/klcap}, and \zn{pareto/deepseek/zyfa-fp8/klcap} times the upper end of the BF16 noise band on Mistral-7B, Qwen3-8B, and DeepSeek-V2-Lite. The KL divergence per token measures how far the next-token distributions move from those of the model with \sys{}'s BF16 attention. AITER's FP8 kernel lowers the NLL of Qwen3-8B below that of BF16 while it has the largest KL divergence of all variants. A lower NLL therefore does not show that a variant is closer to BF16, and we use the KL divergence as the accuracy metric.\footnote{The NLL measures how well a model predicts the text, and a perturbation can lower it by chance. The KL divergence measures how far each next-token distribution moves from that of the BF16 model, and it is zero only if the two models agree.}

\begin{figure*}[tp]
\centering
\includegraphics[width=\textwidth]{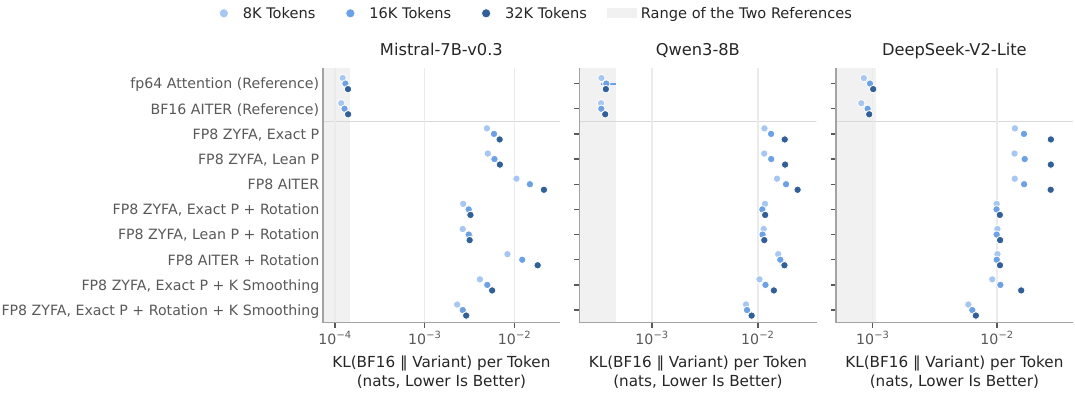}
\caption{KL divergence per token from the model with \sys{}'s BF16 attention, KL(BF16\,$\|$\,variant), when every attention call of the model uses one FP8 kernel variant, at $8$K, $16$K, and $32$K tokens on \zn{lean/e2e/docs} PG-19 documents per model. Lower means closer to the BF16 model. The first two rows are references, attention in fp64 and AITER's BF16 kernel, which show how far two correct kernels differ through rounding alone, and the gray band covers their range.}
\label{fig:stage0}
\end{figure*}

\paragraph{Lean P and exact P}
Figure~\ref{fig:stage0} gives the KL divergence of every FP8 variant on \zn{lean/e2e/docs} PG-19 test documents per model, split over the three lengths, from a run that preceded the evaluation on \zn{pareto/docs} documents in Section~\ref{sec:res-pareto}. On the \zn{pareto/docs} evaluation documents of Mistral-7B, lean P raises the NLL relative to exact P by \zne{paired/mistral/zyfa-fp8-vs-zyfa-fp8exact/all/diff} nats per token, with a $95\%$ interval from \zne{paired/mistral/zyfa-fp8-vs-zyfa-fp8exact/all/lo} to \zne{paired/mistral/zyfa-fp8-vs-zyfa-fp8exact/all/hi}.

\begin{figure*}[tp]
\centering
\includegraphics[width=\textwidth]{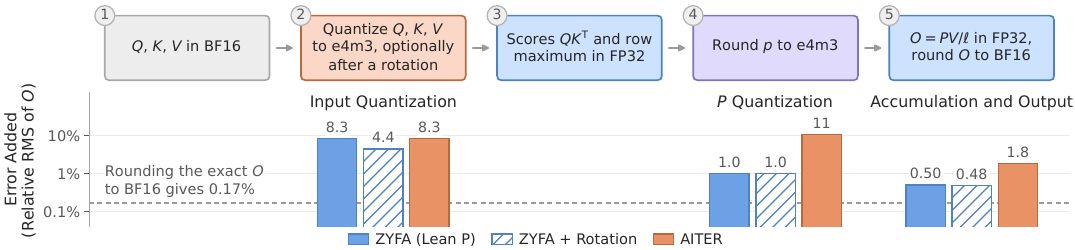}
\caption{Quantizing the inputs $Q$, $K$, and $V$ to FP8 adds most of the error of \sys{}'s FP8 attention, and rounding the probabilities $p$ adds much less. The top row shows the stages of one FP8 attention call. Each bar below is the error that a stage adds on Mistral-7B (\zn{budget/docs} PG-19 documents, log scale), as the RMS of the change that the stage makes to the attention output $O$, divided by the RMS of fp64 attention on the BF16 inputs, in percent, pooled over all layers, heads, and rows. A relative RMS error of $r\%$ is a signal-to-noise ratio (SNR) of $-20\log_{10}(r/100)$\,dB, and the stages of \sys{} add in quadrature to its total error. Bars show \sys{} with lean P, \sys{} with a Hadamard rotation of $Q$ and $K$, and AITER's FP8 kernel. The dashed line is the error of rounding the exact output $O$ to BF16. The error of the running maximum and of the FP32 sums is counted in the last stage.}
\label{fig:errflow}
\end{figure*}

\paragraph{Error added by each stage}
We measure the error that each stage of one attention call adds (Figure~\ref{fig:errflow}, method in Section~\ref{app:e2emethods}). On Mistral-7B, quantizing $Q$, $K$, and $V$ adds \zn{budget/mistral/inoverleanP} times the error of quantizing $p$ with lean P, and for \sys{} the input term (the error from quantizing $Q$, $K$, and $V$) is also the largest on the other two models. Quantizing only $Q$, only $K$, or only $V$ gives a relative error of \zne{budget/mistral/in_qonly}, \zne{budget/mistral/in_konly}, and \zne{budget/mistral/in_vonly} on Mistral-7B, against \zne{budget/mistral/in_ph} for all three. An FP8 KV cache with BF16 attention would therefore already add much of this error, and $V$ adds the least. The exponential in the softmax explains why $Q$ and $K$ add more error than $V$. A rounding error in $q$ or $k$ changes a score by an amount that grows with the magnitude of $q$ and $k$, and the softmax turns an error of $\delta$ in a score, in base-2 units, into a relative error of about $\delta\ln 2$ in $p$.\footnote{Because $p=2^{x}$, an error $\delta$ in $x$ multiplies $p$ by $2^{\delta}\approx1+\delta\ln2$.} Rounding $p$ itself adds a relative error that does not depend on the magnitude of the scores. The e4m3 signal-to-noise ratio of $Q$ and $K$ is \zn{budget/qk/snrdb}\,dB in $90\%$ of heads, the value that a 3-bit mantissa gives in the normal range. A floating-point format has the same relative precision over its whole normal range (a scale moves values within the range but does not change how finely the format resolves them), so a separate scale for each head would not raise this signal-to-noise ratio.

\begin{figure*}[tp]
\centering
\includegraphics[width=\textwidth]{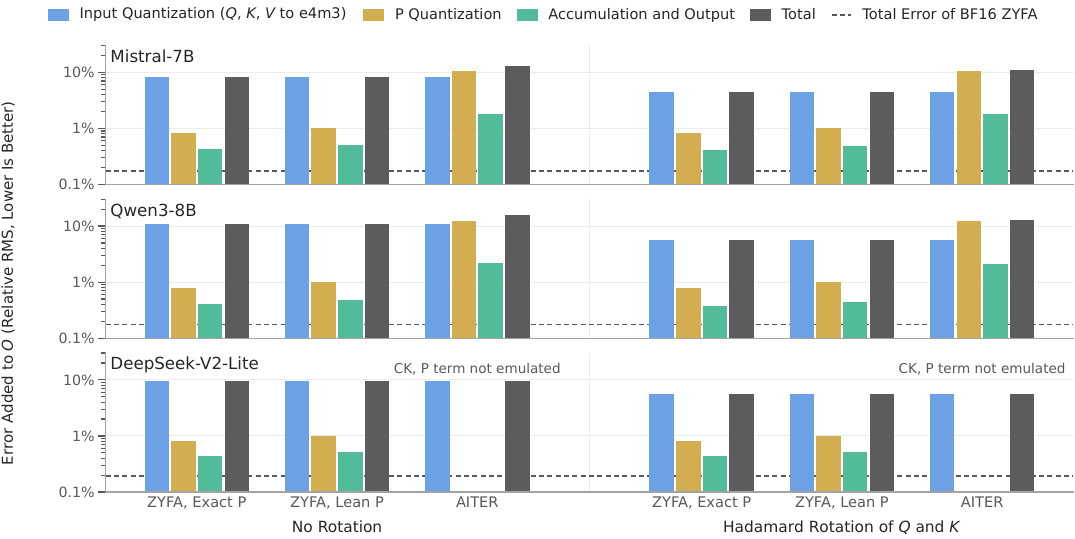}
\caption{Error that each stage of one FP8 attention call adds to the attention output, as relative RMS error against fp64 attention, on all three models, without and with a Hadamard rotation of $Q$ and $K$. The stages are quantizing $Q$, $K$, and $V$ (input), quantizing the probabilities ($P$), and the rest of the kernel's arithmetic (Section~\ref{app:e2emethods}). The dashed line is the total error of \sys{}'s BF16 kernel. At $\dqk{=}192$ AITER runs Composable Kernel (CK), whose $P$ arithmetic is not emulated, so only its input term and total are shown.}
\label{fig:budget}
\end{figure*}

\paragraph{Error of one call on every model}
Figure~\ref{fig:budget} extends Figure~\ref{fig:errflow} to all three models, with and without the rotation. For \sys{} the three error terms (input quantization, $P$ quantization, and the rest of the kernel's arithmetic) add in quadrature to the total within \zn{budget/quadrature/zyfa}\%, so they behave as independent errors.\footnote{Independent errors add like the sides of a right triangle, so the total RMS error is the square root of the sum of the squared RMS errors of the stages.} For AITER at $\dqk{=}128$ the sum in quadrature exceeds the total by \zn{budget/quadrature/aiter}\%, because its $P$ term and its last term partly cancel.

\paragraph{Rotation and K smoothing}
A Hadamard rotation of $Q$ and $K$ over the head dimension before quantization~\citep{shah2024flash3,ashkboos2024quarot} leaves $QK^{\top}$ unchanged. The rounding error of a score $q k^{\top}$ grows with $\sum_i q_i^2 k_i^2$, which is large when a few channels of $q$ and $k$ are large together, and the rotation lowers this sum by spreading each channel over all channels.\footnote{In the extreme case where all the energy of $q$ and $k$ lies in one shared channel, the rotation spreads it evenly over all $128$ channels and lowers $\sum_i q_i^2k_i^2$ by a factor of $128$. Real heads have milder outliers.} On Mistral-7B the rotation halves the input error and lowers the KL divergence from \zn{pareto/mistral/zyfa-fp8/klcap} to \zn{pareto/mistral/zyfa-fp8-rot/klcap} times the upper end of the BF16 noise band. Subtracting the mean key over the sequence before the rotation (K smoothing~\citep{zhang2024sage}) lowers the KL divergence further on all three models. On Qwen3-8B rotation alone raises the NLL significantly, and with K smoothing the increase is no longer significant (Table~\ref{tab:rotks}).
The rotation adds little time. It needs $2D$ FLOPs per element of $Q$ and $K$, while causal attention needs about $2S$ FLOPs per element of $Q$, so the rotation adds at most $2D/S$ to the FLOPs. On its own, the rotation takes the time of an extra pass over $Q$ and $K$, and inside the fused quantization pass it adds almost no time (Table~\ref{tab:rotks}). The fused pass uses a static scale, which lets it quantize $Q$ without first computing its maximum. Static scales from the calibration documents give the same KL divergence as per-call scales. For all-FP8 on Mistral-7B the paired difference in KL divergence between static and per-call scales is \zne{pairedkl/mistral/zyfa-fp8-st-vs-zyfa-fp8/all/diff} per token, with a $95\%$ interval from \zne{pairedkl/mistral/zyfa-fp8-st-vs-zyfa-fp8/all/lo} to \zne{pairedkl/mistral/zyfa-fp8-st-vs-zyfa-fp8/all/hi}.

\begin{table*}[tp]
\centering
{\footnotesize
\begin{tabular}{@{}lrrrrrrr@{}}
\toprule
 & \multicolumn{2}{c}{error of one call} & & & \multicolumn{2}{c}{time / all-FP8} & transform \\
\cmidrule(lr){2-3}\cmidrule(lr){6-7}
transform & input & total & KL & dNLL [95\% interval] & fused & separate & ms at 16K \\
\midrule
\multicolumn{8}{@{}l}{\emph{Mistral-7B}} \\
none & 8.27e-02 & 8.32e-02 & 6.05e-03 & +5.4e-03 [+5.0e-03, +5.8e-03] & 1.000 & 1.000 & -- \\
rotation & 4.41e-02 & 4.52e-02 & 2.90e-03 & +2.6e-03 [+2.4e-03, +2.9e-03] & 1.001 & 1.089 & 2.51 \\
rotation, K smoothing & 4.55e-02 & 4.65e-02 & 2.62e-03 & +2.2e-03 [+2.0e-03, +2.5e-03] & 1.006 & 1.165 & 3.97 \\
\midrule
\multicolumn{8}{@{}l}{\emph{Qwen3-8B}} \\
none & 1.09e-01 & 1.09e-01 & 1.53e-02 & +3.9e-03 [+3.2e-03, +4.6e-03] & 1.000 & 1.000 & -- \\
rotation & 5.64e-02 & 5.72e-02 & 1.12e-02 & +6.0e-03 [+5.1e-03, +7.0e-03] & 1.001 & 1.086 & 2.79 \\
rotation, K smoothing & 5.32e-02 & 5.41e-02 & 8.29e-03 & +4.3e-03 [+3.5e-03, +5.1e-03] & 1.005 & 1.158 & 4.44 \\
\midrule
\multicolumn{8}{@{}l}{\emph{DeepSeek-V2-Lite}} \\
none & 9.49e-02 & 9.53e-02 & 2.17e-02 & +1.7e-02 [+1.5e-02, +1.9e-02] & 1.000 & 1.000 & -- \\
rotation & 5.58e-02 & 5.64e-02 & 9.98e-03 & +7.2e-03 [+6.6e-03, +7.7e-03] & 1.034 & 1.176 & 2.74 \\
rotation, K smoothing & 5.30e-02 & 5.36e-02 & 6.27e-03 & +3.8e-03 [+3.5e-03, +4.1e-03] & 1.039 & 1.370 & 5.23 \\
\bottomrule
\end{tabular}
}
\caption{Effect of rotating $Q$ and $K$ with a Hadamard matrix (rotation) and of subtracting the mean key of each head (K smoothing) before FP8 quantization, for all-FP8 \sys{} attention. The error of one call is the relative RMS error of the attention output $O$ against fp64 attention, from quantizing the inputs alone (input) and in total. KL is the per-token KL divergence from the model with BF16 attention, and dNLL the change in negative log-likelihood per token, both on the \zn{pareto/docs} PG-19 evaluation documents. Times are relative to all-FP8 attention without transforms. The fused column includes one quantization pass that also applies the transforms. The separate column runs the transforms as PyTorch operations and includes no quantization. The last column is the time of the separate transforms per forward pass at $16$K.}
\label{tab:rotks}
\end{table*}

\paragraph{Choosing heads for BF16}
\label{sec:res-where}
A mixed-precision policy (a fixed assignment of each head or layer to BF16 or FP8) keeps some heads in BF16, and it should keep those whose FP8 error changes the output of the model most. The local error of a head measures how much FP8 changes its attention output, but its effect on the model output also depends on the later layers. We compare rankings by the KL divergence that remains when the top $10\%$ of heads by a score run in BF16, divided by the KL divergence of all-FP8 attention, on $12$ evaluation documents per model (Table~\ref{tab:klleft}). The worst $10\%$ of heads by local error carry \zn{conc/mistral/head/10}\% of the squared output error on Mistral-7B, yet running them, or the worst $10\%$ of $128$-row blocks, in BF16 leaves \zn{anat/localoracle10/klleft} of the all-FP8 KL divergence on the three models. Local error is largest in the last quarter of layers, while the gradient-weighted score below is largest in the first quarter, whose errors pass through every later layer.

We therefore rank heads by a gradient-weighted score, ge2, named for the product $g^2e^2$ that it sums. A change $e$ in the output of a head changes the loss by about $g\cdot e$, where $g$ is the gradient of the loss with respect to that output. On \zn{ge2/calibdocs} calibration documents, half at $8$K and half at $16$K tokens, one forward and backward pass of the BF16 model gives the difference $e$ between the FP8 and BF16 outputs of each head and the gradient $g$ of the summed next-token NLL, taken at the attention output before the output projection. The FP8 outputs come from an exact-P build of the FP8 kernel on the BF16 inputs of each layer. The score of a head is the sum of $g^2e^2$ over documents, rows, and channels.\footnote{Squaring keeps errors that raise the loss on some tokens and lower it on others from cancelling, and summing $g^2e^2$ over channels instead of squaring $g\cdot e$ ignores the correlations between channels.} This score is the diagonal Fisher approximation used to assign bit widths in mixed-precision quantization~\citep{kim2024squeezellm,dong2019hawq}. Running the top $10\%$ of heads by ge2 in BF16 leaves \zn{anat/ge2_10/klleft/mistral}, \zn{anat/ge2_10/klleft/qwen3}, and \zn{anat/ge2_10/klleft/deepseek} of the all-FP8 KL divergence on the three models. Policies with rotation use the same ranking and rotate the heads that remain in FP8.

The layer KL score ranks whole layers instead, by how much FP8 attention in one layer alone changes the output of the model. The score of layer $L$ compares the BF16 model with the model that runs \sys{}'s FP8 attention in every head of layer~$L$ and BF16 attention in all other layers. The score is the mean over positions of the KL divergence between the next-token distributions of the two models, summed over \znp{layerkl/calibdocs}{4} calibration documents at $8$K tokens. The policy keeps the layers with the largest score in BF16, as whole layers and part of one layer at the boundary. As for ge2, the FP8 kernel in this calibration is the exact-P build, and the evaluated policies use lean P. Each policy ranks heads or layers by one score alone.

\begin{figure*}[tp]
\centering
\includegraphics[width=\textwidth]{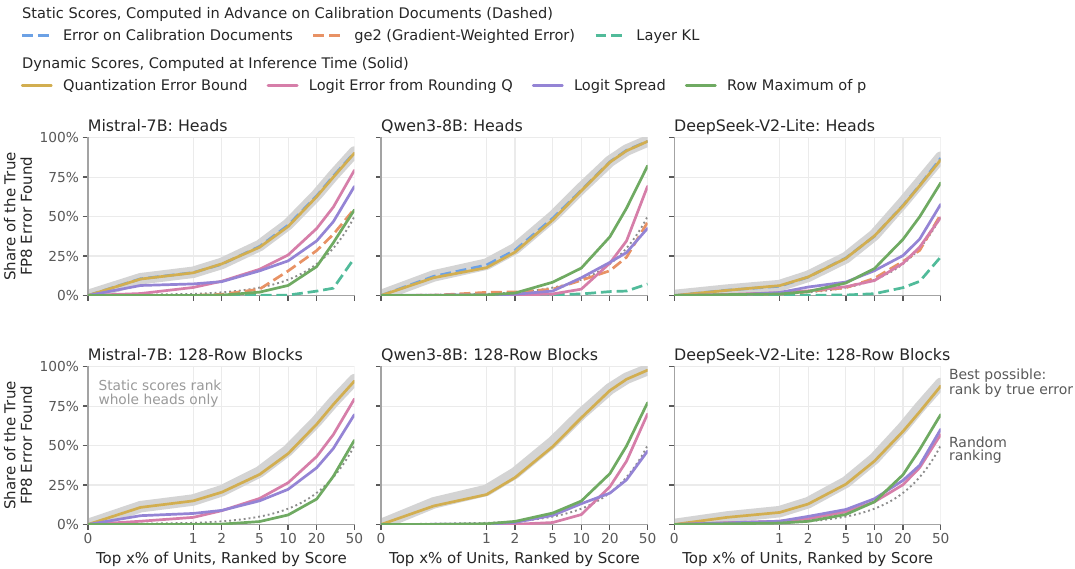}
\caption{Scores that measure the local FP8 error (the error of each unit's own attention output), such as the error on calibration documents and the quantization error bound, find almost as much of it as ranking by the true error (gray band). ge2 (the gradient-weighted error of a head) and layer KL (how much FP8 attention in one layer alone changes the model output) find little of it, because they rank units by their effect on the model output, which works better end to end (Table~\ref{tab:klleft}). Each line gives the share of the true FP8 error energy in the heads (top) or $128$-row blocks (bottom) that a score ranks highest. Static scores are computed in advance on calibration documents and dynamic scores at inference time, and the dotted line ranks the heads or row blocks in random order.}
\label{fig:capture}
\end{figure*}

\paragraph{Scores computed at inference time}
A score that needs no gradient could choose heads during inference. Four such dynamic scores ran end to end: a $3\sigma$ estimate of the output error from rounding $Q$, $K$, and $V$ to e4m3, the attention-weighted mean absolute logit error from rounding $Q$, the largest absolute logit error from rounding $Q$ over the keys of a row, and the spread of the logits. With the top $10\%$ of heads or $128$-row blocks by one of these scores in BF16, these scores leave \zn{anat/dyn10/klleft} of the all-FP8 KL divergence. The row maximum of $p$ was computed for Figure~\ref{fig:capture}, but no policy ranked by it ran end to end. Figure~\ref{fig:capture} ranks heads and $128$-row blocks by each dynamic score and by three static scores, which are computed in advance (the error energy on the calibration documents, ge2, and the layer KL), and gives the share of the true FP8 error energy in the top $x\%$. The dynamic scores track the local error rather than the effect of the error on the model output. Table~\ref{tab:klleft} gives the KL divergence that remains when each score chooses the units that run in BF16.

\begin{figure*}[tp]
\centering
\includegraphics[width=0.9\textwidth]{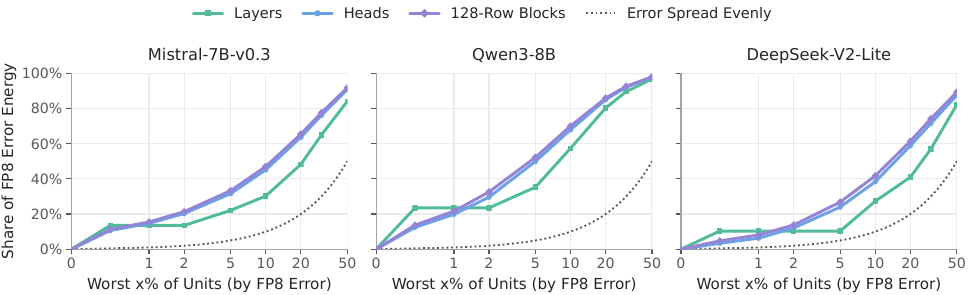}
\caption{Share of the FP8 error energy of the attention outputs that falls in the worst $x\%$ of layers, heads, and $128$-row blocks, ranked by their own FP8 error, for the median document. The dotted curve is the share if the error were spread evenly, $x\%$ of the error energy in $x\%$ of the units. The horizontal axis is linear below $1\%$ and logarithmic above.}
\label{fig:conc}
\end{figure*}

\paragraph{Concentration of the error}
The last quarter of the layers holds the largest share of the local FP8 error, \zn{anat/local/lastq/mistral}, \zn{anat/local/lastq/qwen3}, and \zn{anat/local/lastq/deepseek}\% on Mistral-7B, Qwen3-8B, and DeepSeek-V2-Lite. The first quarter holds the largest share of the ge2 score, \zn{anat/ge2/firstq/mistral}, \zn{anat/ge2/firstq/qwen3}, and \zn{anat/ge2/firstq/deepseek}\%. The heads with the largest ge2 score carry small shares of the local error, so a ranking by local error selects different heads. Figure~\ref{fig:conc} gives the share of the FP8 error energy in the worst layers, heads, and $128$-row blocks. Tables~\ref{tab:heads-mistral} to~\ref{tab:heads-deepseek} list the heads with the largest ge2 score on each model with their attention statistics.

\begin{table*}[tp]
\centering
{\footnotesize
\gdef\klleftmaxdiff{0.02}
\begin{tabular}{llrrrrrrrrr}
\toprule
 & & \multicolumn{3}{c}{Mistral-7B} & \multicolumn{3}{c}{Qwen3-8B} & \multicolumn{3}{c}{DeepSeek-V2-Lite} \\
\cmidrule(lr){3-5}\cmidrule(lr){6-8}\cmidrule(lr){9-11}
score & unit & 5\% & 10\% & 20\% & 5\% & 10\% & 20\% & 5\% & 10\% & 20\% \\
\midrule
local-error oracle & heads & 0.94 & 0.91 & 0.83 & 0.94 & 0.91 & 0.83 & 0.98 & 0.94 & 0.87 \\
local-error oracle & row blocks & 0.93 & 0.89 & 0.82 & 0.93 & 0.90 & 0.81 & 0.95 & 0.91 & 0.84 \\
ge2 & heads & 0.52 & 0.43 & 0.31 & 0.66 & 0.57 & 0.43 & 0.41 & 0.34 & 0.25 \\
layer KL & layers & 0.67 & 0.61 & 0.45 & 0.74 & 0.66 & 0.54 & 0.85 & 0.64 & 0.33 \\
\midrule
logit spread & heads & 0.95 & 0.92 & 0.98 & 0.87 & 0.79 & 0.66 & 0.86 & 0.70 & 0.58 \\
logit spread & row blocks & 0.97 & 0.98 & 0.89 & 0.86 & 0.76 & 0.67 & 0.84 & 0.71 & 0.50 \\
$3\sigma$ error estimate & heads & 0.94 & 0.90 & 0.83 & 0.94 & 0.91 & 0.84 & 0.97 & 0.94 & 0.85 \\
$3\sigma$ error estimate & row blocks & 0.94 & 0.89 & 0.82 & 0.94 & 0.90 & 0.82 & 0.95 & 0.90 & 0.83 \\
weighted logit error of $Q$ & heads & 0.95 & 0.91 & 0.84 & 0.94 & 0.86 & 0.70 & 0.73 & 0.64 & 0.28 \\
weighted logit error of $Q$ & row blocks & 0.94 & 0.90 & 0.83 & 0.91 & 0.81 & 0.67 & 0.74 & 0.61 & 0.39 \\
maximum logit error of $Q$ & heads & 0.93 & 0.89 & 0.81 & 0.87 & 0.78 & 0.64 & 0.85 & 0.76 & 0.54 \\
maximum logit error of $Q$ & row blocks & 0.93 & 0.89 & 0.85 & 0.84 & 0.75 & 0.62 & 0.83 & 0.70 & 0.52 \\
\bottomrule
\end{tabular}
}
\caption{How well each score chooses the units (heads, $128$-row blocks, or layers) that run in BF16. Each value is the KL divergence from the BF16 model that remains when the top $5$, $10$, or $20\%$ of units by the score run in BF16, divided by the KL divergence of all-FP8 attention (lower is better), on $12$ evaluation documents per model with the exact-P build of \sys{}'s FP8 kernel. The local-error oracle ranks units by the true FP8 error of their own attention output. ge2 ranks heads by their gradient-weighted error $g^2e^2$, and layer KL ranks layers by how much FP8 attention in that layer alone changes the output of the model. Both use calibration documents (Section~\ref{sec:res-where}), and the scores below the rule are computed at inference time. A row block has $128$ query rows, and its BF16 share of the causal work differs slightly from the nominal share. On the \zn{pareto/docs} documents of Figure~\ref{fig:pareto} with lean P, ge2 and layer KL give values within \klleftmaxdiff{} of these.}
\label{tab:klleft}
\end{table*}

\begin{table*}[tp]
\centering
{\footnotesize
\begin{tabular}{@{}rrrrrrrrrrl@{}}
\toprule
 & & \multicolumn{2}{c}{share of} & \multicolumn{2}{c}{attention weight on} & mean & & \multicolumn{2}{c}{top-channel share} & \\
\cmidrule(lr){3-4}\cmidrule(lr){5-6}\cmidrule(lr){9-10}
rank & layer.head & ge2 & local error & key 0 & last 64 & distance & entropy & $Q$ & $K$ & type \\
\midrule
1 & 4.9 & 6.6\% & 0.01\% & 0.49 & 0.25 & 3108 & 1.87 & 0.12 & 0.18 & long-range \\
2 & 4.8 & 6.2\% & 0.01\% & 0.51 & 0.24 & 3159 & 1.87 & 0.14 & 0.18 & sink \\
3 & 4.11 & 2.3\% & 0.01\% & 0.55 & 0.20 & 3322 & 1.65 & 0.12 & 0.18 & sink \\
4 & 5.11 & 2.0\% & 0.02\% & 0.23 & 0.06 & 2547 & 4.85 & 0.14 & 0.15 & long-range \\
5 & 2.21 & 1.6\% & 0.00\% & 0.49 & 0.03 & 3138 & 2.41 & 0.11 & 0.09 & long-range \\
6 & 4.23 & 1.5\% & 0.00\% & 0.74 & 0.04 & 3908 & 1.18 & 0.28 & 0.31 & sink, outlier \\
7 & 1.7 & 1.2\% & 0.00\% & 0.48 & 0.49 & 2050 & 2.26 & 0.11 & 0.11 & long-range \\
8 & 3.22 & 1.1\% & 0.02\% & 0.43 & 0.30 & 2852 & 2.04 & 0.18 & 0.26 & long-range, outlier \\
9 & 2.22 & 1.0\% & 0.00\% & 0.58 & 0.11 & 3515 & 1.65 & 0.09 & 0.09 & sink \\
10 & 3.20 & 1.0\% & 0.01\% & 0.39 & 0.36 & 2534 & 2.68 & 0.19 & 0.26 & long-range, outlier \\
11 & 0.31 & 0.9\% & 0.00\% & 0.03 & 0.91 & 207 & 2.37 & 0.08 & 0.09 & local \\
12 & 1.0 & 0.9\% & 0.00\% & 0.54 & 0.41 & 2308 & 1.49 & 0.07 & 0.07 & sink \\
13 & 0.29 & 0.9\% & 0.00\% & 0.13 & 0.65 & 724 & 2.78 & 0.14 & 0.09 & local \\
14 & 4.10 & 0.8\% & 0.01\% & 0.49 & 0.25 & 3075 & 2.01 & 0.14 & 0.18 & long-range \\
15 & 5.10 & 0.7\% & 0.01\% & 0.33 & 0.03 & 3033 & 3.95 & 0.22 & 0.15 & long-range, outlier \\
16 & 18.9 & 0.6\% & 0.51\% & 0.26 & 0.55 & 1706 & 1.85 & 0.28 & 0.22 & local, outlier \\
\bottomrule
\end{tabular}
}
\caption{Heads of Mistral-7B with the largest ge2 score, the gradient-weighted FP8 error $g^2e^2$ of a head (Section~\ref{sec:res-where}), on the calibration documents. The share columns give the head's share of the ge2 score and of the local FP8 error. The attention-weight columns give the mean weight on key~$0$ and on the last $64$ keys over the query rows from $256$ on, on two calibration documents. Qwen3-8B's documents have no BOS token, so its key~$0$ is the first text token. The mean distance between query and key is weighted by attention and given in tokens, and the entropy of the attention row is given in nats. The top-channel share is the share of the energy of post-RoPE $Q$ and $K$ in their largest channel, pooled over the tokens of a document and averaged over eight calibration documents. A head is a sink head with weight on key~$0$ of at least $0.5$, local with weight on the last $64$ keys of at least $0.5$, long-range with neither and a mean distance of at least $1024$, and an outlier head with a top-channel share of at least $0.2$.}
\label{tab:heads-mistral}
\end{table*}

\begin{table*}[tp]
\centering
{\footnotesize
\begin{tabular}{@{}rrrrrrrrrrl@{}}
\toprule
 & & \multicolumn{2}{c}{share of} & \multicolumn{2}{c}{attention weight on} & mean & & \multicolumn{2}{c}{top-channel share} & \\
\cmidrule(lr){3-4}\cmidrule(lr){5-6}\cmidrule(lr){9-10}
rank & layer.head & ge2 & local error & key 0 & last 64 & distance & entropy & $Q$ & $K$ & type \\
\midrule
1 & 0.6 & 3.3\% & 0.00\% & 0.00 & 0.88 & 122 & 3.26 & 0.08 & 0.97 & local, outlier \\
2 & 0.11 & 2.5\% & 0.00\% & 0.00 & 0.87 & 138 & 3.14 & 0.07 & 0.97 & local, outlier \\
3 & 7.7 & 2.0\% & 0.01\% & 0.69 & 0.02 & 3396 & 1.17 & 0.28 & 0.27 & sink, outlier \\
4 & 10.2 & 1.9\% & 0.07\% & 0.36 & 0.56 & 1570 & 1.93 & 0.31 & 0.66 & local, outlier \\
5 & 7.5 & 1.7\% & 0.01\% & 0.76 & 0.08 & 3281 & 0.92 & 0.24 & 0.27 & sink, outlier \\
6 & 10.1 & 1.6\% & 0.05\% & 0.41 & 0.49 & 1808 & 2.41 & 0.30 & 0.66 & long-range, outlier \\
7 & 0.23 & 1.5\% & 0.00\% & 0.00 & 0.67 & 299 & 3.87 & 0.07 & 0.96 & local, outlier \\
8 & 3.7 & 1.2\% & 0.00\% & 0.02 & 0.21 & 2055 & 3.17 & 0.19 & 0.21 & long-range, outlier \\
9 & 34.20 & 1.1\% & 1.71\% & 0.80 & 0.07 & 3652 & 0.69 & 0.27 & 0.36 & sink, outlier \\
10 & 14.3 & 1.0\% & 0.06\% & 0.32 & 0.63 & 1378 & 1.83 & 0.12 & 0.43 & local, outlier \\
11 & 5.10 & 1.0\% & 0.00\% & 0.00 & 0.68 & 830 & 3.31 & 0.27 & 0.15 & local, outlier \\
12 & 11.19 & 1.0\% & 0.03\% & 0.32 & 0.62 & 1431 & 1.94 & 0.35 & 0.39 & local, outlier \\
13 & 2.31 & 0.8\% & 0.00\% & 0.01 & 0.09 & 1675 & 4.06 & 0.17 & 0.20 & long-range, outlier \\
14 & 10.3 & 0.8\% & 0.05\% & 0.59 & 0.37 & 2549 & 1.80 & 0.31 & 0.66 & sink, outlier \\
15 & 10.13 & 0.8\% & 0.02\% & 0.39 & 0.55 & 1673 & 2.90 & 0.36 & 0.67 & local, outlier \\
16 & 11.10 & 0.8\% & 0.03\% & 0.59 & 0.07 & 2626 & 1.78 & 0.28 & 0.34 & sink, outlier \\
\bottomrule
\end{tabular}
}
\caption{Heads of Qwen3-8B with the largest ge2 score (gradient-weighted FP8 error), with the columns of Table~\ref{tab:heads-mistral}.}
\label{tab:heads-qwen3}
\end{table*}

\begin{table*}[tp]
\centering
{\footnotesize
\begin{tabular}{@{}rrrrrrrrrrl@{}}
\toprule
 & & \multicolumn{2}{c}{share of} & \multicolumn{2}{c}{attention weight on} & mean & & \multicolumn{2}{c}{top-channel share} & \\
\cmidrule(lr){3-4}\cmidrule(lr){5-6}\cmidrule(lr){9-10}
rank & layer.head & ge2 & local error & key 0 & last 64 & distance & entropy & $Q$ & $K$ & type \\
\midrule
1 & 0.9 & 5.4\% & 0.04\% & 0.00 & 0.95 & 114 & 0.89 & 0.34 & 0.55 & local, outlier \\
2 & 2.9 & 5.1\% & 0.05\% & 0.00 & 0.96 & 93 & 2.96 & 0.21 & 0.65 & local, outlier \\
3 & 1.8 & 5.0\% & 0.01\% & 0.00 & 0.94 & 65 & 3.23 & 0.24 & 0.67 & local, outlier \\
4 & 4.7 & 3.6\% & 0.14\% & 0.41 & 0.56 & 1746 & 2.11 & 0.35 & 0.57 & local, outlier \\
5 & 5.2 & 2.7\% & 0.61\% & 0.17 & 0.74 & 1073 & 1.16 & 0.30 & 0.68 & local, outlier \\
6 & 0.8 & 2.4\% & 0.02\% & 0.00 & 0.94 & 174 & 2.39 & 0.31 & 0.55 & local, outlier \\
7 & 5.3 & 1.3\% & 0.34\% & 0.28 & 0.57 & 1752 & 2.09 & 0.33 & 0.67 & local, outlier \\
8 & 0.6 & 1.3\% & 0.01\% & 0.00 & 0.87 & 352 & 2.78 & 0.28 & 0.55 & local, outlier \\
9 & 11.14 & 1.3\% & 1.42\% & 0.15 & 0.73 & 1055 & 1.61 & 0.33 & 0.43 & local, outlier \\
10 & 1.13 & 1.2\% & 0.03\% & 0.00 & 0.93 & 98 & 2.21 & 0.18 & 0.66 & local, outlier \\
11 & 8.13 & 1.2\% & 0.60\% & 0.17 & 0.72 & 1135 & 1.63 & 0.31 & 0.67 & local, outlier \\
12 & 4.10 & 1.2\% & 0.07\% & 0.31 & 0.47 & 1402 & 3.51 & 0.38 & 0.57 & long-range, outlier \\
13 & 5.13 & 1.1\% & 0.31\% & 0.40 & 0.48 & 2106 & 1.58 & 0.30 & 0.67 & long-range, outlier \\
14 & 5.14 & 1.0\% & 0.13\% & 0.35 & 0.48 & 2043 & 2.48 & 0.27 & 0.68 & long-range, outlier \\
15 & 4.5 & 1.0\% & 0.18\% & 0.39 & 0.57 & 1709 & 1.52 & 0.26 & 0.57 & local, outlier \\
16 & 2.4 & 0.8\% & 0.02\% & 0.01 & 0.92 & 194 & 1.28 & 0.08 & 0.65 & local, outlier \\
\bottomrule
\end{tabular}
}
\caption{Heads of DeepSeek-V2-Lite with the largest ge2 score (gradient-weighted FP8 error), with the columns of Table~\ref{tab:heads-mistral}.}
\label{tab:heads-deepseek}
\end{table*}

\begin{figure*}[tp]
\centering
\includegraphics[width=\textwidth]{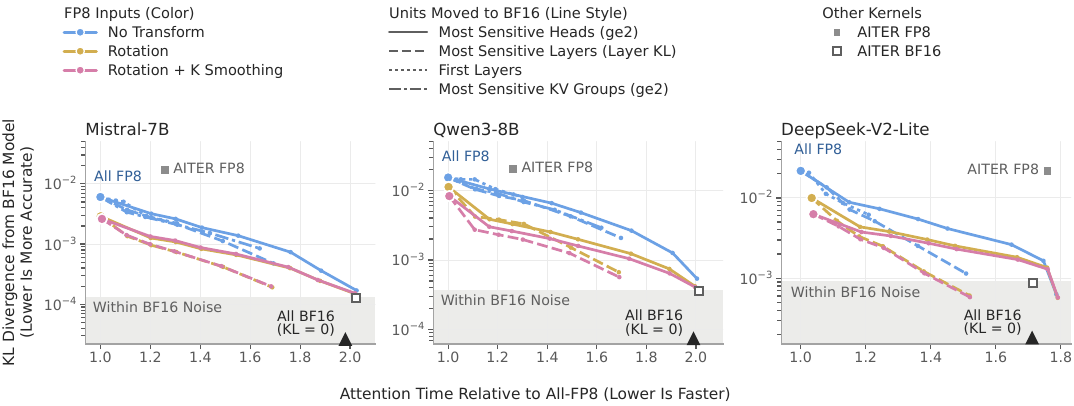}
\caption{Rotating $Q$ and $K$ with a Hadamard matrix before FP8 quantization buys more accuracy per unit of time than running heads or layers in BF16. Each line starts from all-FP8 attention with one input transform (large dot) and moves a growing share of heads or layers to BF16. Time is attention time inside the model relative to all-FP8 \sys{}, including one fused quantization pass. The KL divergence measures how far the next-token distributions move from those of the BF16 model, on \zn{pareto/docs} PG-19 documents per model. On Mistral-7B and Qwen3-8B, only all-BF16 attention reaches the gray BF16 noise band, where a policy (a choice of the heads or layers that run in BF16) is as close to the BF16 model as a second, equally correct BF16 kernel would be.}
\label{fig:pareto}
\end{figure*}

\begin{figure}[t]
\centering
\includegraphics[width=\columnwidth]{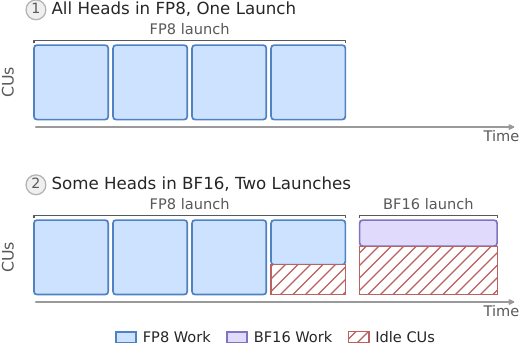}
\caption{Occupancy of the compute units (CUs) of the GPU over time for one attention layer (schematic). (1) With all heads in FP8, one kernel launch runs whole rounds of query-tile pairs, the units of work of the kernel, on every CU. (2) With some heads in BF16, the layer needs two launches. The FP8 launch ends with a partial round, in which only some CUs still have work, and the BF16 launch has too few pairs to occupy all CUs. On Mistral-7B with $5\%$ of heads in BF16, attention takes \zn{pareto/mistral/static-ge2-5/tfused} times the all-FP8 time, against \zn{pareto/mistral/static-ge2-5/tfusedlin} times by its share of the work. Both times exclude the gather and scatter copies of Table~\ref{tab:mixcopies}.}
\label{fig:mixed}
\end{figure}

\paragraph{Time and accuracy of mixed-precision policies}
\label{sec:res-pareto}
Figure~\ref{fig:pareto} places each policy by its attention time and its KL divergence. Attention time includes one fused pass that quantizes $Q$, while $K$ and $V$ are quantized as they are written to the cache (Section~\ref{app:e2emethods}). With this pass, rotation takes \zn{pareto/mistral/zyfa-fp8-rot/tfused} times the all-FP8 time on Mistral-7B, and all-BF16 attention takes \zn{pareto/mistral/zyfa-bf16/tfused} times. A subset of heads in BF16 costs more than its share of the work. A layer with heads in both precisions needs two kernel launches, and the BF16 launch covers few heads and leaves most of the $256$ CUs idle (Figure~\ref{fig:mixed}). With $5\%$ of Mistral-7B's heads in BF16, attention takes \zn{pareto/mistral/static-ge2-5/tfused}\x{} the all-FP8 time, against \zn{pareto/mistral/static-ge2-5/tfusedlin}\x{} if each head cost its share of the work. These times exclude the copies that gather the heads of each precision and scatter their outputs back into $O$, which add \zn{mixcopy/all/16k/range}\% to the attention time of a mixed policy at $16$K, the most for policies with many mixed layers (Table~\ref{tab:mixcopies}). The times of mixed policies are therefore lower bounds.

\begin{table*}[tp]
\centering
{\footnotesize
\begin{tabular}{@{}lrrrrrrrr@{}}
\toprule
 & & \multicolumn{3}{c}{all policies with mixed layers} & \multicolumn{4}{c}{$10\%$ of heads in BF16 by ge2} \\
\cmidrule(lr){3-5}\cmidrule(lr){6-9}
model & $S$ & policies & copy ms & \% of attention & mixed layers & copy ms & attention ms & \% \\
\midrule
Mistral-7B & 8K & 28 & 0.16--5.38 & 1.2--35.2 & 23 & 3.49 & 11.36 & 30.8 \\
Mistral-7B & 16K & 28 & 0.31--10.77 & 0.7--22.3 & 23 & 7.15 & 35.68 & 20.0 \\
Mistral-7B & 32K & 28 & 0.60--21.18 & 0.3--12.1 & 23 & 13.85 & 129.21 & 10.7 \\
Qwen3-8B & 8K & 28 & 0.15--5.39 & 1.0--31.9 & 20 & 3.04 & 12.40 & 24.5 \\
Qwen3-8B & 16K & 28 & 0.31--11.01 & 0.5--20.5 & 20 & 6.24 & 39.90 & 15.6 \\
Qwen3-8B & 32K & 28 & 0.60--21.51 & 0.3--11.1 & 20 & 12.10 & 145.76 & 8.3 \\
DeepSeek-V2-Lite & 8K & 20 & 0.08--2.53 & 1.4--28.4 & 12 & 1.14 & 6.49 & 17.5 \\
DeepSeek-V2-Lite & 16K & 20 & 0.18--4.78 & 0.8--17.2 & 12 & 2.25 & 20.68 & 10.9 \\
DeepSeek-V2-Lite & 32K & 20 & 0.33--9.14 & 0.4--9.1 & 12 & 4.27 & 75.72 & 5.6 \\
\bottomrule
\end{tabular}
}
\caption{Time per forward pass of the copies that gather the heads of each precision into contiguous tensors and scatter their outputs back, in mixed-precision layers (layers with heads in both BF16 and FP8). The left group covers every policy with at least one mixed layer and gives the range over policies. The right group is the policy with the $10\%$ of heads with the largest ge2 score in BF16. The attention times of Section~\ref{sec:res-pareto} exclude the copies.}
\label{tab:mixcopies}
\end{table*}

Keeping the most sensitive heads in BF16 is not a useful middle ground on Mistral-7B and Qwen3-8B. No policy short of all-BF16 reaches the BF16 noise band there, with or without rotation. With $90\%$ of heads in BF16 the KL divergence is \zn{pareto/mistral/static-ge2-90/klcap} and \zn{pareto/qwen3/static-ge2-90/klcap} times the upper end of the BF16 noise band without rotation and \zn{pareto/mistral/static-ge2-on-zyfa-fp8rot-90/klcap} and \zn{pareto/qwen3/static-ge2-on-zyfa-fp8rot-90/klcap} times with it, and the rotated policies take \zn{pareto/mistral/static-ge2-on-zyfa-fp8rot-90/tfused}\x{} and \zn{pareto/qwen3/static-ge2-on-zyfa-fp8rot-90/tfused}\x{} the all-FP8 time, more than all-BF16 attention (\zn{pareto/mistral/zyfa-bf16/tfused}\x{} and \zn{pareto/qwen3/zyfa-bf16/tfused}\x{}). The heads with the largest ge2 score remove a large part of the KL divergence (Table~\ref{tab:klleft}), but the rest is spread over most of the other heads and still lies above the BF16 noise band, so even if each head cost only its share of the work, $90\%$ of the heads in BF16 would take \zn{pareto/mistral/static-ge2-90/tfusedlin}\x{} and \zn{pareto/qwen3/static-ge2-90/tfusedlin}\x{} the all-FP8 time, close to all-BF16 attention. On DeepSeek-V2-Lite mixed policies do reach the BF16 noise band. Without rotation, $90\%$ of heads in BF16 reaches it at \zn{pareto/deepseek/static-ge2-90/tfused}\x{} the all-FP8 time, more than all-BF16 at \zn{pareto/deepseek/zyfa-bf16/tfused}\x{} because of its second launch. With rotation, keeping the $70\%$ of layers with the largest layer KL in BF16 reaches it at \zn{pareto/deepseek/rotmix/inband}\x{}, but its dNLL interval excludes zero, so it matches BF16 on KL divergence only. Rotation is also a better use of time than BF16 heads. Without rotation, every policy with BF16 heads or layers takes longer than rotation on Mistral-7B and Qwen3-8B, and on DeepSeek-V2-Lite the most accurate such policy that takes no longer than rotation has \zn{eqt/deepseek/fused/rot/klratio} times the KL divergence of rotation.

\begin{figure*}[tp]
\centering
\includegraphics[width=\textwidth]{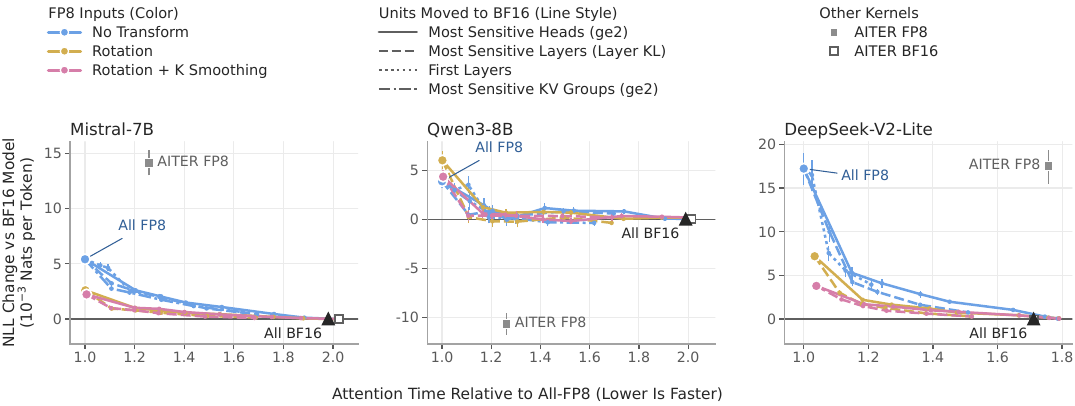}
\caption{On the NLL, most FP8 policies also differ from the BF16 model, and on Qwen3-8B rotation raises the NLL although it lowers the KL divergence. Each point is the change in negative log-likelihood per token against the BF16 model, with its $95\%$ interval over PG-19 books, for the mixed-precision policies of Figure~\ref{fig:pareto} (choices of the heads or layers that run in BF16), against attention time relative to all-FP8 \sys{}. A lower NLL does not mean that a policy is closer to BF16, as AITER's FP8 kernel on Qwen3-8B shows.}
\label{fig:pareto-dnll}
\end{figure*}

\paragraph{NLL}
Figure~\ref{fig:pareto-dnll} repeats Figure~\ref{fig:pareto} with dNLL, the change in negative log-likelihood per token against the BF16 model, on the vertical axis. The dNLL test for matching BF16 is two-sided, so a variant whose NLL is significantly lower than that of BF16 also fails it. Exact fp64 attention fails it on dNLL for Mistral-7B, where it lowers the NLL by \zne{pareto/mistral/fp64/dnll} nats per token.

\begin{figure*}[tp]
\centering
\includegraphics[width=\textwidth]{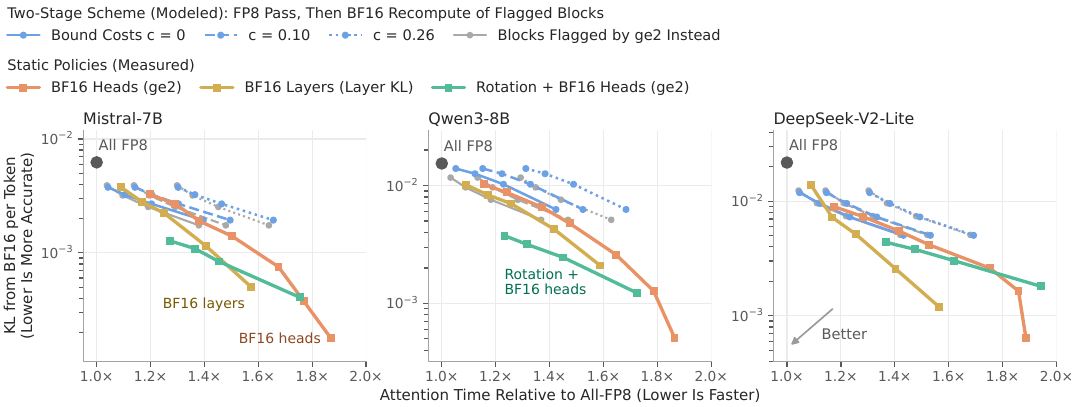}
\caption{Once its error bound has a cost, an in-kernel two-stage scheme is less accurate (it has a higher KL divergence from the BF16 model) than a static policy that takes the same time. The two-stage scheme runs attention in FP8 with an error bound per block and recomputes the flagged blocks in BF16, at a modeled time of $1+c+r\,w$, where $c$ is the cost of the bound, $w$ the share of recomputed blocks, and $r$ the ratio of BF16 to FP8 time. Static policies keep fixed heads or layers in BF16, and their time and KL divergence are measured. The data predate lean P.}
\label{fig:twostage}
\end{figure*}

\paragraph{Error checks at run time}
Instead of choosing heads in advance, a kernel can detect large errors while it runs. Figure~\ref{fig:twostage} models a kernel that computes an error bound for each block in the FP8 pass and recomputes the flagged blocks in BF16. Once the error bound costs $10\%$ of the FP8 time or more, every point of the modeled two-stage scheme lies above the curves of the static policies, which keep fixed heads or layers in BF16, so in this model the two-stage scheme is less accurate than a static policy that takes the same time. Only a bound that costs nothing comes out ahead, and only when it recomputes few blocks.

\paragraph{Retrieval test}
A retrieval test plants a $5$-digit passcode at $10$ to $90\%$ depth in the second half of a test book and uses a cluster-robust $t$ interval over books. Its intervals overlap for all policies, so we do not use it to select policies.

\section{Related Work}
\label{sec:related}

FlashAttention~\citep{dao2022flash} computes attention block by block with a running softmax, so the score matrix never leaves on-chip memory, and FlashAttention-2~\citep{dao2023flash2} assigns tiles of query rows to groups of threads to parallelize over the sequence. FlashAttention-3~\citep{shah2024flash3} uses the asynchronous copies and matrix instructions of Hopper GPUs. It adds FP8 attention with block quantization (one scale factor per block of each tensor) and with a Hadamard transform of $Q$ and $K$, which it calls incoherent processing. FlashAttention-4~\citep{shah2026flash4} targets Blackwell GPUs, where matrix throughput grew faster than the throughput of the exponential unit. It computes part of the exponentials with a polynomial on the FMA units. \sys{}'s lean P, which computes each FP8 probability from its score with one multiply-add and an integer conversion, removes the exponential from the FP8 kernel and uses the interpolation of Schraudolph's fast exponential~\citep{schraudolph1999exp} to write the e4m3 code, the bit pattern of the FP8 value, directly. Softermax~\citep{stevens2021softermax} and I-BERT~\citep{kim2021ibert} also compute the exponential of the softmax with fixed-point or integer arithmetic. SageAttention~\citep{zhang2024sage,zhang2025sage2} quantizes $Q$ and $K$ to 8 or 4 bits after subtracting the mean of $K$ and computes $PV$ in FP8 or FP16. SageAttention3~\citep{zhang2025sage3} normalizes each row of probabilities and quantizes them to FP4 with a scale per $16$ values, the closest prior work to per-block scaling of probabilities. The microscaling formats~\citep{rouhani2023mx} share one power-of-two scale among each block of $32$ values, and the scaled MFMA (matrix fused multiply-add instruction) of CDNA4 applies such scales in hardware, which \sys{} uses to remove the per-block factor at no cost.

HipKittens~\citep{hipkittens} brings the tile abstractions of ThunderKittens~\citep{spector2024thunderkittens} to CDNA3 and CDNA4 GPUs and studies two wave schedules for them. The design of \sys{}'s BF16 kernel at head dimension $128$ started from the HipKittens attention kernel. Rotations~\citep{ashkboos2024quarot} and smoothing~\citep{xiao2023smoothquant} reduce the quantization error of activations with outliers (Section~\ref{sec:res-e2e}). Linear attention~\citep{katharopoulos2020linear} and state-space models~\citep{gu2024mamba} replace the softmax with operations whose cost grows linearly with the sequence length, but attention remains a core building block of frontier models.

\section{Discussion}
\label{sec:discussion}
\label{sec:recs}

\paragraph{Lean P and the number format}
Lean P, \sys{}'s computation of each FP8 probability with one multiply-add and an integer conversion instead of an exponential, replaces the evaluation of a function with a property of the number format. A floating-point code already stores the base-2 logarithm of its value in its exponent bits and interpolates linearly between powers of two in its mantissa bits, so the code of $2^x$ needs only the linear map of Equation~\ref{eq:lean} and a rounding to an integer. FlashAttention-4 also avoids the hardware exponential, but it evaluates a polynomial in FP32 and still converts the result to the output format~\citep{shah2026flash4}. Lean P skips both steps, because the output format itself is the approximation. Its error is therefore set by the coarseness of the format. After rounding, lean P is within \zn{lean/pwl/relerr}\% of $p$, against \zn{leanmath/exact/max}\% for exact P, which rounds the exact $2^x$ to e4m3, and in full models this difference is small next to the error of quantizing $Q$, $K$, and $V$ (Section~\ref{sec:res-e2e}). The same construction applies to any format whose code grows by a fixed step each time its value doubles, which includes e5m2 and, with coarser steps, FP4, although we have not tested these formats.

\paragraph{Remaining limits of FP8 attention on the MI355X}
At $32$K tokens on the Gaussian inputs of kernel benchmarks, \sys{}'s FP8 kernel at $\dqk{=}128$ reaches \zn{tfs/mha/fp8-d128/32k/zyfa}\,TFLOP/s, \zn{peakpct/mha/fp8-d128/32k/zyfa}\% of the datasheet peak and \zn{mixpct/mha/fp8-d128/32k}\% of a synthetic loop with its instruction mix. Three gaps remain, and each points to a different improvement. The first is the clock. Under sustained load, after seconds of back-to-back launches, the kernel runs at about \zn{hl/fp8/real/zyfa/sclk/min}\,MHz on captured activations (the $q$, $k$, and $v$ of real model layers) at $16$K tokens, against the $2.4$\,GHz that the datasheet peak assumes, and even a loop of FP8 MFMAs alone settles at \zn{ceil/fp8-32x32/sclk}\,GHz. Most of the distance to the datasheet peak therefore comes from the limit on current that sets the clock (Section~\ref{sec:res-clock}) rather than from the schedule, and kernels that switch fewer bits per FLOP could gain directly. The second is the transitions between query tiles. The matrix pipe is busy for \zn{asm/lean/mfmabusy}\% of the cycles in steady state, in the middle of the key loop of one query tile, but for \zn{asm/lean/mfmabusy/loop}\% over the whole key loop, so overlapping the end of one tile with the start of the next could recover much of this difference. The third is the work that remains on the MFMA wave, the one of the two waves on a SIMD that issues the matrix instructions in a phase and, with lean P, sets the length of the phase. That wave also computes the row maximum and the rescale, so layers that rescale often take longer (Appendix~\ref{app:layers}), and a schedule that moves this work to the softmax wave, which computes the probabilities, would make the kernel less sensitive to the data.

\paragraph{Comparison with NVIDIA GPUs}
We did not measure NVIDIA GPUs, so this comparison uses published numbers, which come from different inputs and clocks. FlashAttention-4 reports up to $1613$\,TFLOP/s for BF16 attention on the B200, $71\%$ of its BF16 peak~\citep{shah2026flash4}, and FlashAttention-3 reports close to $1.2$\,PFLOP/s for FP8 attention on the H100~\citep{shah2024flash3}. On Gaussian inputs at $32$K tokens, \sys{} reaches \zn{tfs/mha/bf16-d128/32k/zyfa}\,TFLOP/s in BF16 (\zn{peakpct/mha/bf16-d128/32k/zyfa}\% of the BF16 peak of the MI355X) and \zn{tfs/mha/fp8-d128/32k/zyfa}\,TFLOP/s in FP8 (\zn{peakpct/mha/fp8-d128/32k/zyfa}\%). Part of the lower share of peak on the MI355X comes from its clock, which falls from the $2.4$\,GHz of the datasheet peak to \zn{hl/fp8/real/zyfa/sclk/min}\,MHz for \sys{}'s FP8 kernel under sustained load. Both vendors face the same imbalance, in which matrix throughput has grown faster than the throughput of the exponential, and FlashAttention-4 and lean P both answer it by avoiding the hardware exponential. They differ in how the softmax overlaps the matrix work. Blackwell GPUs track asynchronous matrix instructions in hardware and keep accumulators in a separate tensor memory, while on the MI355X the instruction schedule alone provides the overlap. A comparison on the same inputs at sustained clocks, where the data-dependent clock of Section~\ref{sec:limits-clock} would apply on both sides, remains future work.

\paragraph{Recommendations}
GPU kernel developers, especially those working on numerically sensitive operations like quantized model blocks, should develop with real model activations and at sustained clocks, the lower steady clock that the GPU reaches after seconds of back-to-back launches, because the clock depends on data distribution and its rate of change. To lower quantization error, target the quantization of the inputs, for example with a Hadamard rotation of $Q$ and $K$ and K smoothing, which cost one extra pass over $Q$ and $K$. Because $V$ adds the least input error, keeping $QK^{\top}$ in BF16 and running only $PV$ in FP8 would remove most of the input error for three quarters of the matrix cycles of BF16. Reduce the vector work per score before adding matrix throughput. A timing proxy that issues the $QK^{\top}$ MFMAs in FP4 and halves the LDS reads of $K$ raises the throughput by at most \zn{fp4/qkproxy/maxgain}\% at $32$K. Keep packed FP32 instructions out of a softmax that runs next to MFMAs. Use head dimensions of at least $128$, because the matrix cycles per block grow with $\dqk{+}\dv$ while the softmax cycles do not, so at head dimension $64$ even lean P keeps the vector pipe about as busy as the matrix pipe.

\paragraph{Limitations}
\label{app:limits}
The kernels compute the causal forward pass for equal query and key lengths that are multiples of the tile size. They cover dense prefill and do not cover decoding, including the form of MLA used in decoding, paged caches, or packed batches of variable length (Table~\ref{tab:support}). All experiments use one GPU model, one software stack, and one node. The comparisons with AITER describe version~0.1.19, and at $\dqk{=}192$ its FP8 kernel issues older matrix instructions with half the throughput per clock. The end-to-end study uses three models, PG-19, and no downstream tasks, and its attention times come from kernel-only windows inside the model. 
The instruction costs and the data-dependent clock follow from the MI355X and its platform settings, and the clock may differ with other firmware or power settings. The FP8 kernels propagate a NaN input to more rows than the reference (Section~\ref{app:nan}), which in packed batches would reach other requests. With few query heads per GPU, as under tensor parallelism, a call at short and medium lengths has too few tile pairs to fill the $256$ CUs, a case that our benchmarks with at least $32$ heads do not cover. K smoothing with a per-sequence mean cancels in the softmax only if every key of a row uses the same mean, which chunked prefill and prefix caching break unless the mean is fixed in advance. On the MI300X, which uses the FNUZ variant of e4m3 and has no scaled MFMA, lean P needs a different offset and per-block scaling needs extra vector work.\footnote{The FNUZ variant has an exponent bias of $8$ instead of $7$, so the same code stands for half the value, and lean P must add $8$ to its offset.} We did not compare the ping-pong schedule with schedules in which each wave interleaves its own MFMAs and softmax~\citep{hipkittens}.

\section{Conclusion}
\label{sec:conclusion}

On the MI355X, FP8 attention is limited by the vector work of the softmax and by its clock speed, which depends on the input data. By writing each FP8 probability directly from its score and hiding the softmax behind matrix instructions, \sys{}'s FP8 kernels reach \zn{hl/fp8/real/ratio/min}--\zn{hl/fp8/real/ratio/max}\x{} the throughput of the default kernels of AITER, AMD's library of inference kernels, on activations captured from real models and timed at sustained clocks, after seconds of back-to-back launches. In full models most of the FP8 error comes from quantizing the inputs and is spread over most heads, so keeping the most sensitive heads in BF16 is not a useful middle ground, while a Hadamard rotation of $Q$ and $K$ roughly halves the error from quantizing the inputs at almost no cost.

\section*{Acknowledgements}
We thank Beren Millidge for detailed feedback on drafts of this paper, and the developers of AITER, Composable Kernel, and HipKittens, whose open-source kernels served as our baselines and as starting points for our BF16 kernels. All experiments ran on an AMD Instinct MI355X node rented from Vultr.

\section*{Author Contributions}
\noindent\textbf{Quentin Anthony} designed and wrote the FP8 kernels, including the assembly generator and its instruction schedule. He carried out the numerical analysis and wrote the paper.

\noindent\textbf{Ganesh Bikshandi} carried out the initial exploration and laid the groundwork for the BF16 kernels.

\noindent\textbf{Praneeth Medepalli} contributed foundational notes on the numerics of FP8 attention, on which the numerical analysis of this paper builds.

\noindent\textbf{Stephen Ebert} contributed foundational notes on the numerics of FP8 attention, on which the numerical analysis of this paper builds.

\bibliographystyle{plainnat}
\bibliography{refs}

\clearpage
\appendices

\section{Speed of Every Layer}
\label{app:layers}

\begin{figure*}[t]
\centering
\includegraphics[width=\textwidth]{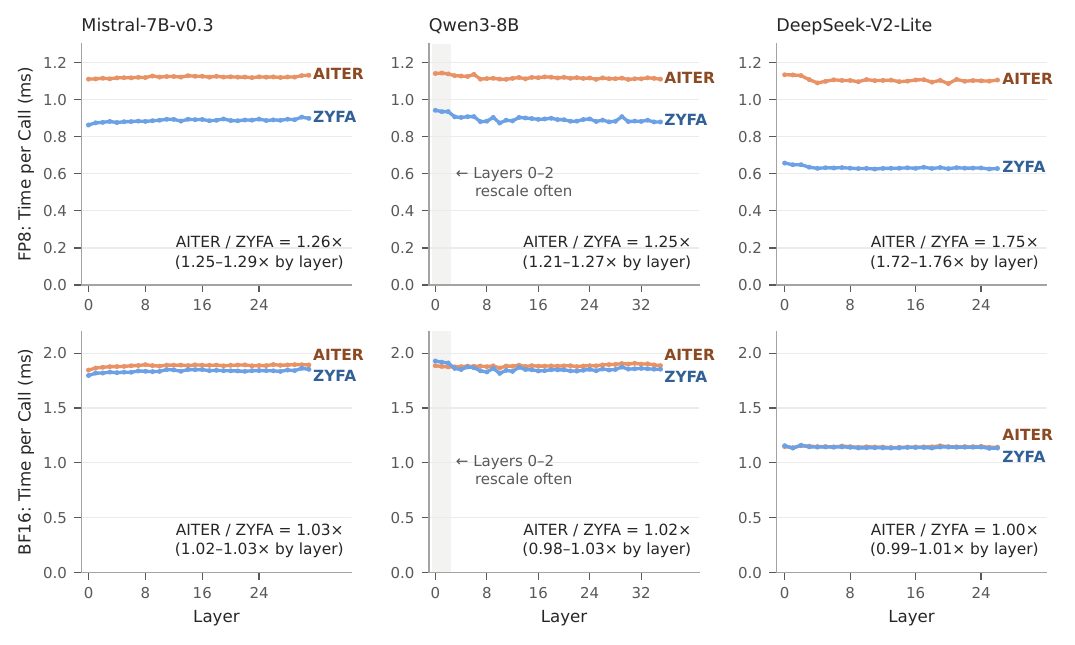}
\caption{The speedup of \sys{} over AITER barely changes between layers. Each line is the time of one attention call of \sys{} or of AITER's default kernel in every layer, measured inside a full forward pass of each model at sequence length $S{=}16$K on two PG-19 documents. The label in each panel gives AITER's time divided by \sys{}'s time, as the median and the range over layers, so a value above $1$ means that \sys{} takes less time. In the shaded first layers of Qwen3-8B the running row maximum keeps growing, so \sys{} rescales its output accumulator far more often than in the other layers.}
\label{fig:layers}
\end{figure*}

Table~\ref{tab:real} times one captured layer per model, from the $q$, $k$, and $v$ of one real layer. To check that these layers are representative, we timed every layer of each model inside a full forward pass, with timing windows that cover only the attention kernels (Section~\ref{app:e2emethods}), on two PG-19 documents at $8$K, $16$K, and $32$K tokens. We also timed four or five layers per model alone at sustained clocks, after seconds of back-to-back launches of one kernel, on captured activations, with the protocol of Table~\ref{tab:real}. Figure~\ref{fig:layers} shows the time of every layer at $16$K, and Table~\ref{tab:layers} summarizes all three lengths.

\begin{table}[t]
\centering
{\footnotesize
\setlength{\tabcolsep}{3pt}
\begin{tabular}{llccc}
\toprule
model & & $8$K & $16$K & $32$K \\
\midrule
Mistral-7B & FP8 & \zn{lyr/mistral/fp8/8k/med} (\zn{lyr/mistral/fp8/8k/min}--\zn{lyr/mistral/fp8/8k/max}) & \zn{lyr/mistral/fp8/16k/med} (\zn{lyr/mistral/fp8/16k/min}--\zn{lyr/mistral/fp8/16k/max}) & \zn{lyr/mistral/fp8/32k/med} (\zn{lyr/mistral/fp8/32k/min}--\zn{lyr/mistral/fp8/32k/max}) \\
Qwen3-8B & FP8 & \zn{lyr/qwen3/fp8/8k/med} (\zn{lyr/qwen3/fp8/8k/min}--\zn{lyr/qwen3/fp8/8k/max}) & \zn{lyr/qwen3/fp8/16k/med} (\zn{lyr/qwen3/fp8/16k/min}--\zn{lyr/qwen3/fp8/16k/max}) & \zn{lyr/qwen3/fp8/32k/med} (\zn{lyr/qwen3/fp8/32k/min}--\zn{lyr/qwen3/fp8/32k/max}) \\
DeepSeek & FP8 & \zn{lyr/deepseek/fp8/8k/med} (\zn{lyr/deepseek/fp8/8k/min}--\zn{lyr/deepseek/fp8/8k/max}) & \zn{lyr/deepseek/fp8/16k/med} (\zn{lyr/deepseek/fp8/16k/min}--\zn{lyr/deepseek/fp8/16k/max}) & \zn{lyr/deepseek/fp8/32k/med} (\zn{lyr/deepseek/fp8/32k/min}--\zn{lyr/deepseek/fp8/32k/max}) \\
\midrule
Mistral-7B & BF16 & \zn{lyr/mistral/bf16/8k/med} (\zn{lyr/mistral/bf16/8k/min}--\zn{lyr/mistral/bf16/8k/max}) & \zn{lyr/mistral/bf16/16k/med} (\zn{lyr/mistral/bf16/16k/min}--\zn{lyr/mistral/bf16/16k/max}) & \zn{lyr/mistral/bf16/32k/med} (\zn{lyr/mistral/bf16/32k/min}--\zn{lyr/mistral/bf16/32k/max}) \\
Qwen3-8B & BF16 & \zn{lyr/qwen3/bf16/8k/med} (\zn{lyr/qwen3/bf16/8k/min}--\zn{lyr/qwen3/bf16/8k/max}) & \zn{lyr/qwen3/bf16/16k/med} (\zn{lyr/qwen3/bf16/16k/min}--\zn{lyr/qwen3/bf16/16k/max}) & \zn{lyr/qwen3/bf16/32k/med} (\zn{lyr/qwen3/bf16/32k/min}--\zn{lyr/qwen3/bf16/32k/max}) \\
DeepSeek & BF16 & \zn{lyr/deepseek/bf16/8k/med} (\zn{lyr/deepseek/bf16/8k/min}--\zn{lyr/deepseek/bf16/8k/max}) & \zn{lyr/deepseek/bf16/16k/med} (\zn{lyr/deepseek/bf16/16k/min}--\zn{lyr/deepseek/bf16/16k/max}) & \zn{lyr/deepseek/bf16/32k/med} (\zn{lyr/deepseek/bf16/32k/min}--\zn{lyr/deepseek/bf16/32k/max}) \\
\bottomrule
\end{tabular}}
\caption{AITER's time divided by \sys{}'s time for the attention of every layer, measured inside a full forward pass of the model at each sequence length, as the median and, in parentheses, the range over the layers of each model, on two PG-19 documents. Values above $1$ mean that \sys{} takes less time. DeepSeek is DeepSeek-V2-Lite.}
\label{tab:layers}
\end{table}

\paragraph{Small differences between layers}
Every layer of a model computes attention of the same shape, so in every layer each kernel issues the same instructions and moves the same bytes. Only the data differ, and the data can change the time of a kernel in two ways. They set how often the running maximum of a row grows, which decides how often the kernel rescales its output (Section~\ref{sec:common}), and they set how much the values in the arithmetic units change from one instruction to the next, which decides the clock (Section~\ref{sec:limits-clock}). In most layers both effects are small, so the speedup is close to a constant of the model. At $16$K, \sys{}'s FP8 kernel achieves \zn{lyr/mistral/fp8/16k/min}--\zn{lyr/mistral/fp8/16k/max}\x{} the throughput of AITER's across the layers of Mistral-7B, \zn{lyr/qwen3/fp8/16k/min}--\zn{lyr/qwen3/fp8/16k/max}\x{} across those of Qwen3-8B, and \zn{lyr/deepseek/fp8/16k/min}--\zn{lyr/deepseek/fp8/16k/max}\x{} across those of DeepSeek-V2-Lite. The small differences that remain are partly systematic rather than noise. Measured separately on the two documents, the per-layer speedups have a Pearson correlation of \zn{lyr/docrho/qwen3} on Qwen3-8B, \zn{lyr/docrho/mistral} on Mistral-7B, and \zn{lyr/docrho/deepseek} on DeepSeek-V2-Lite, whose layers differ the least.

\paragraph{The first layers of Qwen3-8B}
The exception is the first three layers of Qwen3-8B. There \sys{}'s FP8 kernel achieves only \zn{lyr/qwen3/fp8/first3/rmin}--\zn{lyr/qwen3/fp8/first3/rmax}\x{} the throughput of AITER's, and its BF16 kernel takes longer than AITER's (\zn{lyr/qwen3/bf16/first3/rmin}\x{}). These layers rescale the output accumulator far more often than the others. Both kernels of \sys{} visit the keys of a row from key~$0$ toward the diagonal. In most layers an early key, such as an attention sink on the first token, holds the largest score of a row, so the running maximum is set almost at once and rarely grows afterwards. When the largest scores of a row lie close to the query instead, for example in heads that attend mostly to nearby tokens, the maximum keeps growing until the kernel reaches them. In these three layers the FP8 kernel rescales on \zn{lyr/qwen3/rescale/first3/min}--\zn{lyr/qwen3/rescale/first3/max}\% of its tiles, against a median of \zn{lyr/qwen3/rescale/rest/med}\% in the other layers of the model. Lazy rescaling therefore turns a fixed cost into a cost that depends on the data. Lazy rescaling saves time when an early key sets the maximum, and it loses that saving when the maximum keeps growing. On these layers \sys{}'s FP8 kernel takes \zn{lyr/qwen3/fp8/first3/zyfa}\% longer than its median over the other layers, against \zn{lyr/qwen3/fp8/first3/aiter}\% for AITER's, and in BF16 \sys{}'s kernel takes \zn{lyr/qwen3/bf16/first3/zyfa}\% longer while AITER's time does not change (\zn{lyr/qwen3/bf16/first3/aiter}\%). The FP8 kernel is this sensitive because with lean P the rescale lies on its critical path. The MFMA wave, the wave that issues the matrix instructions in a phase, computes the rescale between its matrix instructions, and with lean P that wave sets the length of every phase (Section~\ref{sec:asm}), so each rescale delays the matrix pipe directly. Across the layers of Qwen3-8B, the FP8 speedup and the share of rescaled tiles have a Pearson correlation of \zn{lyr/qwen3/fp8/rescalecorr}. The first layer of DeepSeek-V2-Lite, its slowest, also rescales unusually often, on \zn{lyr/deepseek/rescale/l0}\% of its tiles. A larger rescaling threshold for such layers, or a schedule that moves the rescale off the MFMA wave, could recover the difference.

\paragraph{Inside the model and alone}
Timed alone at sustained clocks, after seconds of back-to-back launches of one kernel, the FP8 speedup of the same layers is lower on Mistral-7B and Qwen3-8B, \zn{lyr/alone/mistral/fp8/min}--\zn{lyr/alone/mistral/fp8/max}\x{} and \zn{lyr/alone/qwen3/fp8/min}--\zn{lyr/alone/qwen3/fp8/max}\x{} against \zn{lyr/mistral/fp8/16k/med}\x{} and \zn{lyr/qwen3/fp8/16k/med}\x{} inside the model, and about the same on DeepSeek-V2-Lite, \zn{lyr/alone/deepseek/fp8/min}--\zn{lyr/alone/deepseek/fp8/max}\x{} against \zn{lyr/deepseek/fp8/16k/med}\x{}. The two settings differ in three ways. Their inputs come from different text, PG-19 documents inside the model and WikiText-2 alone. Inside the model each call follows the matrix multiplications of its layer and a short busy-wait, while alone it follows the previous launch of the same kernel. Finally, only the sustained runs keep one kernel busy for seconds, long enough for the clock to settle at the level that its own operands allow (Section~\ref{sec:limits-clock}). We have not yet determined which of these differences raises the speedup inside the model. For Mistral-7B and Qwen3-8B the sustained speedups of Table~\ref{tab:real} are the lower of the two measurements, so the table understates rather than overstates the speedup that these models see.

\paragraph{Reproducibility}
Measured again alone at sustained clocks in a later session, the layers of Table~\ref{tab:real} give \zn{lyr/alone/mistral/paperlayer}\x{}, \zn{lyr/alone/qwen3/paperlayer}\x{}, and \zn{lyr/alone/deepseek/paperlayer}\x{} in FP8, against \zn{lyr/paper/mistral}\x{}, \zn{lyr/paper/qwen3}\x{}, and \zn{lyr/paper/deepseek}\x{} in the table. Each of these layers lies within the range of the other layers of its model timed alone, so the layers reported in Section~\ref{sec:res-real} are representative of their models.

\end{document}